\documentclass[reprint,aps,pra,superscriptaddress,noeprint]{revtex4-2}

\usepackage{array}
\newcolumntype{C}[1]{>{\centering\arraybackslash}m{#1}}

\usepackage{physics}
\usepackage[normalem]{ulem}
\usepackage{siunitx}
\usepackage{amsmath}
\usepackage{amsbsy}
\usepackage{amssymb} 
\usepackage{graphicx}
\usepackage{float}
\usepackage{bm}
\usepackage{placeins}

\usepackage{comment}
\usepackage[caption=false]{subfig}
\usepackage{hyperref}
\hypersetup{
	colorlinks=true,
	linkcolor=blue,
	citecolor=green,
	filecolor=magenta,
	urlcolor=blue
}
\usepackage{siunitx}
\usepackage[dvipsnames]{xcolor}
\usepackage{booktabs}
\usepackage{multirow}

\usepackage{xcolor}
\usepackage{tikz}

\definecolor{RegIcol}{HTML}{7B2CBF}
\definecolor{RegIIcol}{HTML}{6E6E6E}
\definecolor{RegIIIcol}{HTML}{FB9A99}

\hypersetup{
	colorlinks=true,
	linkcolor=blue,
	citecolor=purple,
	filecolor=darkgreen,
	urlcolor=blue
}

\begin{document}
\title{Dissipative generation of spin squeezing in the resolved vacuum Rabi splitting limit}

\author{Edwin Chaparro}
\email[Corresponding author: ]{Edwin.Chaparro@colorado.edu}
\affiliation{JILA, University of Colorado and National Institute of Standards and Technology, and Department of Physics, University of Colorado, Boulder, Colorado 80309, USA}
\affiliation{Center for Theory of Quantum Matter, University of Colorado, Boulder, Colorado 80309, USA}

\author{Eric Yilun Song}
\affiliation{JILA, University of Colorado and National Institute of Standards and Technology, and Department of Physics, University of Colorado, Boulder, Colorado 80309, USA}
\affiliation{Center for Theory of Quantum Matter, University of Colorado, Boulder, Colorado 80309, USA}

\author{Diego Barberena}
\affiliation{JILA, University of Colorado and National Institute of Standards and Technology, and Department of Physics, University of Colorado, Boulder, Colorado 80309, USA}
\affiliation{Center for Theory of Quantum Matter, University of Colorado, Boulder, Colorado 80309, USA} 
\affiliation{T.C.M. Group, Cavendish Laboratory, University of Cambridge, J.J. Thomson Avenue, Cambridge CB3 0US, UK}

\author{James K. Thompson}
\affiliation{JILA, University of Colorado and National Institute of Standards and Technology, and Department of Physics, University of Colorado, Boulder, Colorado 80309, USA}

\author{Ana Maria Rey}
\affiliation{JILA, University of Colorado and National Institute of Standards and Technology, and Department of Physics, University of Colorado, Boulder, Colorado 80309, USA}
\affiliation{Center for Theory of Quantum Matter, University of Colorado, Boulder, Colorado 80309, USA}

\author{Jeremy T. Young}
\affiliation{Institute of Physics, University of Amsterdam, 1098 XH Amsterdam, the Netherlands}
\affiliation{JILA, University of Colorado and National Institute of Standards and Technology, and Department of Physics, University of Colorado, Boulder, Colorado 80309, USA}
\affiliation{Center for Theory of Quantum Matter, University of Colorado, Boulder, Colorado 80309, USA}

\date{\today}

\begin{abstract}
Harnessing dissipation in the presence of strong symmetries has recently emerged as a promising route for generating entanglement in atomic clocks. However, previous proposals relied on regimes where cavity photons can be adiabatically eliminated, significantly 
limiting their applicability to experimentally relevant cavity-QED regimes that lie  in or near the resolved vacuum Rabi splitting regime. Here we show that symmetry-protected dissipative spin squeezing can  be realized even when cavity photons actively participate in the dynamics, extending the experimental relevance of the protocol. We study a three-level ensemble of 
$^{87}\mathrm{Sr}$ atoms coupled to an optical cavity in the resolved vacuum Rabi splitting regime and demonstrate that, with smooth ramps of the drive amplitude and detunings, the driven-dissipative dynamics enters a stable low-photon regime in which nonadiabatic cavity excitations and sector-resolving photon leakage can be controlled.
Within this low-photon operating window, sector-resolving photon leakage is
suppressed and the sector-dependent geometric phase realizes effective
one-axis twisting. At the end of the protocol the entanglement can also be efficiently transferred  directly onto the long-lived clock states by turning  the drive  off. For experimentally realistic parameters,  we theoretically show  that  more than $25\,\mathrm{dB}$
of squeezing  can be generated   for 
$10^5$ atoms, closely saturating the ideal one-axis twisting scaling 
$\xi_{\min}^2 \propto N^{-2/3}$. At fixed cooperativity, the optimized squeezing remains broadly comparable to
the unresolved-regime implementation, while the resolved-regime implementation
reaches comparable squeezing on a substantially shorter physical timescale.
These results establish symmetry-protected dissipative dynamics as a practical route to beyond  the standard-quantum-limit performance in optical-clock platforms.

\end{abstract}

\pacs{}

\maketitle

\section{Introduction}

Quantum metrology leverages entangled many-particle states to surpass the standard quantum limit (SQL). For \(N\) uncorrelated atoms, the phase uncertainty scales as \(\Delta\phi_{\mathrm{SQL}}\!\sim\!1/\sqrt{N}\), whereas suitably entangled states can approach the Heisenberg limit \(\Delta\phi_{\mathrm{HL}}\!\sim\!1/N\) \cite{Wineland1992,Kitagawa1993,Wineland1994,Pezze2018}. Optical lattice clocks based on neutral strontium have already demonstrated fractional frequency uncertainties at the \(10^{-19}\) level by interrogating the ultra-narrow \({}^{1}\mathrm{S}_{0}\leftrightarrow{}^{3}\mathrm{P}_{0}\) transition in large ensembles of uncorrelated atoms \cite{Bloom2014,Ludlow2015,Schulte2020,Aeppli2024,Fortier2026}.
However, these state-of-the-art systems are approaching the regime in which quantum projection noise becomes a key limitation to further improvements in clock precision. Advancing the next generation of optical clocks therefore requires techniques that generate non-classical spin-squeezed states \cite{Kitagawa1993,Wineland1992,Wineland1994}, and reduce the phase variance below the SQL \cite{Pedrozo-Penafiel2020,yang2025clockprecisionstandardquantum,Robinson2024}.

Cavity quantum electrodynamics (QED) provides a powerful platform for engineering 
collective atom--photon interactions \cite{Kimble1998}. In optical-clock 
platforms, these interactions have enabled several routes to metrological gain: 
light-shift-induced shearing, or one-axis twisting, of the collective spin 
\cite{Leroux2010,Schleier-Smith2010,Greve2022}; coherent photon-mediated 
exchange interactions on optical transitions \cite{Norcia2018,LewisSwan2018}; 
and quantum nondemolition (QND) measurements in optical cavities 
\cite{Chen2014,Cox2016,Hosten2016,yang2025clockprecisionstandardquantum,
Robinson2024,Hobson2019}. A complementary route uses unmonitored  dissipation
\cite{DallaTorre2013,Lee2014a,Fernandez-Lorenzo2017,Groszkowski2022,Pavlov2023,Barberena2023,GonzlezTudela2013,Wolfe2014DrivenSuperradiance,Somech2024,Anjun2025} to drive the system into a squeezed steady state.

It was recently proposed that by exploiting strong symmetries of the driven-dissipative dynamics, a collective spin can be protected in the steady state and detuning the drive realizes a dissipative Berry phase that effectively implements one-axis twisting for the protected collective spin \cite{JeremyOAT}. 
Here, by a strong symmetry \cite{Prosen2010,Buca2012,Lieu2020} we mean a conserved quantity that commutes with both the Hamiltonian and the jump operators of the Lindblad master equation, so that the open-system dynamics decomposes into symmetry sectors that do not mix under quantum jumps. Crucially, in contrast to other dissipative proposals \cite{Barberena2019,Somech2024,DallaTorre2013,Lee2014a,Fernandez-Lorenzo2017,Groszkowski2022,Pavlov2023,Barberena2024,GonzlezTudela2013,Wolfe2014DrivenSuperradiance}, this protocol can  operate away from critical points, avoiding critical slowdown. This enables rapid state preparation and direct mapping of the generated entanglement onto long-lived clock states simply by turning on and off the drive and detunings.

In the original proposal \cite{JeremyOAT}, photons were assumed to leave the
cavity faster than the collective spin dynamics, allowing the cavity mode to be
adiabatically eliminated.
However, this assumption is not necessarily satisfied for the
\({}^{1}\mathrm{S}_{0}\) to \({}^{3}\mathrm{P}_{1}\) transition in
\({}^{87}\mathrm{Sr}\), one of the current candidates for setting frequency
standards \cite{Cline2025,Eric_CRF}. The
\({}^{1}\mathrm{S}_{0}\)--\({}^{3}\mathrm{P}_{1}\) transition is convenient for
cavity dynamics because it is about three orders of magnitude narrower than
typical dipole-allowed transitions, while at the same time being six orders of
magnitude faster than the clock transition, thus allowing for convenient state
manipulation and rapid generation of entanglement.

The trade-off is that, even with a moderate atom number, the cavity field remains an active dynamical degree of freedom \cite{Winchester2017,Cline2025,Eric_CRF}, placing the system in or near the resolved vacuum Rabi splitting (RVRS) regime, where the resulting polaritonic dynamics is not, in general, captured by a dynamical adiabatic-elimination description. This raises the central question of this work: whether the strong-symmetry squeezing mechanism of Ref.~\cite{JeremyOAT} can remain effective when the photons that mediate the interaction also participate coherently in the dynamics and can carry sector-dependent information into the output field. 

We show that smooth ramps of the drive amplitude and detunings keep the dynamics in a stable low-photon operating regime,
suppressing nonadiabatic spin--photon excitations and sector-resolving photon leakage while preserving the effective OAT mechanism.
Although utilizing the direct coupling between the cavity and the clock transition would allow the cavity to be adiabatically eliminated and eliminate the need for ramps, the squeezing timescales would be several orders of magnitude longer. Due to the longer timescales, the effects of various decoherence sources, such as atomic interactions, magnetic field fluctuations, Stark shifts, and light scattering from the lattice, would be much more severe, strongly limiting or preventing the generation of squeezing.

To make this question concrete, we consider a three-level ensemble in which the
relevant strong symmetry is associated with conservation of the population of
the driven \({}^{1}\mathrm{S}_{0}\)--\({}^{3}\mathrm{P}_{1}\) manifold, denoted by \(N_J\). This
conservation places the initial ground--clock coherent spin state into a
coherent superposition of \(N_J\) sectors. Because the driven steady state
depends on \(N_J\), different sectors accumulate different geometric phases;
the nonlinear dependence of this phase on \(N_J\) produces an effective
one-axis twisting interaction, as in Ref.~\cite{JeremyOAT}.

The role of the ramps is to make this geometric mechanism compatible
with RVRS dynamics. Smooth changes of the drive amplitude and detunings
keep the system on the low-photon, spin-polarized branch, suppressing
nonadiabatic spin--photon excitations and reducing the distinguishability
of the sector-conditioned output fields. This is the central RVRS
constraint: the resolved regime does not break the strong symmetry,
so \(N_J\) remains conserved, but cavity memory can allow the leaking
field to partially resolve different \(N_J\) sectors and dephase the
inter-sector coherences needed for geometric OAT. Within the low-photon
operating window, turning off the drive maps the generated squeezing
back onto the long-lived ground--clock manifold without additional
control pulses, following the same storage principle as in the UVRS
protocol~\cite{JeremyOAT}.

We analyze this operating window using a dissipative truncated-Wigner
approximation (DissTWA) applied to the full three-level atom--cavity
dynamics~\cite{Polkovnikov2010,Schachenmayer2015,Huber2022,Mink2022,Mink2023,Hosseinabadi2025},
and benchmark it against a Holstein--Primakoff effective
description~\cite{Holstein1940,Emary2003,Kurucz2010,JeremyOAT}.
This comparison determines when the full RVRS dynamics reduce to the
effective OAT picture used below to optimize the squeezing.

For experimentally relevant parameters in \(^{87}\mathrm{Sr}\) cavity-QED platforms, the protocol theoretically yields \(26\)--\(28\) dB of squeezing at \(N=10^5\). For larger ensembles, it approaches the characteristic one-axis twisting scaling \(\xi_{\min}^2 \propto N^{-2/3}\), as expected when collective coherent dynamics dominates over both collective dissipation and local dephasing. The vacuum-resolved limit is particularly favorable because, for realistic cavity linewidths and detunings, it shifts the balance between unitary squeezing and the relevant dephasing channels, including those not directly associated with cavity dissipation or spontaneous emission, thereby improving the attainable correlations while also reducing the time required to reach the optimal squeezing. 
As a result, the relevant state-preparation times remain short and compatible
with state-of-the-art experimental implementations, while sustaining beyond-SQL
performance across an experimentally relevant region of parameter space.

\subsection*{Organization of the paper}

The remainder of the paper is organized as follows. Section~\ref{sec:Model} introduces the three-level atom--cavity model,
describes  the hierarchy of energy scales  used throughout the paper,
and characterizes the unresolved vacuum Rabi splitting (UVRS) and resolved
vacuum Rabi splitting (RVRS) regimes.
Section~\ref{sec:strong_weak_symmetry}  reviews the distinction between strong and weak symmetries in Lindblad dynamics and identifies the conservation laws relevant to enforce the strong symmetry in our system.
Section~\ref{sec:Multistage} presents the multistage symmetry-protected squeezing protocol  and its geometric interpretation following Ref.~\cite{JeremyOAT}.
It  clarifies 
the role of dynamical cavity photons  in the RVRS regime 
and identifies the low-photon operating window used in the simulations in the following sections.
Section~\ref{sec:HP_effective} summarizes the Holstein--Primakoff effective spin description and gives the effective coherent squeezing and collective decoherence rates. Section~\ref{sec:Spin_squeezing_dyn} benchmarks the effective description against DissTWA simulations of the squeezing dynamics. It compares ramped and quenched protocols and identifies  optimal operating regimes. Section~\ref{sec:single_particle_decoh} incorporates single-particle decoherence and evaluates the optimized squeezing at fixed cooperativity. Finally, Section~\ref{outlook} summarizes our main results and discusses future directions.

\section{Model and Setup}
\label{sec:Model}

\begin{figure}[!t]
    \centering
    \includegraphics[scale=.295]{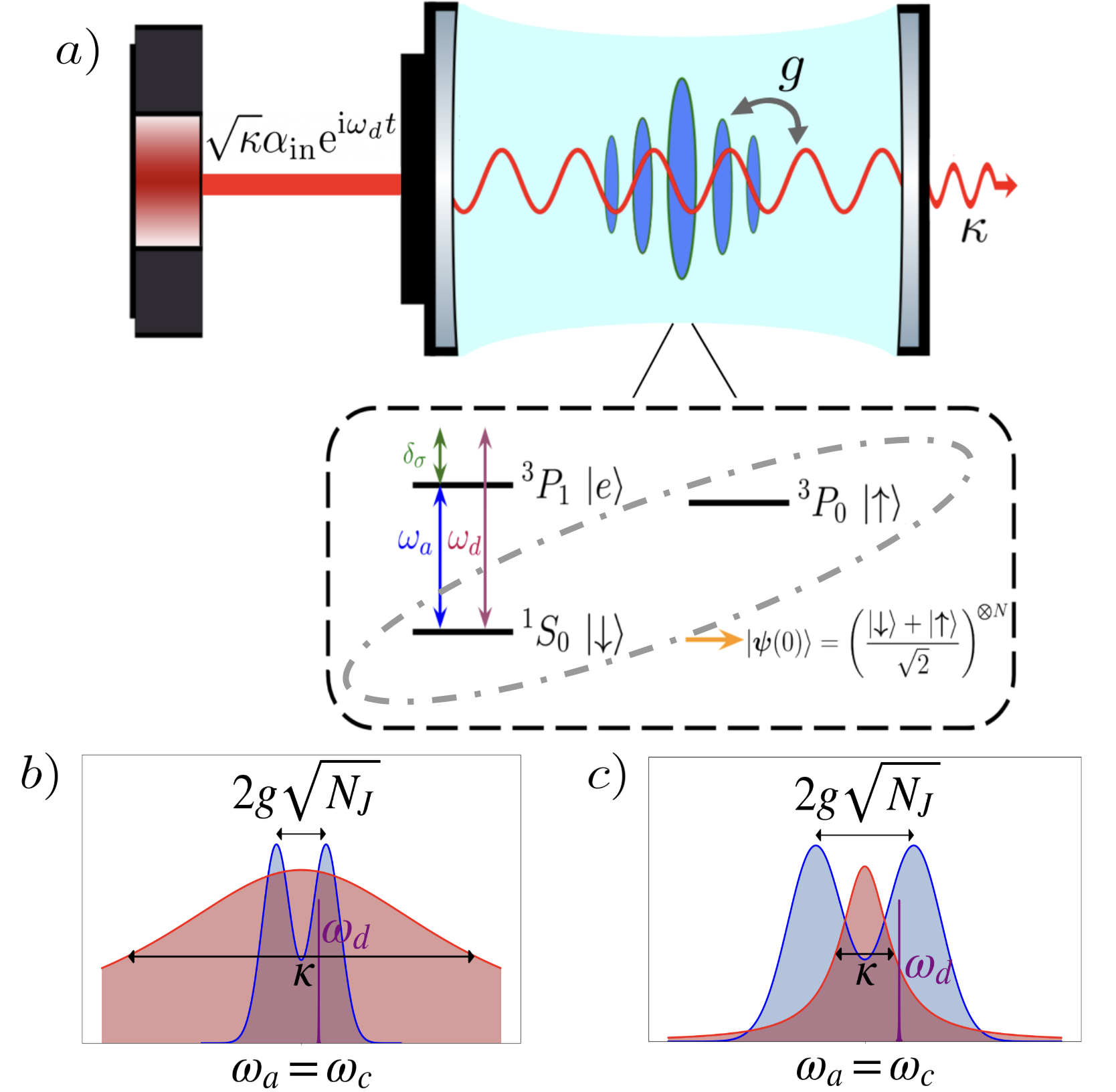}
    \caption{{\bf Cavity-QED platform and vacuum Rabi regimes.}
    (a) An ensemble of three-level atoms (ground \(\ket{\downarrow}\), clock \(\ket{\uparrow}\), optical \(\ket{e}\)) couples collectively to a single-mode cavity at rate \(g\). The input drive at frequency \(\omega_d\) and amplitude \(\sqrt{\kappa}\alpha_{\rm in}\) populates the mode, which loses photons at rate \(\kappa\) (equivalently, has linewidth \(\kappa\)).
    We take \(\omega_c=\omega_a\) unless stated otherwise; atoms are initially prepared in a coherent ground--clock superposition while the cavity starts in vacuum. 
    (b)--(c) The hybridization of the collective atomic excitation with the cavity photon produces polaritons split by the low-excitation vacuum Rabi frequency $(\Omega_{\mathrm{VRS}}=2g\sqrt{N_J})$. In the \textbf{unresolved vacuum Rabi splitting (UVRS)} limit (left), $(\kappa\gg\Omega_{\mathrm{VRS}})$, the broadened peaks overlap and the cavity acts as a fast, lossy mediator; in the \textbf{Resolved vacuum Rabi splitting (RVRS)} limit (right), $(\kappa\!\sim\!\Omega_{\mathrm{VRS}})$, the two peaks are spectrally separated and the cavity retains memory, modifying both steady states and transients.
    }
    \label{fig:lab_fig_energy_levels}
\end{figure}

We consider an ensemble of three-level atoms coupled to a single-mode optical cavity. Each atom \(i\) has a ground state \(\ket{\downarrow}_i\), a long-lived clock state \(\ket{\uparrow}_i\), and an additional excited state \(\ket{e}_i\). A relevant example is provided by the \({}^{1}\mathrm{S}_{0}\) (ground),
\({}^{3}\mathrm{P}_{0}\) (clock), and \({}^{3}\mathrm{P}_{1}\) (excited)
levels of alkaline-earth atoms
\cite{Cline2025}. The cavity mode with frequency \(\omega_c\) couples the   \(\ket{\downarrow}\!\leftrightarrow\!\ket{e}\)  transition. It leaks photons at rate \(\kappa\) and is driven at frequency \(\omega_d\) by an input field with complex amplitude \(\alpha_{\rm in}\), normalized so that
\(|\alpha_{\rm in}|^2\) is the incident photon flux and \(\sqrt{\kappa}\alpha_{\rm in}\) is the corresponding drive amplitude in the cavity equation of motion. 
The excited state $\ket{e}$ also decays spontaneously to $\ket{\downarrow}$ at rate $\gamma_{e\downarrow}$. The single-photon Rabi frequency on the \(\ket{\downarrow}\!\leftrightarrow\!\ket{e}\) transition is \(2g\), and the corresponding collective vacuum Rabi splitting of the \(\{|{\downarrow}\rangle,|e\rangle\}\) manifold, evaluated around the low-excitation state with the \(N_J\) atoms predominantly in \(\ket{\downarrow}\), is \(2g\sqrt{N_J}\), where \(N_J\) counts the atoms in the \(\{|{\downarrow}\rangle,|e\rangle\}\) manifold, as shown in Fig.~\ref{fig:lab_fig_energy_levels}(a). We choose the quantization axis and cavity polarization such that the driven cavity mode couples the $\left|\downarrow\right> \leftrightarrow |e\rangle$ transition, while the clock state $\left|\uparrow\right>$ remains effectively decoupled from the cavity dynamics \cite{PineiroOrioli2022}.

Under these conditions and setting $\hbar=1$, the dynamics of the density matrix of the full system, $\hat \rho$, is  governed by a master equation, 
\begin{subequations}
\label{eq:Hamil_cavity}
\begin{gather}
\partial_{t}\hat{\rho} = -{\rm i}[\hat{H},\hat{\rho}]+ \kappa\mathcal{D}[\hat{a}]\hat{\rho}+\gamma_{e\downarrow}\sum_i\mathcal{D}[\ket{\downarrow_i}\!\bra{e_i}]\hat{\rho},\\
\begin{aligned}
\hat{H} =~& g \left(\hat{J}^{+} \hat{a}  + \hat{J}^{-}\hat{a}^{\dagger}\right)
-{\rm i}\sqrt{\kappa}\alpha_{\mathrm in} \left(\hat{a}e^{{\rm i}\omega_d t}-\text{h.c.}\right) +\\
&\omega_c \hat{a}^\dagger\hat{a}+\omega_a \hat{J}^z,
\end{aligned}
\end{gather}
\end{subequations}
where $\mathcal{D}[\hat{O}]\hat{\rho}=\hat{O}\hat{\rho}\hat{O}^{\dagger}-\frac{1}{2}\{\hat{O}^{\dagger}\hat{O},\hat{\rho}\}$, and
$\hat{J}^{+}=\hat{J}^{x}+ i \hat{J}^{y}=\sum_i\ket{e_i}\!\bra{\downarrow_i}$,
 \(\hat{J}^{z}=\tfrac{1}{2}\sum_i\big(\ket{e_i}\!\bra{e_i}-\ket{\downarrow_i}\!\bra{\downarrow_i}\big)\) are   collective operators that  satisfy \([\hat{J}^\alpha,\hat{J}^\beta]={\rm i}\epsilon_{\alpha\beta\gamma}\hat{J}^\gamma\), where $\epsilon_{\alpha\beta\gamma}$ is the Levi-Civita symbol. The operators  \(\hat{a}^{\dagger}(\hat{a})\) are  the photon creation (annihilation) operators   of the relevant cavity mode. In what follows, we work in the bad-cavity regime, \(\kappa \gg \gamma_{e\downarrow}\).
We neglect decay and dephasing of the clock state $\left|\uparrow\right>$, whose rates are
expected to be much smaller than the cavity decay rate and the
$\left|e\right>\to \left|\downarrow\right>$ spontaneous-emission rate and negligible over the
protocol durations considered here. We first focus on the
idealized limit \(\gamma_{e\downarrow}=0\) and defer the effects of finite spontaneous
emission, \(\gamma_{e\downarrow}\neq0\), to Sec.~\ref{sec:single_particle_decoh}.

The interaction \(g(\hat{J}^+\hat{a}+\text{h.c.})\) is the Tavis--Cummings coupling on the optical \({}^{1}\mathrm{S}_{0}\leftrightarrow{}^{3}\mathrm{P}_{1}\) transition.
To account for the fact that the photons are driven, it is convenient to  transform to a frame rotating at \(\omega_d\), generated by \(\hat{H}_{\rm ref}=\omega_d(\hat{a}^\dagger\hat{a}+\hat{J}^z)\). In this rotating frame the Hamiltonian becomes time-independent and is given by: 
\begin{equation}
\begin{aligned}
\hat{H}=&~ g\big(\hat{J}^+\hat{a}+\text{h.c.}\big)
-{\mathrm i}\sqrt{\kappa}\alpha_{\mathrm in}\!\left(\hat{a}-\hat{a}^{\dagger}\right)+ \\
&-\Delta_c\hat{a}^\dagger\hat{a}-\delta_{\sigma}\hat{J}^z,
\end{aligned}
\label{eq:originalHamiltRWA}
\end{equation}
with cavity detuning \(\Delta_c=\omega_d-\omega_c\) and atomic detuning \(\delta_{\sigma}=\omega_d-\omega_a\). 
Unless stated otherwise, we take the cavity and atomic transition to be resonant,
\(\omega_c=\omega_a\), so that \(\Delta_c=\delta_\sigma\) throughout the protocol
and numerical simulations.

We assume  the ensemble of $N$ atoms is  initialized
in a coherent superposition of the ground \({}^{1}\mathrm{S}_{0}\)
(\(\ket{\downarrow}\)) and clock \({}^{3}\mathrm{P}_{0}\)
(\(\ket{\uparrow}\)) states,
as shown in Fig.~\ref{fig:lab_fig_energy_levels}(a),
\begin{equation}
\label{eq:init_state_spin}
    \ket{\psi(t=0)} = \left(\frac{\ket{\downarrow} + \ket{\uparrow}}{\sqrt{2}}\right)^{\otimes N},
\end{equation}
while the bosonic cavity mode is initialized in (or very close to) its vacuum state $\ket{0}$.

The clock state \(\ket{\uparrow}\) is chosen to be decoupled from the
cavity--spin \({}^{1}\mathrm{S}_{0}\leftrightarrow{}^{3}\mathrm{P}_{1}\)
dynamics described by Eq.~\eqref{eq:originalHamiltRWA}. As a result, the projector $\hat N_\uparrow=\sum_j \ket{\uparrow_j}\!\bra{\uparrow_j}$
commutes with both the Hamiltonian and with all Lindblad jump operators, so it defines a
strong symmetry of the Liouvillian \cite{Buca2012,Lieu2020}. Consequently, the $\ket{\uparrow}$ population
is conserved under both the
Hamiltonian evolution and the quantum jumps, and the dynamics separates into sectors labeled by
\(N_\uparrow\), or equivalently \(N_J\equiv N-N_\uparrow\), where $N_{\uparrow}$ corresponds to an eigenvalue of $\hat{N}_{\uparrow}$. This property enables squeezing on the clock transition ($|{\downarrow}\rangle\xrightarrow{}|{\uparrow}\rangle$) when  initialized   in a superposition on that transition since it  spans sectors with different \(N_J\). 

Since the driven $\{\ket{\downarrow},\ket{e}\}$ dynamics (and its steady state) depend on $N_J$
through the collectively enhanced coupling $g\sqrt{N_J}$, each sector follows a slightly different
driven-dissipative evolution and accumulates a distinct phase, which is crucial for generating
spin squeezing on the clock transition. The intracavity photon field mediates the coherent dynamics within the $\{\ket{\downarrow},\ket{e}\}$ manifold, while photon leakage through the cavity damps that field and limits the effective interaction time. The balance between the spin--photon interaction rate
and the photon lifetime sets key qualitative features of the evolution, including the steady-state behavior and the build-up of correlations, and motivates characterizing the parameter space in terms of the ratio $g\sqrt{N_J}/\kappa$.

\subsection{Unresolved and Resolved  Vacuum Rabi Splitting Regimes}
\label{sec:VacRabi}
The atom-light interaction, described by the first term in  Eq.~(\ref{eq:originalHamiltRWA}),
 hybridizes the cavity photons with the collective atomic excitations. Around the low-excitation state in which the \(N_J\) atoms in the driven manifold are predominantly in \(\ket{\downarrow}\), the two resulting normal modes (polaritons) have a frequency separation at cavity--atom resonance, \(\omega_a=\omega_c\), as depicted in Fig.~\ref{fig:lab_fig_energy_levels}(b,c), known as the collective vacuum Rabi splitting:
\begin{equation}
\Omega_{\mathrm{VRS}} = 2g\sqrt{N_J} .
\end{equation} 
The distinguishability  of the  two polariton modes   depends on the ratio of the cavity loss rate \(\kappa\) to \(\Omega_{\mathrm{VRS}}\).

\paragraph*{Unresolved vacuum Rabi splitting (UVRS).}
If
\begin{equation}
  \kappa \gg \Omega_{\mathrm{VRS}},
\end{equation}
the normal-mode splitting is small compared with the dissipative broadening
set by the cavity linewidth. The two transmission features are therefore not
spectrally resolved and appear as a single broad response
[Fig.~\ref{fig:lab_fig_energy_levels}(b)].
 In this unresolved limit, photons leak from the cavity on a timescale much shorter than it can be coherently absorbed/emitted by the atoms. In this regime, the cavity field adiabatically follows the collective dipole and can be eliminated from the dynamics.

To perform the adiabatic elimination,  we allow for nonzero
atom--cavity detuning ($\omega_c\neq \omega_a$) and derive  the Heisenberg equations of motion for the cavity mode and the collective spin operators, 
\begin{subequations}
\label{eq:UVRS_dynamics}
\begin{align}
\partial_t \hat a(t) &= 
-\mathrm{i} g\hat J^-(t)
+\Bigl(\mathrm{i}\Delta_c - \tfrac{\kappa}{2}\Bigr)\hat a(t)
+\sqrt{\kappa}\alpha_{\rm in},
\label{eq:UVRS_a}\\
\partial_t \hat J^-(t) &= 
+2\mathrm{i} g\hat a(t)\hat J^z(t)
+ \mathrm{i}\delta_{\sigma}\hat J^-(t),
\label{eq:UVRS_Sm}\\
\partial_t \hat J^z(t) &= 
-\mathrm{i} g\bigl[\hat a(t)\hat J^+(t)
- \hat a^{\dagger}(t)\hat J^-(t)\bigr].
\label{eq:UVRS_Sz}
\end{align}
\end{subequations}
Here, we focus on the deterministic part of the dynamics and omit the Langevin noise terms associated with the cavity input field. For the steady-state mean-field analysis below, the relevant relation is the algebraic constraint obtained by setting the deterministic cavity equation to zero. From Eq.~\eqref{eq:UVRS_a}, this gives
\begin{equation}
\hat a(t)\simeq
\frac{\sqrt{\kappa}\alpha_{\rm in}-\mathrm{i}g\hat J^-(t)}
{\kappa/2-\mathrm{i}\Delta_c}
=
\alpha
-\frac{\mathrm{i}g}{\kappa/2-\mathrm{i}\Delta_c}\hat J^-(t),
\label{eq:a_in_Jminus}
\end{equation}
where
\begin{equation}
\alpha=
\frac{\sqrt{\kappa}\alpha_{\rm in}}
{\kappa/2-\mathrm{i}\Delta_c}
\end{equation}
is the coherent intracavity amplitude set by the external drive.

Eq.~\eqref{eq:a_in_Jminus} separates the cavity field into a drive-induced coherent component and a collective-dipole-induced component. In the UVRS regime, where the cavity decay is the fastest timescale, this relation can be interpreted as the leading-order adiabatic elimination of the cavity field. In the RVRS regime, the cavity and collective atomic excitation hybridize and the cavity cannot generally be eliminated from the full dynamics. Nevertheless, Eq.~\eqref{eq:a_in_Jminus} still applies as a stationary mean-field constraint obtained from \(\partial_t \hat a=0\). In the RVRS regime, this constraint should not be interpreted as a dynamical adiabatic elimination, since transient spin--photon dynamics remain explicit.

Substituting Eq.~\eqref{eq:a_in_Jminus} into the full atom--cavity master
equation, Eq.~\eqref{eq:originalHamiltRWA}, gives the leading UVRS spin-only description derived in
Appendix~\ref{app:uvrs_reduction}:

\begin{subequations}
\label{eq:Hamil_spin}
\begin{align}
\partial_{t}\hat{\rho}
&= -{\rm i}[\hat{H}_{\rm eff},\hat{\rho}]
+ \Gamma \mathcal{D}[\hat{J}^-]\hat{\rho},
\label{eq:Hamil_spin_master}\\[4pt]
\hat{H}_{\rm eff}
&= \chi\hat{J}^{+}\!\hat{J}^{-}
\!\!-\! \delta_{\sigma}\hat{J}^{z}
\!+\! \left(\frac{\Omega_{\mathrm{eff}}}{2}\hat{J}^{+}\!\!+\!\mathrm{h.c.}\right)
\label{eq:Hamil_spin_Heff}
\end{align}
\end{subequations}
where
\(\chi\), $\Omega_{\mathrm{eff}}$ and $\Gamma$ are the collective spin-exchange rate, the effective Rabi frequency and the collective decay rate, respectively. $\chi=\frac{g^2\Delta_c}{\Delta_c^2+\kappa^2/4}$ determines the strength of the coherent part of the cavity-mediated spin interaction evaluated at the
cavity--pump detuning \(\Delta_c\).
It quantifies an effective elastic spin--exchange interaction generated by virtual exchange
of photons between atoms. This interaction shifts the atomic energies and correlates
their phases without changing the total number of spin excitations. In
contrast,
\(\Gamma=\frac{g^2\kappa}{\Delta_c^2+\kappa^2/4}\)
sets the rate of collective photon emission into the lossy cavity mode; because
all atoms radiate into the same mode, this decay is enhanced and appears as a
superradiant relaxation channel for the ensemble \cite{Dicke1954,GrossHaroche1982}. We also define the resonant single-atom cavity decay scale
\(\Gamma_0 \equiv 4g^2/\kappa\), which is used below to normalize detunings and times.

The effective 
complex Rabi frequency $\Omega_{\mathrm{eff}}$  between the  
 \(\{\ket{\downarrow},\ket{e}\}\) levels  sets the drive rate of the collective spin. It is  defined as 
\begin{equation}
\Omega_{\rm eff}= 2g \frac{\sqrt{\kappa}\alpha_{\rm in}}{\kappa/2-\mathrm{i}\Delta_c}
= 2g\frac{\sqrt{\kappa}\alpha_{\rm in}\left(\frac{\kappa}{2}+\mathrm{i}\Delta_c\right)}{\Delta_c^{2}+\kappa^{2}/4}.
\label{eq:Omega_eff}
\end{equation}
In the UVRS regime,
the dynamics are captured by the spin-only description in Eq.~(\ref{eq:Hamil_spin}) with effective Rabi frequency of magnitude
$|\Omega_{\rm eff}|
=
2g\sqrt{\kappa}|\alpha_{\rm in}|/\sqrt{\Delta_c^{2}+\kappa^{2}/4}$ which reduces to $|\Omega_{\rm eff}|\approx 2g\sqrt{\kappa}|\alpha_{\rm{in}}|/(\kappa/2)$ when $\kappa\gg|\Delta_c|$.
This expression makes explicit how the external cavity drive effectively drives the atomic spin since  \(\sqrt{\kappa}|\alpha_{\rm in}|\) sets the intracavity field amplitude generated by the drive while the ratio \(2g/(\kappa/2)\) quantifies how efficiently that field couples to the atoms.

\paragraph*{Resolved vacuum Rabi splitting (RVRS).}
When
\begin{equation}
\kappa \lesssim \Omega_{\mathrm{VRS}},
\end{equation}
the system enters the collective strong-coupling regime, where the normal-mode
splitting becomes spectrally resolvable in cavity transmission,
cf.~Fig.~\ref{fig:lab_fig_energy_levels}(c). Near atom--cavity resonance, the dressed-mode resonance frequencies differ by \(\Omega_{\rm VRS}\), with linewidths set by the combined cavity
and atomic decay rates; for \(\gamma_{e\downarrow}\ll\kappa\), each polariton
has a linewidth of order \(\kappa/2\).

In contrast to the UVRS limit, the cavity field no longer adiabatically follows the atoms but instead remains a genuine dynamical degree of freedom over several exchange cycles \cite{Norcia2016sr}. As photons are emitted into and reabsorbed from the cavity, they act back on the collective spin. This interaction potentially modifies both steady states and transients and also generates appreciable spin--photon correlations.

The relevance of these two limits is not merely conceptual. In the UVRS domain, it has been shown that dissipative protocols based on effective spin models can already generate substantial metrological gain in theory \cite{JeremyOAT}. However, many state-of-the-art cavity-QED experiments naturally operate with \(\kappa \sim g\sqrt{N_J}\), i.e., in or near the RVRS regime \cite{Pedrozo-Penafiel2020,Robinson2024,Eric_CRF}, where photon dynamics play a more relevant role in determining the achievable spin squeezing. 
We organize our analysis around these two regimes and investigate how the protocol of \cite{JeremyOAT}, originally formulated in the UVRS limit, can be extended into the RVRS regime while remaining effective across the intermediate crossover region.

\section{Strong and Weak Symmetries}
\label{sec:strong_weak_symmetry}

The distinction between strong and weak symmetries is central to our protocol because the squeezing mechanism relies not only on conserving the sector label \(N_J\), but also on preserving relative phases between different \(N_J\) sectors. A strong symmetry guarantees that \(N_J\) is conserved under both Hamiltonian evolution and individual quantum jumps, so dissipation can relax the driven manifold without transferring population between sectors\cite{Prosen2010,Buca2012,AlbertJiang2014,SanchezMunoz2019,Lieu2020,Liu2023}. This conservation remains exact in the RVRS regime. What changes in RVRS is more subtle: because the cavity field has memory, different \(N_J\) sectors can generate distinguishable transient output fields. If the output is unobserved, tracing over it produces an effective collective dephasing.

In contrast,  a weak symmetry means that the conservation law only holds at the level of the density matrix. In that case, the ensemble-averaged  state \(\hat{\rho}(t)\) respects the symmetry in the symmetry-broken phase, but individual quantum jumps can move population between different sectors, and the environment gains knowledge about the symmetry sector, precluding the preservation of coherences between them \cite{Lieu2020,Liu2023}.

More formally, we consider a Lindblad master equation

\begin{equation}
\label{eq:general_Lindblad}
\partial_t \hat{\rho}
= \mathcal{L}[\hat{\rho}]=
-i[\hat{H},\hat{\rho}]
+ \sum_{\mu}\mathcal{D}[\hat{L}_\mu]\hat{\rho}
\end{equation}
A strong symmetry is present when there exists an operator \(\hat{Q}\) such that \cite{Prosen2010,Buca2012,AlbertJiang2014,SanchezMunoz2019}
\begin{equation}
  [\hat{Q},\hat{H}]=0,
  \qquad
  [\hat{Q},\hat{L}_\mu]=0
  \quad\forall\mu,
  \label{eq:strong_symmetry_commutators_general}
\end{equation}
so that \(\hat{Q}\) is conserved not only by the coherent dynamics but also by every quantum jump.
Since the Liouvillian \(\mathcal{L}\) is a superoperator acting on \(\hat{\rho}\), its block structure can be  expressed in terms of the left- and right-multiplication superoperators \(\mathcal{Q}_L[\hat{\rho}]\equiv \hat{Q}\hat{\rho}\) and \(\mathcal{Q}_R[\hat{\rho}]\equiv \hat{\rho}\hat{Q}^\dagger\), which satisfy \([\mathcal{L},\mathcal{Q}_L]=[\mathcal{L},\mathcal{Q}_R]=0\). Equivalently, \(\hat{\rho}\) decomposes into blocks \(\hat{\Pi}_q \hat{\rho}\hat{\Pi}_{q'}\) (with \(\hat{\Pi}_q\) the projector onto the \(\hat{Q}\) eigenspace \(q\)), and each \((q,q')\) block evolves independently under \(\mathcal{L}\)~\cite{Buca2012,Lieu2020}. 

By contrast, a weak symmetry only requires the existence of a unitary symmetry operator \(\hat P\) such that the
Liouvillian commutes with the corresponding symmetry superoperator
\(\mathcal{P}[\hat{\rho}]=\hat P\hat{\rho}\hat P^\dagger\), i.e.,
\([\mathcal{L},\mathcal{P}]=0\),
 while the jump operators \(\hat{L}_\mu\) themselves need not commute with \(\hat{P}\). In this case, the symmetry is still respected in the sense that \(\mathcal{L}\) commutes with the symmetry superoperator \(\mathcal{P}\), but individual trajectories can hop between different symmetry sectors,
and coherence between the sectors is not protected by the symmetry (and will dephase).

In the three-level cavity-QED setup of Ref.~\cite{JeremyOAT} and the present work,
the relevant strong symmetry corresponds to global phase rotations of the driven manifold \(\{|{\downarrow}\rangle,|e\rangle\}\). It is generated by the conserved population operator
\begin{equation}
  \hat Q = \hat{N}_J
  =
  \sum_i
  \Bigl(
    |{\downarrow}_i\rangle\langle{\downarrow}_i|
    + |e_i\rangle\langle e_i|
  \Bigr).
  \label{eq:NJ_def_main}
\end{equation}
Because the total atom number \(N\) is fixed, using \(\hat N_J\) is entirely equivalent to using the conserved spectator-state population \(\hat N_\uparrow=N-\hat N_J\). In what follows, we use \(\hat N_J\) as the primary sector label.

In the full atom--cavity model, \(\hat N_J\) commutes with the Hamiltonian and
with the microscopic jump operators appearing in Eq.~\eqref{eq:Hamil_cavity}. The
cavity-loss jump \(\hat a\) acts only on the cavity Hilbert space, while the
spontaneous-emission jump \(\ket{\downarrow_i}\!\bra{e_i}\) maps
\(\ket{e_i}\) to \(\ket{\downarrow_i}\) within the same driven manifold. After
adiabatic elimination in the UVRS limit, the corresponding effective
collective jump operator \(\hat J^-\) also preserves \(\hat N_J\). Thus, \(\hat N_J\) is conserved along each quantum trajectory.
As a consequence, the Liouvillian decomposes into sectors of fixed eigenvalues \(N_J\) \cite{Prosen2010,Buca2012,AlbertJiang2014,SanchezMunoz2019}, and, in the symmetry-broken (spin--polarized) phase, photon-emission records generally do not fully
resolve which \(N_J\) sector the system occupies, so superpositions of different \(N_J\) components can retain their relative phase coherence \cite{Lieu2020,Liu2023,JeremyOAT}.
In RVRS, however, this protection can be reduced because different \(N_J\) sectors generate different transient cavity fields over the longer cavity-memory timescale. Photon leakage then imprints partial $N_J$ sector information on the output field; when that output is unobserved, tracing it out suppresses the off-diagonal coherences between different \(N_J\) sectors, acting as an effective decoherence
mechanism in the reduced encoded-spin description \cite{Barberena2024}.
\section{Squeezing Protocol}
\label{sec:Multistage}

\begin{figure}[t]  

    \centering

    \includegraphics[width=1.02\linewidth]{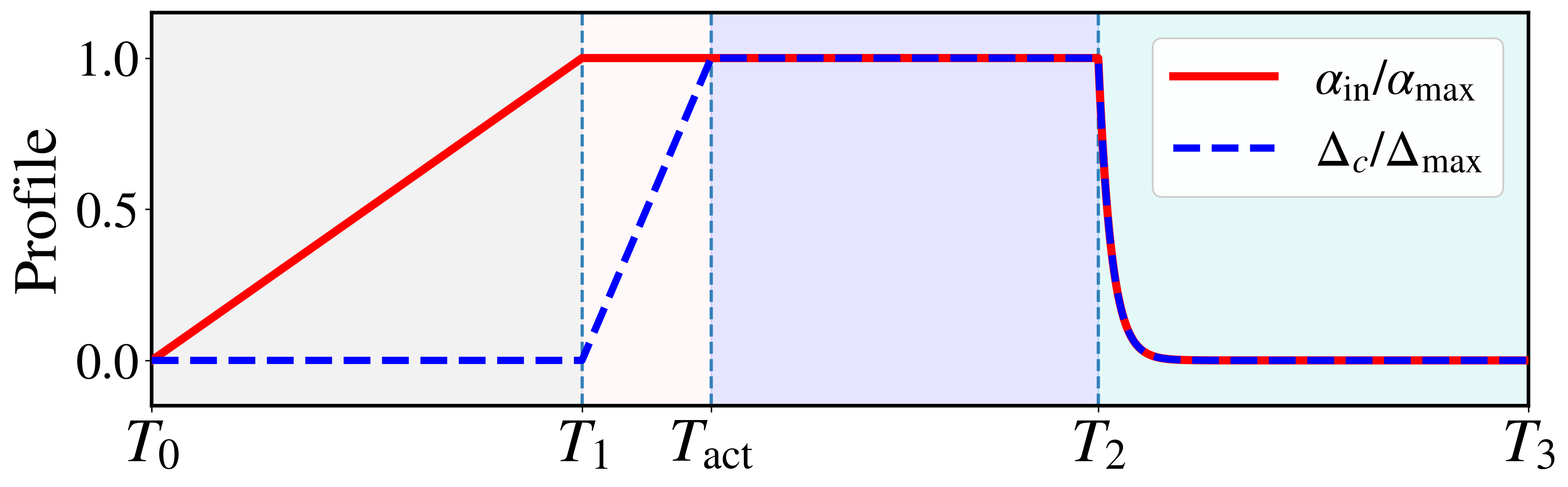}\\[0.6em]
    
        \setlength{\tabcolsep}{4pt}
        \renewcommand{\arraystretch}{0.9}

    \setlength{\tabcolsep}{6pt}
    \renewcommand{\arraystretch}{1.3}

    \begin{tabular}{|c|c|c|}
        \hline
        \textbf{Stage I} & \textbf{Stage II} & \textbf{Stage III} \\  
        \hline
        
        \includegraphics[width=0.28\linewidth]{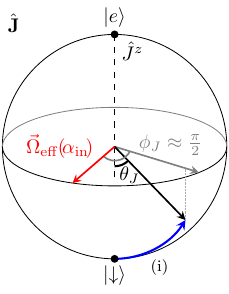} & 
        \includegraphics[width=0.28\linewidth]{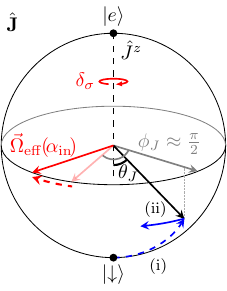} & 
        \includegraphics[width=0.28\linewidth]{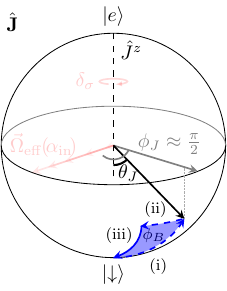} \\
        \hline
        
        \includegraphics[width=0.28\linewidth]{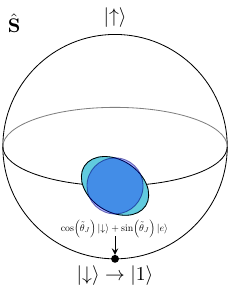} & 
        \includegraphics[width=0.28\linewidth]{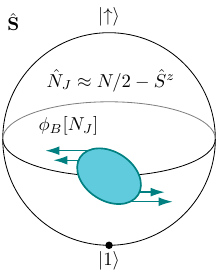} & 
        \includegraphics[width=0.28\linewidth]{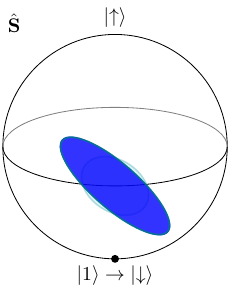} \\
        \hline
        
        \includegraphics[width=0.28\linewidth]{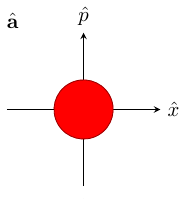} & 
        \includegraphics[width=0.28\linewidth]{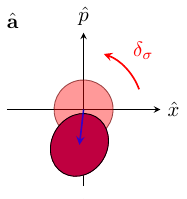} & 
        \includegraphics[width=0.28\linewidth]{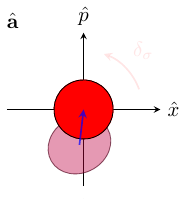} \\
        \hline
    \end{tabular}
    
    \caption{{\bf Multistage protocol for generating and storing spin squeezing.}  The top panel  shows  a schematic of the  time-dependent profiles of the  detuning (blue dashed line) and the drive amplitude (solid red line). The panel below describes the protocol.  Stages~I--III are displayed in separate columns; the  top rows show the Bloch sphere of the driven manifold \(\{\ket{\downarrow},\ket{e}\}\) with collective spin \(\hat{J}\),  the  middle row  the  Bloch sphere of the encoded spin \(\hat{S}\) spanned by \(\ket{0}\equiv\ket{\uparrow}\) and \(\ket{1}\equiv \alpha\ket{\downarrow}+\beta\ket{e}\), and the bottom row has a  phase-space representation of the cavity mode \(\hat{a}\).   In Stage~I, a resonant drive rotates \(\hat{J}\) into a spin-polarized steady state, \(\ket{1}\), while the cavity relaxes to  the vacuum state. In Stage~II, a finite detuning \(\delta_\sigma\) causes the \(\hat{J}\) Bloch vector to precess, imprinting a sector-dependent phase \(\phi_B[N_J]\) on \(\hat{S}\) and generating spin squeezing, while the cavity field becomes displaced and slightly squeezed. In Stage~III, the drive and detunings are turned off, the excited population decays back to \(\ket{\downarrow}\), the squeezing is stored in the \(\{\ket{\uparrow},\ket{\downarrow}\}\) manifold of \(\hat{S}\), and the cavity field relaxes back to vacuum.}
    \label{fig:multi_level_system}
\end{figure}

The protocol of Ref.~\cite{JeremyOAT} showed that a driven-dissipative cavity can generate one-axis twisting dynamics and metrologically useful squeezing by exploiting a strong symmetry \cite{Prosen2010,Buca2012}. Fig.~\ref{fig:multi_level_system}
shows the three-stage implementation considered here. The ramp times are
control parameters chosen to suppress nonadiabatic spin--photon excitations:
they should be long compared with the relevant cavity/polariton relaxation
times so that the system follows the low-photon branch, but short compared
with the decoherence times that limit useful squeezing. In the simulations
below, these times are indicated by the shaded regions in the dynamical plots
and reported in units of \(N\Gamma_0 t\).

As sketched in the top of Fig.~\ref{fig:multi_level_system}, the protocol is
composed of three stages. Stage~I uses a resonant drive,
\(\Delta_c(t)=\delta_{\sigma}(t)=0\), to prepare the driven manifold in a
spin-polarized, low-photon steady state that is coherent but essentially
unsqueezed. Stage~II slowly detunes the drive so that each \(N_J\) sector
follows its instantaneous low-photon branch while limiting sector-dependent
cavity displacements. The sectors acquire a sector-dependent geometric phase
that realizes one-axis twisting. Stage~III turns off the drive and detunings,
mapping the generated correlations into the long-lived clock manifold. We
describe the dynamics of each stage below.

\subsection*{Stage I: Ramp the Drive  (\(T_0\to T_1\)).}

In the first stage, we begin by ramping the cavity drive amplitude \(\alpha_{\rm in}(t)\) from zero up to its full value while holding both detunings \(\Delta_c\) and \(\delta_{\sigma}\) at zero, as shown in the top of Fig.~\ref{fig:multi_level_system} in the region between $t\in (T_0,T_1)$.
To connect with the phase accumulation in the later stages, we decompose the initial state into sectors with a fixed number of atoms, denoted by $N_J$ here, in the driven manifold \(\{\ket{\downarrow},\ket{e}\}\). Starting from the initial coherent spin state
\begin{equation}
\left(\frac{\ket{\downarrow} +\ket{\uparrow}}{\sqrt{2}}\right)^{\otimes N}
=
\sum_{N_{J}=0}^{N}\sqrt{\frac{\binom{N}{N_{J}}}{2^{N}}}
\mathcal{S}\Bigl(\ket{\downarrow}^{\otimes N_{J}}\ket{\uparrow}^{\otimes N_{\uparrow}}\Bigr),
\end{equation}
with \(N_\uparrow = N - N_J\).
The dynamics can be organized into sectors labelled by the number of atoms \(N_J\) in the driven manifold,  where \(\mathcal{S}(\cdot)\) denotes symmetrization over all atoms. 

As in Ref.~\cite{JeremyOAT}, the key point is that \(\ket{\uparrow}\) acts as a
spectator state and does not enter the Liouvillian. The total number of atoms in the driven manifold \(\{\ket{\downarrow},\ket{e}\}\) is therefore conserved, so the dynamics decomposes into sectors of fixed \(N_J\). To describe Stage~I,
we may analyze each sector separately as a driven two-level ensemble in the
optical manifold, with collective operators \(\hat J^+\) and \(\hat J^z\)
defined at fixed \(N_J\).

When the drive is resonant with the atomic transition, \(\delta_\sigma = 0\) in  Eq.~(\ref{eq:UVRS_dynamics}), the coupled atom--cavity dynamics in each \(N_J\)  exhibits a nonequilibrium phase transition at~\cite{Barberena2019}
\begin{equation}
\label{eq:OmegacDiego}
    \Omega_c[N_J] = \frac{N_J}{2}\sqrt{\Gamma^2 + 4\chi^2}
    =
    \frac{N_J}{2}\frac{2g^2}{\sqrt{\kappa^2/4+\Delta_c^2}}.
\end{equation}
For \(|\Omega_{\rm eff}| > \Omega_c[N_J]\), the system enters the normal phase, and the density-matrix steady state becomes mixed, with vanishing steady-state inversion, \(\langle \hat{J}_{N_J}^z \rangle_{\rm ss}\to 0\), in the thermodynamic mean-field limit. The collective dipole therefore has no macroscopic expectation value.

For \(|\Omega_{\rm eff}| < \Omega_c[N_J]\), the system is in a spin-polarized (superradiant) phase \cite{Eric_CRF}.
As illustrated in the top of Fig.~\ref{fig:multi_level_system}
the drive and collective dissipation compete, pushing the collective $J$ spin away from the south pole towards \(\ket{e}\). It realizes a steady-state polar angle \(\theta_J\) set by the balance between driving and collective decay, realizing a   nonzero steady-state inversion
\begin{equation}
\langle \hat{J}^z_{N_J} \rangle_{\rm ss}
    =
    -\frac{N_J}{2}\sqrt{1 - \left|\Omega_{\rm eff}/\Omega_c[N_J]\right|^2}.
\end{equation}
 At \(\theta_J[N_J]\), where we have explicitly incorporated the $N_J$ dependence, the field re-radiated by the collective dipole  opposes the injected drive inside the cavity. As a result, the net field inside the cavity  approaches the vacuum state, \(\langle \hat a \rangle_{\rm ss}\approx0\).

For large \(N_J\), as illustrated in the top and bottom panels of Stage~I in
Fig.~\ref{fig:multi_level_system}, the collective Bloch vector points at a fixed polar angle \(\theta_J\), while the steady state of the cavity,  represented by a circular Gaussian Wigner distribution, is closely centered at the origin, indicating zero displacement and mostly  vacuum noise. In the large-\(N_J\) mean-field limit and when the spin steady state is located away from the equator in the $\hat{J}$-sphere, the spin steady state is well
approximated by a symmetric product state

\begin{equation}
    \ket{\psi_{N_J}}_{\rm ss}
    \!\approx\!
    \bigg(
        \cos\frac{\theta_J[N_J]}{2}\ket{\downarrow}
        +\mathrm{e}^{-\mathrm{i}\phi_J[N_J]}\sin\frac{\theta_J[N_J]}{2}\ket{e}
    \bigg)^{\!\otimes N_J},
\end{equation}
where the polar angle is set by the drive via
\(\cos\theta_J[N_J]=
    \sqrt{1 - \bigl|\Omega_{\rm eff}/\Omega_c[N_J]\bigr|^2}\).  When the drive is resonant with the cavity, the azimuthal angle takes the definite value
\(\phi_J[N_J] = \pi/2\) so that the  steady-state Bloch vector lies in the \(y\)--\(z\) plane of the Bloch sphere. This structure motivates  us  to  introduce an effective Bloch sphere whose south pole is aligned with the mean-field Bloch vector of the driven \(\{\ket{\downarrow},\ket{e}\}\) manifold and define

\begin{subequations}
\label{eq:1defm_mean_sector}
\begin{align}
\ket{1}_{N_J}
&\equiv
\cos\frac{\theta_J[N_J]}{2}\ket{\downarrow}
+
\mathrm{e}^{-\mathrm{i}\phi_J[N_J]}
\sin\frac{\theta_J[N_J]}{2}\ket{e},
\label{eq:1defm_mean_sector_1}
\\
\ket{0}
&\equiv
\ket{\uparrow}.
\label{eq:1defm_mean_sector_0}
\end{align}
\end{subequations}
In combination with the strong symmetry, the steady state of the full three-level system is well-approximated by
\begin{equation}
\ket{\psi}_{\text{ss}}
\approx
\sum_{N_J=0}^{N}
\sqrt{\frac{\binom{N}{N_J}}{2^{N}}}
\mathcal{S}\bigl(\ket{1}_{N_J}^{\otimes N_J}\ket{0}^{\otimes N_\uparrow}\bigr).
\label{eq:state_T1_strong}
\end{equation}

For the initial state in Eq.~\eqref{eq:init_state_spin}, one has \(\langle \hat N_J\rangle = N/2\), with sector weights concentrated in a window of width \(\Delta N_J\sim \sqrt{N}\). We denote by \(\hat{\mathbf S}\) the encoded collective spin associated with the two effective states \(\ket{0}\) and \(\ket{1}\). Then when \(|N_J-N/2|\sim \sqrt{N}\), to a good approximation we can  visualize the  \(\hat S\) manifold (middle of Fig.~\ref{fig:multi_level_system}) in terms of a single mean-field  Bloch sphere  spanned by 
\begin{equation}
\label{eq:1defm}
\ket{1} \equiv \cos\!\frac{\tilde {\theta}_J}{2}\ket{\downarrow}
+
\mathrm{e}^{-\mathrm{i}{\tilde \phi}_J}\sin\!\frac{{\tilde \theta}_J}{2}\ket{e},
\qquad
\ket{0}\equiv\ket{\uparrow},
\end{equation}
where \(\tilde{\theta}_J \equiv \theta_J[N_J=\langle \hat N_J\rangle]\) and \(\tilde{\phi}_J \equiv \phi_J[N_J=\langle \hat N_J\rangle]\) are the mean-field polar and azimuthal angles of the \(\{\ket{\downarrow},\ket{e}\}\) Bloch vector evaluated at the mean sector \(N_J \simeq \langle \hat N_J\rangle\). 

In this basis, the operator \(\hat{\mathbf S}\) denotes the collective spin built from \(\ket{0},\ket{1}\). To leading order,

\begin{equation}
\hat S^z \approx \frac{N}{2} - \hat N_J,
\label{eq:Sz_NJ_relation}
\end{equation}  and

\begin{equation}
    |\psi\rangle_{\text{ss}} \approx \left(\frac{\ket{1} +\ket{0}}{\sqrt{2}}\right)^{\otimes N}.
\end{equation}
Hence, the initial coherences between $\ket{\downarrow}$ and $\ket{\uparrow}$ are preserved  during the driven-dissipative dynamics and in the steady state they are   mapped to coherences between $\ket{1}$ and $\ket{0}$. In this basis, the symmetrized state \(\mathcal{S}\bigl(\ket{\psi_{N_J}}_{\rm ss}\ket{\uparrow}^{\otimes N_\uparrow}\bigr)\) is therefore well approximated by a collective Dicke state  with magnetization, $m_S = \frac{N_\uparrow - N_J}{2}$.

\subsection*{Stage II: Detuning the Drive (\(T_1\to T_2\)).}

In the second stage (\(T_1 \leq t \leq T_2\)),
the drive amplitude \(\sqrt{\kappa}\alpha_{\mathrm{in}}(t)\) has reached its maximum value and is held constant while the detuning is ramped on.
In the \(\hat J\) manifold, the detuning corresponds to a rotation about the \(J_z\) axis, which generates  a $N_J$-dependent phase, as sketched in the {top} of Fig.~\ref{fig:multi_level_system}.II. When the corresponding trajectory is completed, this phase admits the usual geometric (Berry-phase) \cite{Berry1984} interpretation in terms of the enclosed solid angle.

This Berry phase picture is dependent on the rotating frame used, which we discuss in detail in Appendix \ref{app:equiv_rotating_frames}. Briefly, the Berry phase description arises when we consider the rotating frame of the atomic transition frequency rather than the drive frequency, which imparts a time-dependent phase on the effective drive through the detuning. The instantaneous steady-state Bloch vector follows this phase, leading to a change in $\phi_J$ during this stage of the protocol.

To understand  this geometric phase more explicitly, we connect to the encoded-spin picture
introduced in Stage~I.
  The Berry phase $\Phi_B(N_J)$ is  the geometric
phase acquired by the $N_J$ sector along the closed path obtained by completing its trajectory on
the Bloch sphere. It becomes physically relevant only through relative phases between sectors (or
relative to the spectator state $|{\uparrow}\rangle$). For  a closed trajectory, it is set by the
enclosed solid angle  \cite{Berry1984}
\begin{equation}
\Phi_B(N_J)=\frac{N_J}{2}\int_{\mathcal{C}} (1-\cos\theta_J)\dot{\phi}_Jdt=\frac{N_J}{2}\Omega_{\mathcal C}(N_J),  
\label{eq:Phi_B_NJ}
\end{equation}
with $\Omega_{\mathcal C}(N_J)$ the enclosed solid angle.
Under the detuning ramp of Stage~II, each sector with fixed \(N_J\) running
phase 
accumulates a Berry phase
\(\phi_B(N_J,t)\) associated with the path traced by its Bloch vector.

The phase relevant for the effective OAT description is the one built up by the end of the detuning
ramp, \(\phi_B(N_J,T_2)\), while the full closed-loop Berry phase \(\Phi_B(N_J)\) is recovered only
after the trajectory is completed during Stage~III.
 The sector-dependent Berry phase \(\phi_B(N_J,t)\) can then be viewed as a phase that depends on the collective magnetization \(m_S \approx \frac{N}{2} - N_J\). It is natural to expand \(\phi_B\) around the mean sector \(N_J=N/2\):

\begin{equation}
\begin{aligned}
\phi_B(N_J,t)\simeq&
\phi_B\!\left(\tfrac{N}{2},t\right)
+\left.\partial_{N_J}\phi_B(N_J,t)\right|_{N_J=N/2}\!\left(N_J-\tfrac{N}{2}\right)\\
&+\frac{1}{2}\left.\partial_{N_J}^2\phi_B(N_J,t)\right|_{N_J=N/2}\!\left(N_J-\tfrac{N}{2}\right)^2
+\cdots .
\end{aligned}
\label{eq:Berry_Taylor_NJ}
\end{equation}

The unitary generated by the Berry phase may thus be written as
\(\hat U_B(t) = \exp\bigl[\mathrm{i}\phi_B(\hat N_J,t)\bigr]\).
Up to the overall phase \(\phi_B^{(0)}(t)\), this is equivalent to evolution under an effective Hamiltonian in the $S$ sphere

\begin{equation}
\hat H_{\rm eff}(t)
=
\omega_z^B\hat S^z
+
\bar\chi(t)(\hat S^z)^2
+\cdots,
\label{eq:Heff_OAT_from_Berry}
\end{equation}
With
\(\hat U_B(t)=\exp[i\phi_B(\hat N_J,t)]\), the corresponding instantaneous
generator is
\[
\hat H_{\rm eff}(t)=-\partial_t\phi_B(\hat N_J,t).
\]

Here the linear term proportional to \(\hat S^z\) generates a collective rotation about the \(z\) axis on the \(\hat S\) Bloch sphere, while the quadratic term proportional to \( (\hat S^z)^2\) realizes one-axis twisting (OAT) dynamics. Physically, the key ingredient is the \emph{nonlinearity} of the sector phase $\phi_B(N_J,t)$ in $N_J$.
The initial coherent spin state on the \(\hat S\) Bloch sphere, shown in the {middle} of Fig.~\ref{fig:multi_level_system}.I,
is sheared by the OAT Hamiltonian \(\propto(\hat S^z)^2\), producing a
spin-squeezed state with reduced variance along the optimal measurement
quadrature, as illustrated in the {middle} of Fig.~\ref{fig:multi_level_system}.II. 

Because the finite detuning \(\delta_\sigma \neq 0\) rotates the Bloch vector of the \(\{\ket{\downarrow},\ket{e}\}\) manifold, the resulting mismatch between the cavity field, the collective dipole, and the input drive generates  a nonzero intra-cavity field, as can be seen from Eq.~(\ref{eq:a_in_Jminus}) and is sketched in the {top} of Fig.~\ref{fig:multi_level_system}.II).
Generically, for  $\delta_\sigma\neq0$, each $N_J$ sector develops a steady coherent cavity amplitude
$\alpha_{N_J}(t)\equiv\langle \hat a\rangle_{N_J}$ whose phase follows the azimuthal orientation of
$\langle\hat J^-\rangle$. This is important because $\alpha_{N_J}(t)$ can become
$N_J$-dependent, especially for RVRS, so
photon leakage at rate $\kappa$ may carry  information on the $N_J$ sector label and dephases inter-sector coherences
$\rho_{N_J,N_J'}$, i.e., it induces collective dephasing of the encoded spin \cite{Barberena2024T}. During an adiabatic ramp the
cavity remains close to a displaced, nearly Gaussian state following $\alpha_{N_J}(t)$.

\subsection*{Stage III: Turn-off and Storage (\(T_2\to T_3\)).}
\label{sec:Stage_III}

Stage~III is designed to transfer the squeezing into the initial state manifold \(\{\ket{\uparrow},\ket{\downarrow}\}\) while removing the excited state and cavity population with minimal additional control.
Operationally, during \(T_2\to T_3\) we simultaneously ramp the detunings \(\Delta_c(t)\), \(\delta_\sigma(t)\) and the drive amplitude \(\alpha_{\rm in}(t)\) back to zero so that the cavity relaxes to vacuum while the driven manifold relaxes back to \(\ket{\downarrow}\), leaving the population stored in the long-lived \(\{\ket{\uparrow},\ket{\downarrow}\}\) manifold. In RVRS, this storage step preserves the squeezing without additional decoherence provided the turn-off is sufficiently smooth so that the cavity displacement and excited-state population relax without leaving significant residual excitation due to cavity back-action. As illustrated in the {top} of Fig.~\ref{fig:multi_level_system}.III, the Bloch vector on the \(\hat J\) Bloch sphere returns to the south pole provided the turn-off is slow enough for the excited-state population to relax.

During this process  each sector with fixed \(N_J\) relaxes effectively independently under \(\mathcal{L}_{\rm cav}\).
In combination with the strong symmetry, a sufficiently smooth turn-off brings the cavity field back close to vacuum so that photon-emission events do not significantly resolve which \(N_J\) sector the system occupies \cite{JeremyOAT}. In other words,  within a given sector, the driven steady state \(\ket{\psi_{N_J}}_{\rm ss}\) relaxes back to \(\ket{\downarrow}^{\otimes N_J}\),
\begin{equation}
\mathcal{S}\bigl(\ket{\psi_{N_J}}_{\rm ss}\ket{\uparrow}^{\otimes N_\uparrow}\bigr)
\longrightarrow
\mathcal{S}\bigl(\ket{\downarrow}^{\otimes N_J}\ket{\uparrow}^{\otimes N_\uparrow}\bigr).
\end{equation}
In the \(\hat S\) language, this means that the OAT generated during Stage~II is mapped from the \(\{\ket{0},\ket{1}\}\) Bloch sphere onto the manifold \(\{\ket{\uparrow},\ket{\downarrow}\}\), as depicted in the {middle} of Fig.~\ref{fig:multi_level_system}.III. 
At the same time,
once the drive is turned off, the 
cavity mode relaxes to the vacuum as illustrated in the {bottom} of Fig.~\ref{fig:multi_level_system}.III. 

While all of the above discussion was explicitly studied in the UVRS \cite{JeremyOAT}, we will show below  that by  ramping the drive on and off  slowly, this behavior can likewise be extended to the RVRS limit
bringing the dynamics into close agreement with the UVRS results.
The remaining question is how the cavity relevance in RVRS modifies this
picture. We address this by examining the cavity quadrature dynamics and then
identifying the low-photon operating window used in the simulations below.

\label{sec:limits}

\subsection*{Role of Photon Dynamics on Protocol}
\label{subsec:RolePhotons}

In this section, we analyze how dynamical cavity photons in the RVRS regime modify the original squeezing protocol defined in the UVRS regime. To quantify how active photons affect the transient and steady-state dynamics,
 we analyze the quadrature operators
\(\hat x = (\hat a + \hat a^{\dagger})/\sqrt{2}\) and
\(\hat p = (\hat a - \hat a^{\dagger})/(\sqrt{2}\mathrm i)\).
We also define the operator \(\hat a_{\mathrm{AE}}\),
and its  adiabatic quadratures,
\(\hat x_{\mathrm{AE}} = (\hat a_{\mathrm{AE}} + \hat a_{\mathrm{AE}}^{\dagger})/\sqrt{2}\) and
$\hat p_{\mathrm{AE}} = (\hat a_{\mathrm{AE}} - \hat a_{\mathrm{AE}}^{\dagger})/\sqrt{2}\mathrm i$, 
to quantify the validity of the adiabatic-elimination approximation $\partial_t \hat{a}=0$ (cf. Eq.~\eqref{eq:a_in_Jminus})
\begin{equation}
    \hat a_{\mathrm{AE}}(t) \equiv
    \frac{\mathrm{i} g\hat J^-(t) - \sqrt{\kappa}\alpha_{\rm in}(t)}
         {\mathrm{i}\Delta_c(t) - \kappa/2},
    \label{eq:a_adiabatic}
\end{equation}
where AE stands for adiabatic elimination.

\begin{figure}[!ht]
  \centering
  \includegraphics[width=1.0\linewidth]{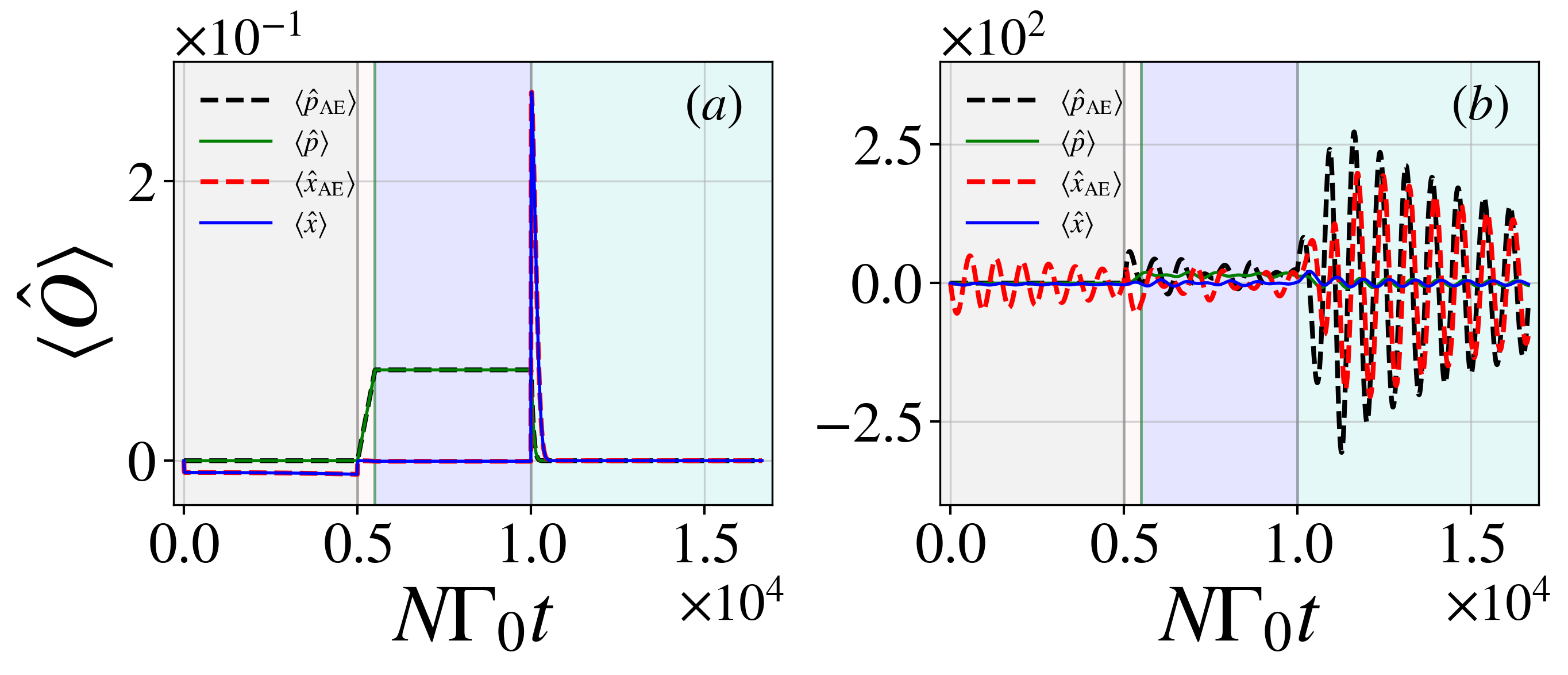}

  \caption{{\bf Dynamics of the cavity-mode quadratures $\hat{x}$ and $\hat{p}$.}
  The solid lines show the full cavity quadratures, while the dashed lines show the
  instantaneous adiabatic-elimination predictions $\hat{x}_{\mathrm{AE}}$ and
  $\hat{p}_{\mathrm{AE}}$ from Eq.~(\ref{eq:a_adiabatic}). The shaded regions
  indicate the stages of the protocol. All curves are computed for $N=10^5$,
  $\delta_{\sigma}/(N\Gamma_0)=\Delta_c/(N\Gamma_0)=1.33\times10^{-3}$, and
  $g=10\,\mathrm{kHz}$. 
  In panel (a), corresponding to the UVRS regime with
  $\kappa=15g\sqrt{N_J}$ and
  $\sqrt{\kappa}\alpha_{\mathrm{in}}=0.5gN_J/2$, the cavity closely follows the
  adiabatic estimate. 
  In panel (b), corresponding to the RVRS regime with
  $\kappa=15g$,
  $\kappa/(g\sqrt{N_J})=6.7\times10^{-2}\ll1$, and
  $\sqrt{\kappa}\alpha_{\mathrm{in}}=0.487gN_J/2$, the full quadratures exhibit
  visible oscillatory deviations from the adiabatic estimate.}
  \label{fig:BC_GC_comparisona}
\end{figure}

The mean-field photon dynamics during each stage of the protocol are shown in Fig.~\ref{fig:BC_GC_comparisona}. In the UVRS regime, the cavity field adiabatically follows the spins, reflecting the fast relaxation of the mode.
By contrast, in the RVRS regime, coherent light--matter oscillations lead to pronounced nonadiabatic dynamics and visible deviations from the adiabatic-elimination prediction. Moreover, for more rapid ramps, the cavity field shows larger deviations from
the instantaneous AE prediction, with quadrature amplitudes reaching values of
order \(10^3\) in the case of quenches. Nevertheless, after the spin--cavity transients damp on a timescale set by the
polariton linewidths, both UVRS and RVRS dynamics approach
\(\langle\hat a\rangle\simeq\langle\hat a_{\mathrm{AE}}\rangle\). The
dimensionless ratio \(\kappa/(g\sqrt{N_J})\) controls whether this relaxation is
overdamped and UVRS-like or retains resolved spin--photon oscillations
characteristic of RVRS.

In general, avoiding large coherent oscillations of the cavity field in the RVRS regime requires all ramps to be sufficiently slow. 
 Otherwise, such   oscillations, combined with the  additional photon loss,  rotate and partially wash out the squeezed quadrature, producing a rapid degradation of the squeezing. 

The quadrature dynamics above show that, in the RVRS regime, cavity photons
affect the transient response of the protocol. The same spin--photon feedback
also determines which quasi-steady branch can be used during the ramps. When
coherent spin--photon exchange is no longer strongly overdamped by cavity loss,
the intracavity field acts back on the collective spin and can reshape the
spin-polarized steady-state response itself. Thus, before deriving an effective squeezing
description, we must identify the region where the driven
\(\{\ket{\downarrow},\ket{e}\}\) manifold remains low-photon and strongly
spin-polarized. This is the operating window in which the sector-dependent
geometric phase can accumulate while avoiding
regions where
cavity backaction causes discontinuous changes in the steady-state response and strongly
reduces the useful spin correlations.

\paragraph*{Mean-field phase diagram.}
We use a mean-field phase diagram as a diagnostic of this operating window.
The detailed mean-field equations, steady-state roots, and bistability
analysis are given in Appendix~\ref{app:mean_field_roots}; here we retain the
physical drive scale needed to organize the operating map. We focus on a single steady-state branch, where by branch we mean a continuous
family of stable steady-state solutions which are followed as the control parameters
are varied, as in standard analyses of optical bistability and cooperative
resonance fluorescence \cite{Bonifacio1978,Carmichael1980, Eric_CRF}. The branch
relevant to the protocol is the low-photon, spin-polarized solution
continuously connected to the resonant spin-polarized state; when
multiple spin-polarized roots coexist, we select the one with the
smallest steady-state mean-field cavity occupation.

\begin{figure}[t!]
    \centering
    \begin{tikzpicture}
        \node[anchor=south west, inner sep=0] (img) at (0,0)
        {\includegraphics[width=\linewidth]{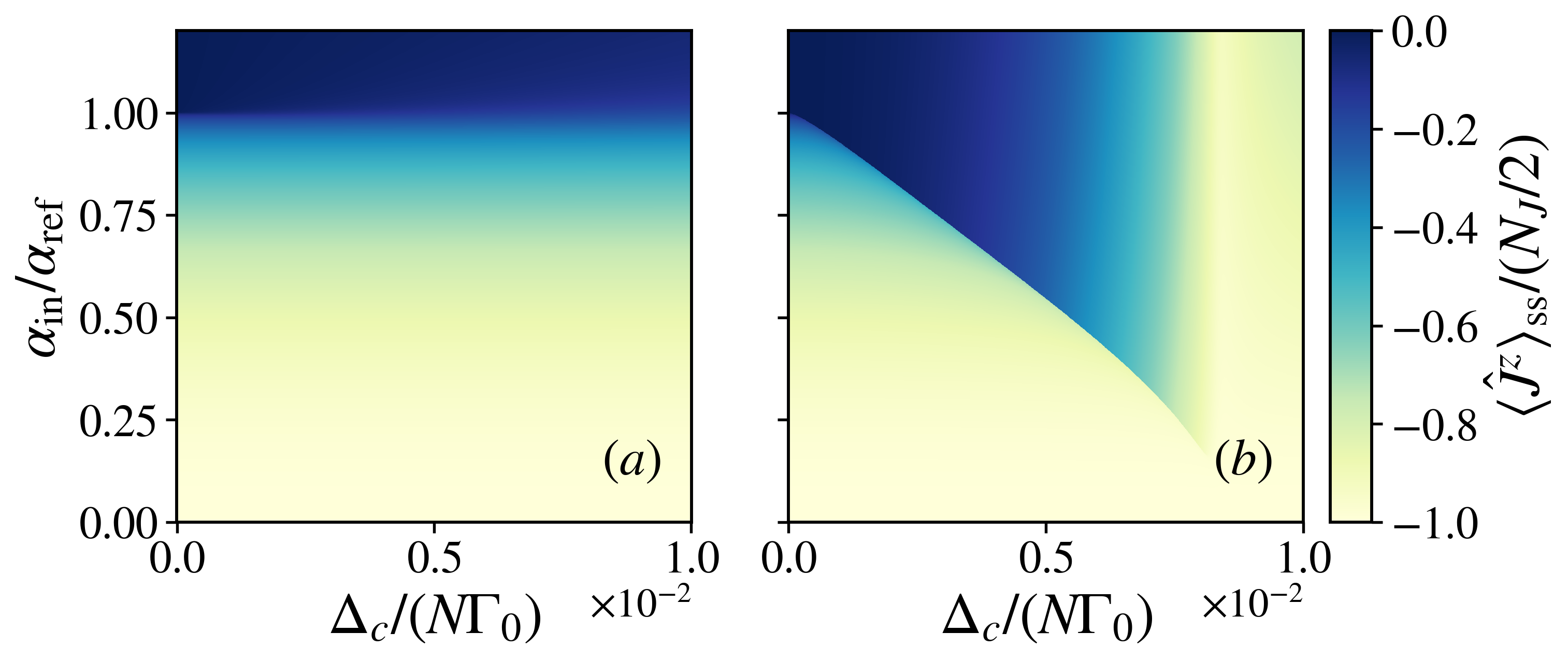}};

        \begin{scope}[x={(img.south east)}, y={(img.north west)}]
            \node[font=\bfseries\small] at (0.285,1.015) {UVRS};
            \node[font=\bfseries\small] at (0.675,1.015) {RVRS};
        \end{scope}
    \end{tikzpicture}

    \caption{{\bf Mean-field phase diagram.}
    Steady-state spin polarization
    \(\langle \hat{J}^z\rangle_{\rm ss}/(N_J/2)\) as a function of scaled
    detuning \(\Delta_c/(N\Gamma_0)\) and normalized drive
    \(\alpha_{\rm in}/\alpha_{\mathrm{ref}}\), for \(\omega_a=\omega_c\).
    The reference amplitude \(\alpha_{\mathrm{ref}}\) is defined in
    Eq.~\eqref{eq:alpha_ref}.
    \textbf{(a) UVRS:} For \(\kappa/(g\sqrt{N_J})=15\gg1\), cavity
    dissipation overdamps the collective spin--photon response, giving a
    smooth operating landscape over the plotted region.
    \textbf{(b) RVRS:} For \(N=10^5\) and \(\kappa=15g\), coherent
    light--matter coupling is stronger relative to cavity dissipation,
    producing a sharper response and a branch-switching boundary.
    The operating points used for the squeezing protocols below are chosen on the
low-photon, strongly spin-polarized branch and away from this switching
boundary.}
    \label{fig:mean_field_phase_diagram}
\end{figure}

On this branch, the injected field is balanced by the collective dipole
radiated by the atoms, with an additional detuning-dependent quadrature set by
the cavity response.
This balance defines a natural reference input amplitude,
\begin{equation}
\sqrt{\kappa}\alpha_{\mathrm{ref}}\equiv
\sqrt{\left(\frac{\kappa\delta_\sigma}{4g}\right)^{2}
+\left(\frac{g N_J}{2}
-\frac{\Delta_c\delta_\sigma}{2g}\right)^{2}},
\label{eq:alpha_ref}
\end{equation}
which reduces to
\(\sqrt{\kappa}\alpha_{\mathrm{ref}}=gN_J/2\) at
\(\delta_\sigma=0\). Thus, on resonance, \(\alpha_{\mathrm{ref}}\) coincides
with the critical drive scale of the spin-polarized solution. At finite
detuning, it provides a local drive scale that absorbs the trivial
detuning dependence of the mean-field response and allows the UVRS and RVRS
operating regions to be compared on common axes.

The steady-state spin polarization plotted in Fig.~\ref{fig:mean_field_phase_diagram}
is obtained from the polar angle of the selected spin-polarized branch,
\(\langle \hat J^z\rangle_{\rm ss}/(N_J/2)=-\cos\tilde{\theta}_J\).
The steady-state angle $\tilde{\theta}_J$ is determined by solutions to the equation
\begin{equation}
\begin{aligned}
(\sqrt{\kappa}\alpha_{\rm in})^2
&=\left(\frac{\delta_\sigma\kappa}{4g}\right)^2
\tan^2\tilde\theta_J \\
&\quad+
\left[
\frac{gN_J}{2}\sin\tilde\theta_J
-\frac{\Delta_c\delta_\sigma}{2g}\tan\tilde\theta_J
\right]^2,
\end{aligned}
\label{eq:app_mf_potentialMain}
\end{equation}
which can permit multiple solutions.
The coexistence of multiple stable mean-field steady-states corresponds to
bistability in a driven-dissipative collective light--matter system
\cite{Bonifacio1978,Carmichael1980}. As the drive or
detuning is varied, the low-photon branch can terminate or switch
discontinuously to another stable solution with larger photon occupation
and reduced spin polarization, which is the mean-field signature of a
first-order transition. The squeezing protocols considered below are
therefore chosen away from these switching boundaries, and more generally
away from phase-transition regions, where large cavity response and
enhanced sector distinguishability would degrade the geometric OAT
dynamics.

Figure~\ref{fig:mean_field_phase_diagram} shows how the role of photons extends from the
transient dynamics to the steady-state operating landscape. In the UVRS
regime, Fig.~\ref{fig:mean_field_phase_diagram}(a), rapid cavity decay smooths the
response of the spin polarization as detuning and drive are varied. In the
RVRS regime, Fig.~\ref{fig:mean_field_phase_diagram}(b), the intracavity field becomes a
stronger self-consistent backaction channel: the collective spin shapes the
optical field, and the optical field in turn modifies the torque acting on the
spin. This feedback sharpens the branch structure and produces the
branch-switching behavior visible  as the sharp color boundary in Fig.~\ref{fig:mean_field_phase_diagram}(b).

For the squeezing protocol, the relevant conclusion is operational. The
trajectories optimized below are chosen on the low-photon, strongly
spin-polarized branch and away from the branch-switching region. In this window, cavity photons remain dynamically active, but they do not induce
excessive sector-resolving photon leakage that would reduce the correlations
generated by the geometric OAT mechanism.

\section{Effective Spin Hamiltonian via Holstein--Primakoff Approximation}
\label{sec:HP_effective}

To quantify the squeezing generated by our protocol and connect the full
cavity--QED dynamics to a simple collective-spin picture, we derive an
effective spin model via a Holstein--Primakoff (HP) expansion about the
macroscopically polarized trajectory of the driven manifold~\cite{Kurucz2010}.
We expand the collective spin and cavity field as weak bosonic fluctuations
around the low-photon, spin-polarized branch followed during the smooth ramp.
The purpose of the ramp is to suppress large nonadiabatic spin--photon
excitations, so that the excited-state population, cavity occupation, and
transverse spin--cavity fluctuations remain small compared with \(N_J\).
Under these conditions, the linearized fluctuation dynamics can be reduced to
an effective description of the encoded \(S\)-spin sector.

This reduction is distinct from the UVRS adiabatic elimination of the full
cavity field. In RVRS, the cavity remains dynamically active and enters the
effective \(S\)-spin model through the linearized spin--cavity response; only
residual perturbative fluctuations around the selected branch are integrated
out. The validity of this reduction is benchmarked against DissTWA simulations
in Sec.~\ref{sec:Spin_squeezing_dyn}. As in the UVRS treatment of
Ref.~\cite{JeremyOAT}, the resulting spin-only description has the form of a
one-axis twisting master equation, with coherent twisting rate
\(\tilde{\chi}\) and collective dephasing rate \(\tilde{\Gamma}\).

Throughout this section, all mean-field quantities are evaluated on the
low-photon, spin-polarized branch identified above. We denote the population fractions by 
$f_J\equiv \frac{N_J}{N}, f_\uparrow\equiv 1-f_J$, which we set to $f_J=1/2$,
and we use the mean-field angles
\((\tilde{\theta}_J,\tilde{\phi}_J)\) associated with the chosen operating
point. These angles specify the polarized direction of the driven
\(\{\ket{\downarrow},\ket{e}\}\) manifold and are determined from the
mean-field steady-state equations, as described in
Appendix~\ref{app:mean_field_roots}. The technical HP construction, including
the rotating frame, the fluctuation basis, the displaced cavity field, and the
nested elimination of residual fluctuation modes, is given in
Appendix~\ref{app:hp_nested_elimination}. The resulting reduced dynamics for
the collective \(S\)-spin is
\begin{equation}
\hat H_{\mathrm{eff}}=\tilde{\chi}\hat S_z^{2},\qquad
\mathcal{L}[\rho]=\tilde{\Gamma}
\Big(\hat S_z\rho\hat S_z-\tfrac{1}{2}\{\hat S_z^{2},\rho\}\Big),
\label{eq:red_spin_HP}
\end{equation}
where \(\tilde{\chi}\) is the coherent OAT rate and
\(\tilde{\Gamma}\) is the collective dephasing rate generated by residual
sector-resolving photon leakage. Physically, \(\tilde{\Gamma}\) measures how
rapidly the unobserved output field acquires information about the conserved
sector label \(N_J\); in the encoded-spin description, this appears as
dephasing generated by \(\hat S_z\).

Both coefficients are determined by the microscopic spin--cavity parameters
and by the mean-field configuration
\((\tilde{\theta}_J,\tilde{\phi}_J)\). The full expressions
are given in Appendix~\ref{app:hp_nested_elimination}. In the small-detuning
regime
\(|\delta_{\sigma}|,|\Delta_c|\ll\{g\sqrt{N_J},\kappa\}\)
and for large \(N_J\), they reduce to
\begin{subequations}
\begin{align}
\tilde{\chi}
&\approx
-\frac{\tan^{2}\tilde{\theta}_{J}}
{2N\cos\tilde{\theta}_{J}}\delta_{\sigma}
\left(
1+\frac{2\Delta_{c}\delta_{\sigma}}{g^{2}N}
\sec^{3}\tilde{\theta}_{J}\right),
\\[6pt]
\tilde{\Gamma}
&\approx
\frac{\tan^{2}\tilde{\theta}_{J}\delta_{\sigma}^{2}}
{g^{2}N^{2}\cos^{4}\tilde{\theta}_{J}}\kappa
\left(
1+\frac{4\Delta_{c}\delta_{\sigma}}{g^{2}N}
\sec^{3}\tilde{\theta}_{J}\right).
\end{align}
\label{eq:tilde-chi-leading}
\end{subequations}
These rates expose the local competition between coherent geometric twisting
and collective dephasing. To leading order,
\begin{equation}
\frac{\tilde{\chi}}{\tilde{\Gamma}}
\approx
-\frac{N g^{2}\cos^{3}\!\tilde{\theta}_J}{2\kappa\delta_{\sigma}}
\left(
1-\dfrac{2\Delta_{c}\delta_{\sigma}}{g^{2}N}
\sec^{3}\!\tilde{\theta}_{J}
\right).
\label{eq:chi_over_Gamma}
\end{equation}
In the small-detuning limit, Eq.~\eqref{eq:tilde-chi-leading} gives
\(\tilde{\chi}\propto\delta_\sigma\) and
\(\tilde{\Gamma}\propto\delta_\sigma^2\). Thus
\(|\tilde{\chi}|/\tilde{\Gamma}\) increases as
\(|\delta_\sigma|\) is reduced. Physically, \(\tilde{\chi}\) is a coherent
phase response and is therefore linear in the sector-dependent detuning-induced
shift. By contrast, \(\tilde{\Gamma}\) is a dephasing rate set by the
distinguishability of the sector-conditioned output fields, which scales as the
square of their small separation. Reducing \(|\delta_\sigma|\) therefore
improves the coherent-to-dissipative ratio, but also reduces the absolute
twisting rate and lengthens the time needed to generate squeezing.

The HP rates therefore provide local intuition, but not by themselves the full
optimization criterion. They assume that the protocol remains on the
low-photon, spin-polarized branch and that residual spin--cavity fluctuations
remain perturbative throughout the ramps. The next section tests these
assumptions dynamically by comparing the HP prediction with full DissTWA
simulations in the UVRS and RVRS regimes. Local single-particle decoherence is
included later in Sec.~\ref{sec:single_particle_decoh}.

\begin{figure*}[t!]
  \centering

  \subfloat[\textbf{UVRS}]{%
    \includegraphics[width=0.47\linewidth]{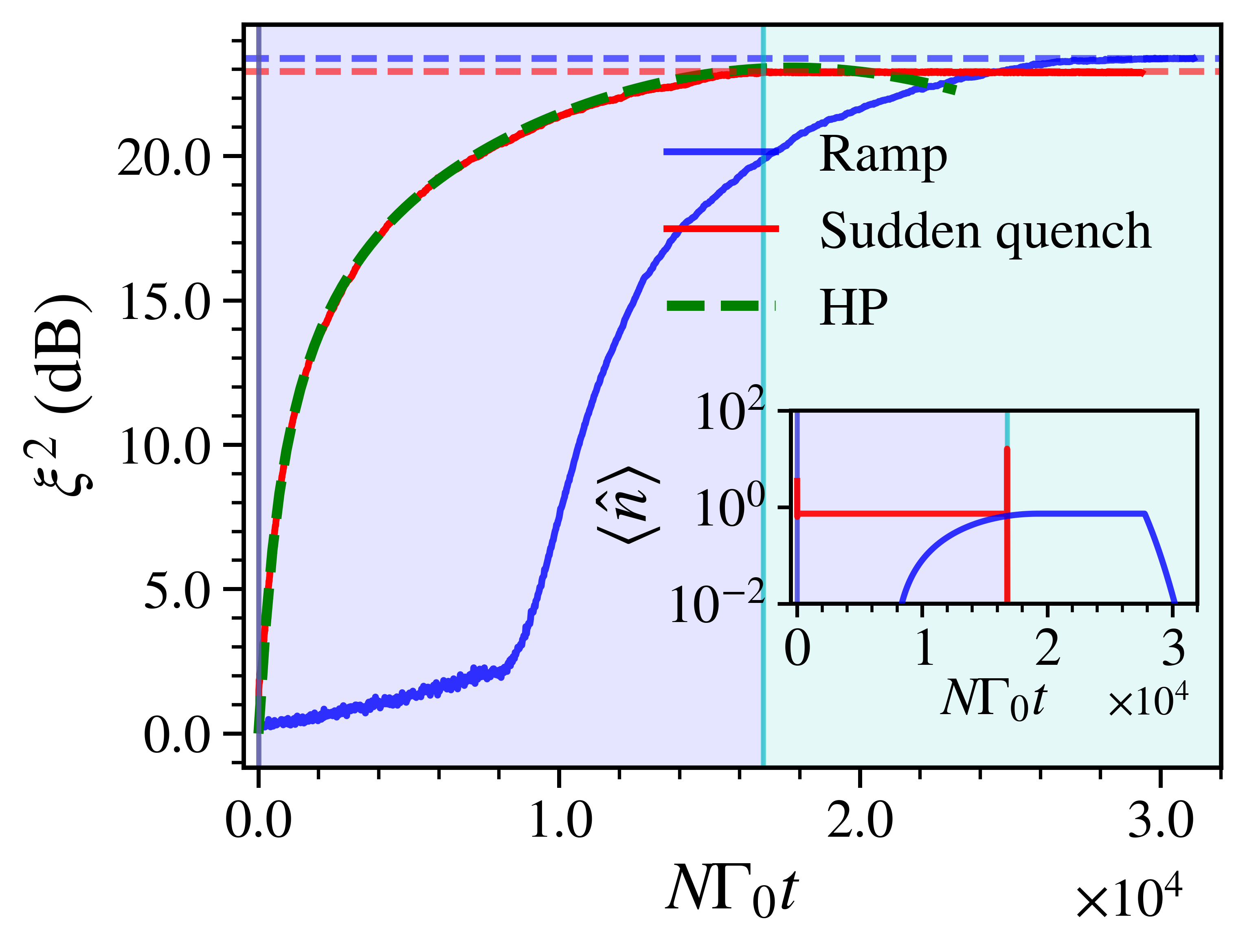}
    \label{fig:quad_uvrs}}
  \quad
  \subfloat[\textbf{RVRS}]{%
    \includegraphics[width=0.47\linewidth]{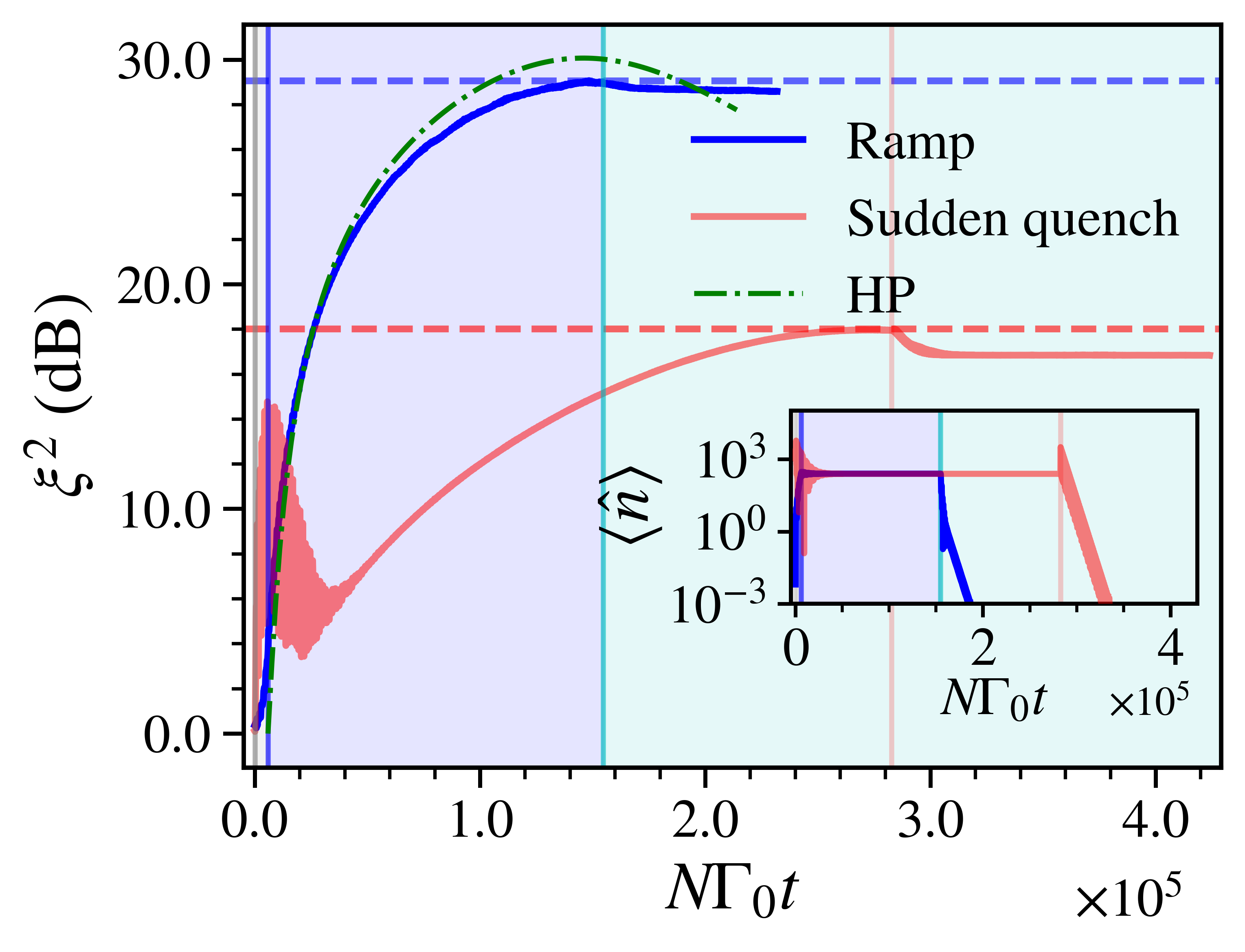}
    \label{fig:spin_sq_quad_rvrs}}

  \caption{{\bf Spin-squeezing dynamics.} Here we compare the ramped protocol
(blue) and a sudden quench (red) with the HP predictions (green). The
background color indicates the three stages of the multistage protocol. The
insets show the intracavity photon number \(\langle \hat n\rangle\), which
remains low for the ramped protocol but can increase transiently after sudden
switching.
In both panels $N=10^5$ and $\alpha_{\rm in}/\alpha_{\rm ref}=0.5$.
The time axis is reported in units of $N\Gamma_0 t$, with $\Gamma_0$ evaluated using the panel-specific cavity linewidth $\kappa$.
(a) UVRS: $\kappa=15g\sqrt{N_J}$ and $\Delta_c=2.5\times10^{-2}N\Gamma_0$.
The fast cavity relaxation makes the final squeezing plateau largely insensitive to abrupt changes;
(b) RVRS: $\kappa=15g$ and $\Delta_c=2\times10^{-3}N\Gamma_0$, $\delta_{\sigma}=\Delta
_c$.
Sudden changes in the drive amplitude excite spin--photon dynamics and produce a large transient photon population, which degrades the squeezing relative to
the smooth ramped protocol.}
  \label{fig:quad_uvrs_rvrs}
\end{figure*}

\section{Spin-squeezing dynamics}
\label{sec:Spin_squeezing_dyn}

In this section we test whether the three-stage protocol generates
metrologically useful squeezing while remaining on the intended low-photon,
spin-polarized branch. We use two complementary descriptions. First, we
simulate the full driven-dissipative multilevel atom--cavity dynamics using
the dissipative truncated-Wigner approximation
(DissTWA)~\cite{Polkovnikov2010,Schachenmayer2015a,Zhu2019,Huber2022,Hosseinabadi2025}.
Second, we compare these results with the Holstein--Primakoff (HP) treatment
of Sec.~\ref{sec:HP_effective}, which provides the effective OAT description
within the low-photon branch. This comparison tests when the effective OAT description captures the full
squeezing dynamics and whether smooth ramps are required to suppress
spin--photon excitations during switching.

To quantify metrological gain, we use the Wineland squeezing parameter
adapted to the instantaneous multilevel spin direction,
\begin{equation}
\xi^2(t)\equiv
\frac{N\,\min_{\perp}\mathrm{Var}\!\left[\hat{\Lambda}_{\perp}(t)\right]}
{\left|\langle \hat{\bm{\Lambda}}_{\rm spin}(t)\rangle\right|^2}.
\label{eq:xi2_eig_form}
\end{equation}
Here \(\hat{\bm{\Lambda}}_{\rm spin}(t)\) is the multilevel generalization of the spin components ($\hat{\bm{\Lambda}}_\perp$ denotes its perpendicular fluctuations). The minimization is analogous to the usual
Wineland minimization over collective quadratures transverse to
\(\langle\hat{\bm{\Lambda}}_{\rm spin}(t)\rangle\), implemented through the
covariance matrix described in Appendix~\ref{app:disstwa_covariance}. This expression reduces to standard two-level Wineland squeezing parameter for $S$ when we restrict it to the subspace \(\{\ket{0},\ket{1(t)}\}\), where
\(\ket{0}\equiv\ket{\uparrow}\) is the spectator clock state and
\(\ket{1(t)}\) is the instantaneous single-particle steady state in the driven
\(\{\ket{\downarrow},\ket e\}\) manifold.  This definition accounts for both the reduction of transverse variance and
the reduction of the mean spin length.
During Stage III, the squeezing
extracted in this basis is mapped onto the long-lived clock manifold
\(\{\ket{\uparrow},\ket{\downarrow}\}\).

\begin{figure*}[t]
    \centering

    \begin{tikzpicture}
        \node[anchor=south west, inner sep=0] (img) at (0,0)
        {\includegraphics[width=0.94\linewidth]{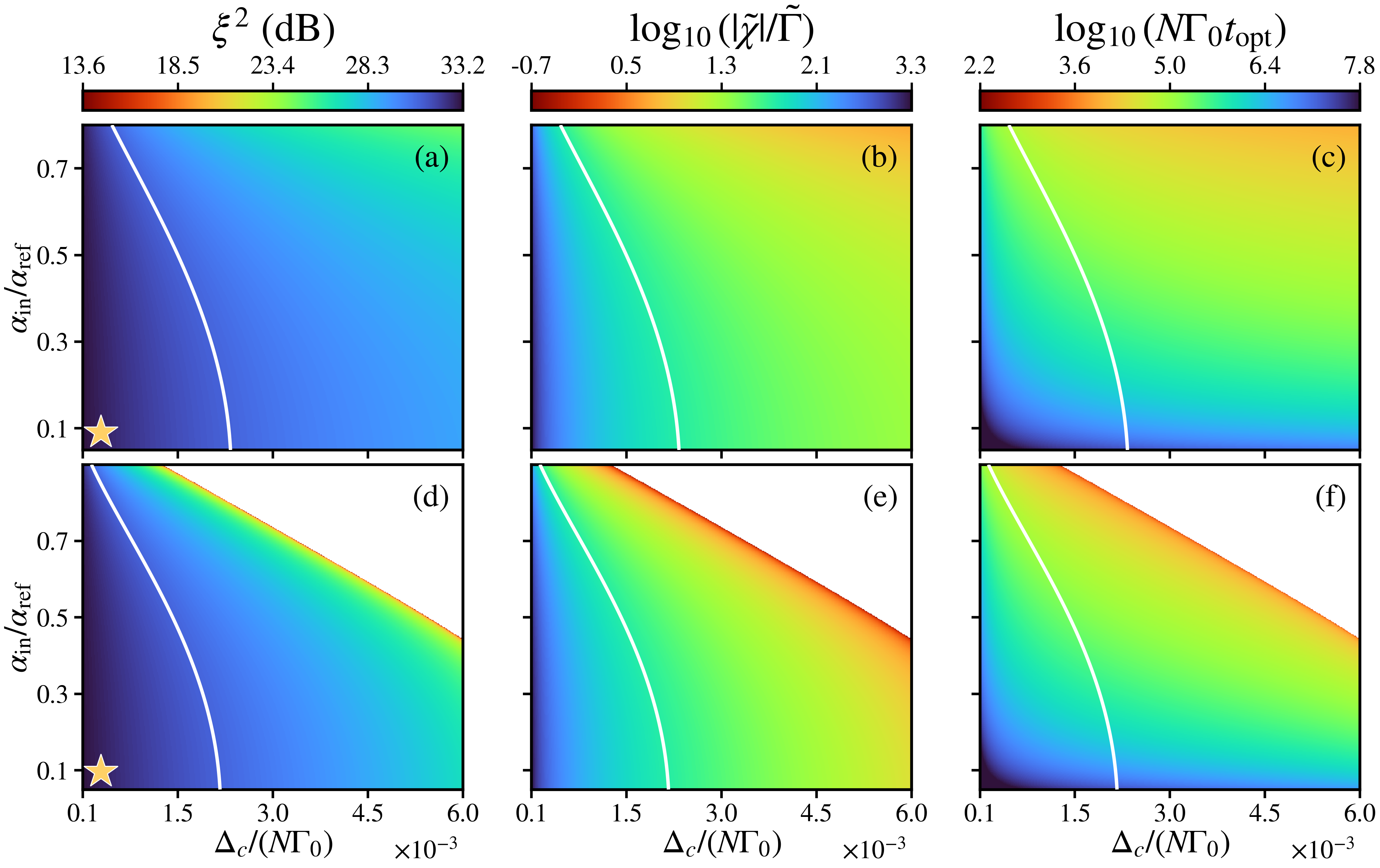}};

        \begin{scope}[x={(img.south east)}, y={(img.north west)}]
            \node[rotate=90, font=\bfseries\large, anchor=center]
            at (-0.045,0.70) {UVRS};

            \node[rotate=90, font=\bfseries\large, anchor=center]
            at (-0.045,0.30) {RVRS};
        \end{scope}
    \end{tikzpicture}

    \caption{{\bf Spin squeezing in the UVRS and RVRS regimes for \(N=10^{5}\) atoms}.
Panels (a,d) show the best squeezing achieved, panels (b,e) show the ratio
\( |\tilde{\chi}|/\tilde{\Gamma} \), and panels (c,f) show the logarithm of the
normalized time \( N\Gamma_0 t_{\mathrm{min}} \) required to reach the optimum
squeezing. The top row corresponds to UVRS and the bottom row to RVRS. The stars mark the optimal spin-squeezing conditions within the displayed
window. When an optimum lies on the lower boundary
\(\alpha_{\rm in}/\alpha_{\rm ref}=0\), the corresponding star is plotted
slightly above the boundary to enhance visibility; its displayed vertical position
does not indicate a finite optimal drive. The white solid contour marks the
crossover between the ideal OAT regime and the collective-dephasing-dominated
regime following Eq.~(\ref{eq:Transition_ideal}). Masked regions lie beyond the
stable spin-polarized branch, i.e.\ beyond the branch-switching region shown in
Fig.~\ref{fig:mean_field_phase_diagram}. Comparisons between UVRS and RVRS highlight how
this boundary compresses and sharpens the useful operating region, while the
optimized squeezing and scaled optimum times remain broadly comparable across
the stable spin-polarized branch (see text).}
    \label{fig:spin_squeezingcoll}
\end{figure*}

We now use $\xi^2(t)$
to study the three-stage dynamics.  In Fig.~\ref{fig:quad_uvrs_rvrs}, we compare \(\xi^2(t)\) obtained from DissTWA
with the HP dynamics generated by the effective rates \(\tilde{\chi}\) and
\(\tilde{\Gamma}\), for which we simulate the corresponding spin-only dynamics
exactly~\cite{Chu2021}.
We also compare the behavior under ramps and quenches for both UVRS and RVRS.
A useful diagnostic of nonadiabatic spin--photon excitation during ramping is
the intracavity photon number
\(n(t)=\langle \hat a^\dagger \hat a\rangle(t)\) (insets of
Fig.~\ref{fig:quad_uvrs_rvrs}). When such excitations occur, they produce a
transient increase in \(n(t)\), increasing photon leakage. 
This additional photon leakage degrades the attainable metrological gain.
The two examples shown in Fig.~\ref{fig:quad_uvrs_rvrs} are chosen as representative operating points within the stable spin-polarized regime identified in 
Fig.~\ref{fig:mean_field_phase_diagram}, so that the comparison isolates the role of cavity dynamics rather than branch- switching near the bistable boundary.

\paragraph*{UVRS dynamics.}
In Fig.~\ref{fig:quad_uvrs_rvrs}(a) we work at detuning $\Delta_c=2.5\times10^{-2}N\Gamma_0$. The cavity is broad,
$\kappa=15g\sqrt{N_J}$, and relaxes rapidly on the timescale $\kappa^{-1}$.  Consequently, once the drive parameters reach Stage~II the field adiabatically follows the atoms, and the squeezing dynamics is set  by the
effective OAT parameters: both the ramp (blue) and sudden quench (red) approach the same long-time squeezing and
are consistent with the HP prediction (green).  The difference between protocols is mainly confined to the driving  protocol: the
quench produces a short spike in $\langle \hat n(t) \rangle$ (inset), i.e.\ a brief period of enhanced photon leakage, but this transient is
quickly damped and leaves little imprint on the later evolution, whereas the ramp smooths the approach and delays the
onset of the strong-shearing portion of Stage~II. In Ref.~\cite{JeremyOAT}, it was proven exactly that the squeezing is fully preserved when the drive is quenched off in the large $N$ limit of the pure spin model.

\paragraph*{RVRS dynamics.}
In Fig.~\ref{fig:quad_uvrs_rvrs}(b) we choose $\Delta_c=2\times10^{-3}N\Gamma_0$. The cavity is narrow, $\kappa=15g$,
so its dynamics remains relevant on timescales comparable to the squeezing evolution.  In this regime, abrupt
changes can strongly excite polariton modes, generating a macroscopic cavity field and associated light--matter
coherence.  This is visible as a large and persistent peak in $\langle \hat n(t)\rangle$ after quenches (red).  Because the
stored field then radiates over an extended time, the system experiences prolonged photon leakage precisely during the
stage where correlations should build,  adding excess fluctuations that reduce the
attainable metrological gain \cite{Gambetta2006,Gambetta2008}. 

The ramped protocol (blue) mitigates this effect by turning the detunings and
drive on and off slowly, keeping the cavity field close to its instantaneous
low-photon response.
Under these conditions, the dynamics is governed primarily by the effective
rates \((\tilde{\chi},\tilde{\Gamma})\), and \(\xi^2(t)\) tracks the HP
prediction much more closely. The remaining small late time deviations can be attributed to effects beyond
the ideal HP/OAT reduction, such as residual nonadiabatic spin--photon
excitations or non-Gaussian correlations not captured by the HP treatment.

Taken together, Fig.~\ref{fig:quad_uvrs_rvrs} shows that ramping is not
critical in the UVRS regime, where photon transients generated during
switching are rapidly damped and leave little imprint on the subsequent
squeezing dynamics. In the RVRS regime, by contrast, ramping is essential: it
suppresses long-lived spin--photon excitations, reduces leakage-induced
dephasing, and restores agreement with the effective HP/OAT description.

\subsection{Numerical optimization of spin squeezing}
\label{subsec:opt_squeezing}

We now optimize the squeezing over detuning and drive amplitude within the
stable spin-polarized branch. Fig.~\ref{fig:spin_squeezingcoll} summarizes the
resulting squeezing, coherent-to-dissipative ratio, and optimal time. We scan
both the scaled detuning
\(\Delta_c/(N\Gamma_0)=\delta_\sigma/(N\Gamma_0)\) (horizontal axis) and the
normalized drive amplitude \(\alpha_{\text{in}}/\alpha_{\rm ref}\) (vertical
axis). The left column [panels (a,d)] reports the optimal squeezing $\xi^2_{\min}$ (in
dB) for a given set of parameters, the middle column [panels (b,e)] shows the diagnostic ratio $\log_{10}(|\tilde{\chi}|/\tilde{\Gamma})$, which benchmarks the competition between coherent elastic
shearing and inelastic collective dephasing, and the
right column [panels (c,f)] gives the corresponding optimal squeezing time $\log_{10}(N\Gamma_0 t_{\min})$.  The top row
corresponds to the UVRS limit and the bottom row to the RVRS limit.
The practical goal of these maps is to identify operating points that combine large coherent shear, weak collective dephasing, short preparation time, and sufficient distance from the branch-switching boundary in the RVRS regime.

We can interpret these trends using the standard analytic approximation for one-axis twisting in the presence of collective dephasing \cite{Schleier-Smith2010,LewisSwan2018,Chu2021,Baamara2022,JeremyOAT},
\begin{equation}
\xi^{2}(t)\approx 
\frac{1+N\tilde{\Gamma}t}{N^{2}\tilde{\chi}^{2}t^{2}}
+\frac{N^{2}\tilde{\chi}^{4}t^{4}}{6}.
\label{eq:Spin_squeez_ideal}
\end{equation}
Optimizing this HP/OAT estimate with respect to time shows that, in the dephasing-dominated regime
$N\tilde{\Gamma}t_{\min}\gg 1$, the optimal squeezing depth and the scaled optimum time are governed by the ratio
$\tilde{\Gamma}/|\tilde{\chi}|$:
\begin{equation}
\begin{aligned}
\xi_{\min}^{2}
\approx
\frac{1.15}{N^{2/5}}
\left(\frac{\tilde{\Gamma}}{|\tilde{\chi}|}\right)^{4/5},\quad
|\tilde{\chi}|t_{\min}
\approx
\frac{1.08}{N^{3/5}}
\left(\frac{\tilde{\Gamma}}{|\tilde{\chi}|}\right)^{1/5}.
\end{aligned}
\label{eq:Analytic_ideal}
\end{equation}

By contrast, when collective dephasing is weak enough that $N\tilde{\Gamma}t_{\min}\ll 1$, the dynamics reduce to the pure OAT limit,
\begin{equation}
\xi_{\mathrm{min}}^{2}=\frac{3^{2/3}}{2N^{2/3}},
\qquad
|\tilde{\chi}|t_{\mathrm{min}}=\frac{3^{1/6}}{N^{2/3}}  .
\label{eq:S40}
\end{equation}
This limit provides the best-case HP benchmark for the protocol: once collective dephasing is included, the attainable squeezing can only worsen relative to the ideal OAT prediction.

To indicate where the optimized dynamics crosses over from the ideal OAT regime to the collective-dephasing-dominated regime, in Fig.~\ref{fig:spin_squeezingcoll} we plot the white contour obtained by equating the two asymptotic expressions for the optimal squeezing. Matching Eq.~(\ref{eq:Analytic_ideal}) to the ideal OAT result, Eq.~(\ref{eq:S40}), gives
\begin{equation}
\frac{\tilde{\Gamma}}{|\tilde{\chi}|}
= 0.882N^{-1/3}  
\label{eq:Transition_ideal}
\end{equation}
We therefore use this line as a guide to the transition between the ideal OAT-like region, where the optimum remains close to the unitary OAT benchmark, and the region where collective dephasing already controls the achievable squeezing.

In the HP/OAT description, regions with larger $|\tilde{\chi}|/\tilde{\Gamma}$ yield a smaller optimal squeezing parameter $\xi_{\min}^2$. The same ratio also controls the scaled time $|\tilde{\chi}|t_{\min}$, while the physical optimum time $t_{\min}$ depends additionally on the absolute magnitude of $|\tilde{\chi}|$. This is consistent with the strong correlation between the squeezing map in Fig.~\ref{fig:spin_squeezingcoll}(a,d) and the ratio map in Fig.~\ref{fig:spin_squeezingcoll}(b,e), while the detailed structure of the time map in Fig.~\ref{fig:spin_squeezingcoll}(c,f) reflects both the ratio $\tilde{\Gamma}/|\tilde{\chi}|$ and the overall timescale set by $\tilde{\chi}$.

\begin{table*}[!htbp]
\centering
\renewcommand{\arraystretch}{1.15}
\setlength{\tabcolsep}{6pt}
\caption{Summary of the asymptotic time-optimized squeezing regimes obtained
from Eq.~\eqref{eq:Spin_squeez_deph}. The first column labels the limiting
solutions, while the second column lists the validity conditions under which
the corresponding expressions for \(t_{\mathrm{min}}\) and
\(\xi_{\mathrm{opt}}^2\) apply.}
\label{tab:regime_summary}

\begin{tabular}{@{}clcc@{}}
\toprule
\toprule
\textbf{Regime}
& \textbf{Validity Conditions}
& \textbf{\(\bm t_{\mathrm{min}}\)}
& \textbf{\(\bm \xi_{\mathrm{opt}}^2\)}
\\
\midrule

\multirow{2}{*}{(I)}
& \(\gamma_d t_{\mathrm{min}} \ll 1\)
& \multirow{2}{*}{\(\displaystyle t_{\mathrm{min}}^{(I)}=\frac{(3N^{2})^{1/6}}{N|\tilde{\chi}|}\)}
& \multirow{2}{*}{\(\displaystyle \xi_{\mathrm{opt},I}^{2}=\frac{3^{2/3}}{2}N^{-2/3}\)}
\\
& \(N\tilde{\Gamma}t_{\mathrm{min}} \ll 1\)
\\
\midrule

\multirow{3}{*}{(II)}
& \(\gamma_d t_{\mathrm{min}} = 1\)
& \multirow{3}{*}{\(\displaystyle t_{\mathrm{min}}^{(II)}=\frac{1}{\gamma_d}\)}
& \multirow{3}{*}{\(\displaystyle \xi_{\mathrm{opt},II}^{2}=e^{2}\!\left(\frac{\gamma_d}{N|\tilde{\chi}|}\right)^{2}\)}
\\
& \(N\tilde{\Gamma} \ll e\gamma_d\)
\\
& \(\gamma_d^{6}/(N^{4}|\tilde{\chi}|^{6}) \gg 1/(6e)\)
\\
\midrule

\multirow{3}{*}{(III)}
& \(\gamma_d t_{\mathrm{min}} \sim 1\)
& \multirow{3}{*}{\(\displaystyle t_{\mathrm{min}}^{(III)} \approx \frac{1}{\gamma_d}\)}
& \multirow{3}{*}{\(\displaystyle \xi_{\mathrm{opt},III}^{2}\approx \frac{3^{2/3}e^{5/3}}{2N^{2/3}}+e\frac{\gamma_d\tilde{\Gamma}}{N|\tilde{\chi}|^{2}}\)}
\\
& \multirow{2}{*}{\(\displaystyle \frac{\gamma_d\tilde{\Gamma}}{N|\tilde{\chi}|^{2}} \gtrsim \frac{3^{2/3}}{2N^{2/3}}\)}
\\
\\

\bottomrule
\bottomrule
\end{tabular}
\end{table*}

In the UVRS regime (top) we observe a smooth landscape.
The dependence on $(\alpha_{\text{in}},\Delta_c)$ is smooth across all panels, with no sharp boundaries.
The best squeezing in Fig.~\ref{fig:spin_squeezingcoll}(a) is confined to the weak-drive, small-detuning region.  This behavior is
directly  correlated with the ratio $\log_{10}(|\tilde{\chi}|/\tilde{\Gamma})$, shown in Fig.~\ref{fig:spin_squeezingcoll}(b). As
$|\tilde{\chi}|/\tilde{\Gamma}$ decreases,  $\xi^2_{\min}$ degrades.
The time map in Fig.~\ref{fig:spin_squeezingcoll}(c) makes the associated cost explicit. The same UVRS region that maximizes $|\tilde{\chi}|/\tilde{\Gamma}$ also tends to produce the largest $t_{\min}$. Thus, already at the HP level, the UVRS regime exhibits a clear speed--performance trade-off: pushing toward the best $\xi_{\min}^2$ generally requires longer preparation times, thereby increasing the sensitivity to additional decoherence channels discussed in Sec.~\ref{sec:single_particle_decoh}.

In the RVRS regime, we observe clear signatures of the branch-switching region
shown in Fig.~\ref{fig:mean_field_phase_diagram}. The RVRS maps (bottom row) display
qualitatively new structure: the degradation of squeezing is accompanied by a
drop in \(|\tilde{\chi}|/\tilde{\Gamma}\), shown in
Fig.~\ref{fig:spin_squeezingcoll}(e), and by a change in the optimal time,
shown in Fig.~\ref{fig:spin_squeezingcoll}(f). This boundary occurs in the same
high-drive/high-detuning region where the mean field displays bistable
behavior. 
Away from this transition line---well inside the stable spin-polarized sector---RVRS supports a broad
band of significant  squeezing as shown  in Fig.~\ref{fig:spin_squeezingcoll}(d). This band is only slight narrower than
in the UVRS case, compressed slightly due to the transition line.

The scaled time $\log_{10}(N\Gamma_0t_{\textrm{min}})$ map in Fig.~\ref{fig:spin_squeezingcoll}(f) indicates that the near-optimal RVRS squeezing region occurs at values of \(N\Gamma_0 t_{\min}\) comparable to those shown in Fig.~\ref{fig:spin_squeezingcoll}(c). 
Thus, the main implication of panels (d)--(f) is not a distinct reduction of the scaled optimal time \(N\Gamma_0 t_{\min}\), but rather the existence of an extended parameter region where strong squeezing and experimentally reasonable preparation times coexist. However, because both the horizontal axis and the time map are expressed in scaled units, this comparison should not be read as a direct comparison of laboratory detuning widths or laboratory preparation times.

\section{Single-particle decoherence effects}
\label{sec:single_particle_decoh}

Having benchmarked the effective OAT description against full spin--cavity
DissTWA dynamics in the low-photon operating regime, we now use this reduced
description to include local single-particle decoherence. So far, we have focused
on the cavity-mediated collective processes that generate metrological gain,
namely the coherent shear \(\tilde{\chi}\) and the collective dephasing
\(\tilde{\Gamma}\). In a realistic implementation, however, these collective
processes compete with local noise acting independently on each atom. Relevant sources include spontaneous emission from the optically excited level, inhomogeneous light shifts, magnetic-field noise, residual Doppler shifts, and weak spatial nonuniformities of the cavity mode. In this section, we focus on the local decoherence mechanism inherent to the implementation considered here and analyze how it reshapes the optimized squeezing.

For the alkaline-earth implementation of interest here
(i.e., \({}^{87}\mathrm{Sr}\) with
\(\ket{\downarrow}\equiv{}^{1}\mathrm{S}_{0}\),
\(\ket e\equiv{}^{3}\mathrm{P}_{1}\), and
\(\ket{\uparrow}\equiv{}^{3}\mathrm{P}_{0}\)), the dominant microscopic decay channel is spontaneous emission from \(\ket e\) back to \(\ket{\downarrow}\),
\begin{equation}
\hat L_{i,\downarrow}=\sqrt{\gamma_{e\downarrow}}\ket{\downarrow_i}\!\bra{e_i}.
\label{eq:micro_jump_dephasing}
\end{equation}
As introduced in Sec.~\ref{sec:Multistage} and Eq.~\eqref{eq:1defm_mean_sector},
the relevant qubit manifold \(\{\ket{0},\ket{1}\}\) is associated with the
\(S\)-spin Bloch sphere. In the present three-level scheme, the encoded state \(\ket{1}\) contains a finite admixture of the optically excited state \(\ket e\), with weight set by the mixing angle \(\tilde{\theta}_J\). 
In the encoded-spin picture, spontaneous emission from the optically excited level acts primarily as local dephasing. The reason is that the dressed state \(|1\rangle\) contains a small \(|e\rangle\) admixture, whereas the spectator state \(|0\rangle=|{\uparrow}\rangle\) does not. A spontaneously emitted photon can therefore reveal whether an atom occupied the driven manifold, suppressing the \(|0\rangle\)--\(|1\rangle\) coherence that carries the metrological squeezing.
The corresponding effective local scattering rate is proportional to the excited-state weight \(\sin^2(\tilde\theta_J/2)\). 
In the reduced description, it acts as an effective local pure-dephasing channel with rate
\begin{equation}
\gamma_d = \gamma_{e\downarrow}\sin^2\!\Big(\frac{\tilde{\theta}_J}{2}\Big).
\label{eq:gamma_d_scaling}
\end{equation}

To capture this competition while retaining the cavity-induced twisting and collective dephasing, we augment the effective master equation with local dephasing only. The resulting master equation is
\begin{align}
\label{eq:LI_local}
\dot{\rho} = -\mathrm{i}[\hat{H},\rho]
+\Gamma_z\mathcal{D}[\hat S_z]\rho
+\gamma_d\sum_{i=1}^{N}\mathcal{D}[\hat\sigma_i^z]\rho .
\end{align}
To access large atom numbers while retaining the essential squeezing physics, we work within the effective HP/OAT description introduced in Sec.~\ref{sec:HP_effective}. Accordingly, \(\hat H\) denotes the effective collective twisting Hamiltonian, and we identify the collective dephasing rate as \(\Gamma_z=\tilde{\Gamma}\). This reduced model is the one used throughout the analysis below.

To make a fair comparison between the UVRS and RVRS regimes in the presence of local decoherence, we later benchmark the time-optimized squeezing at fixed single-atom cavity cooperativity \cite{Kimble1998,TanjiSuzuki2011}. The cooperativity ($C$) is defined as
\begin{equation}
C \equiv \frac{4g^{2}}{\kappa\gamma_{e\downarrow}},
\label{eq:C0_def}
\end{equation}
so that the corresponding collective cooperativity is \(NC\) \cite{Tuchman2006}.
This cooperativity sets the fundamental ratio of cavity-mediated interaction strength to spontaneous-emission loss, and therefore the relevant limitation for optimized squeezing once detuning, drive strength, and evolution time are all allowed to vary~\cite{Barberena2024T}. Fixing both \(C\) and \(N\) therefore enforces the same interaction-to-local-noise budget in both cavity regimes. Any remaining differences in optimized squeezing can then be attributed primarily to how the cavity response reshapes the effective coefficients \((\tilde{\chi},\tilde{\Gamma})\), rather than to a trivial change in the ratio of coherent coupling to single-particle loss.

In the RVRS
implementation, we use the stronger \(\sigma^+\)-polarized
\({}^{1}\mathrm{S}_{0}\)--\({}^{3}\mathrm{P}_{1}\) transition, $\ket{\downarrow}\equiv
\ket{{}^{1}\mathrm{S}_{0},F=9/2,m_F=9/2}$, 
$\ket e\equiv
\ket{{}^{3}\mathrm{P}_{1},F=11/2,m_F=11/2}$, and take the long-lived spectator state to be the clock state $\ket{\uparrow}\equiv
\ket{{}^{3}\mathrm{P}_{0},F=9/2,m_F=9/2}$. This configuration is motivated by Sr cavity-QED experiments operating in or
near the resolved vacuum Rabi splitting regime
\cite{NorciaThompson2016PRA,Eric_CRF}.

For the matched-cooperativity UVRS benchmark, we consider a physically
motivated effective three-level configuration based on the
\({}^{87}\mathrm{Sr}\) clock transition, rather than using the same optical
transition with a different cavity linewidth~\cite{Muniz2021}. Specifically,
we take the driven transition to be the \(\pi\)-polarized clock transition $\ket{\downarrow}\equiv
\ket{{}^{1}\mathrm{S}_{0},F=9/2,m_F=9/2}
\leftrightarrow
\ket e\equiv
\ket{{}^{3}\mathrm{P}_{0},F=9/2,m_F=9/2}$,
which has a favorable Clebsch--Gordan coefficient for the stretched
\(\pi\) transition. We choose the spectator state to be a more separated
ground-manifold Zeeman sublevel, $\ket{\uparrow}\equiv
\ket{{}^{1}\mathrm{S}_{0},F=9/2,m_F=5/2}$.

With this choice, the encoded spin is formed from two long-lived Zeeman
sublevels of the \({}^{1}\mathrm{S}_{0}\) ground manifold, while the cavity
drive resonantly addresses the
\(\ket{\downarrow}\leftrightarrow\ket e\) transition. Coupling of
\(\ket{\uparrow}\) to the corresponding
\(\ket{{}^{3}\mathrm{P}_{0},F=9/2,m_F=5/2}\) clock state is assumed to be suppressed
by the Zeeman splitting and the frequency selectivity of the driven cavity
mode. This more separated state is chosen to suppress unwanted
spin flips. The much weaker dipole matrix element of the
\({}^{1}\mathrm{S}_{0}\)--\({}^{3}\mathrm{P}_{0}\) clock transition then places
this benchmark deep in the UVRS regime even for large \(N\).

\begin{figure*}[t!]
    \centering

    \includegraphics[width=0.98\textwidth]{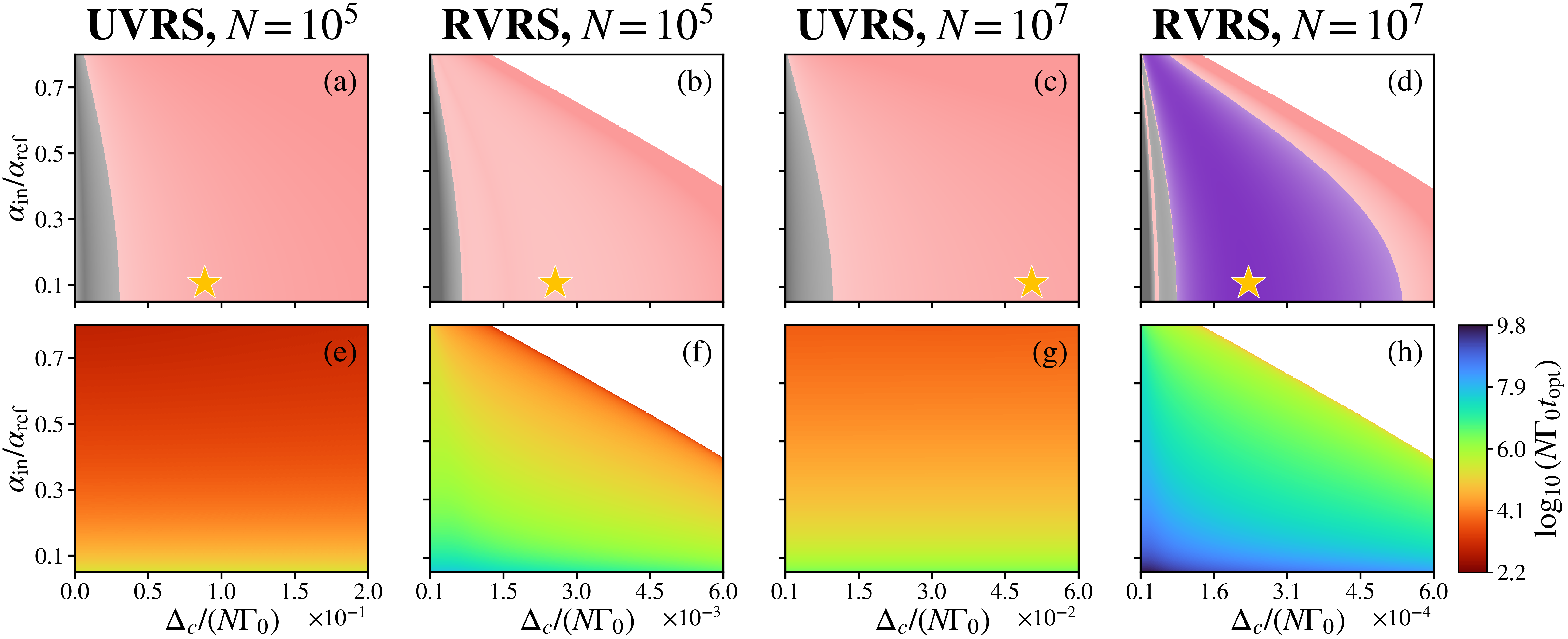}

    \caption{{\bf Spin-squeezing classification and optimal-time maps with local decoherence.}
Columns, from left to right, show UVRS at \(N=10^{5}\) with
\(C=1.59\times10^{-3}\), RVRS at \(N=10^{5}\) with
\(C=0.355\), UVRS at \(N=10^{7}\) with
\(C=1.59\times10^{-4}\), and RVRS at \(N=10^{7}\) with
\(C=0.355\). Top row: classification of the optimized dynamics in the
\((\Delta_c/(N\Gamma_0),\alpha_{\rm in}/\alpha_{\rm ref})\) plane according
to Table~\ref{tab:regime_summary}; purple, gray, and pink denote regimes
I, II, and III, respectively. Bottom row: normalized optimal time
\(\log_{10}(N\Gamma_0 t_{\min})\). The stars mark the optimal conditions within
the displayed window. When an optimum lies on the lower boundary
\(\alpha_{\rm in}/\alpha_{\rm ref}=0\), the corresponding star is plotted
slightly above the boundary to enhance visibility; its displayed vertical position
does not indicate a finite optimal drive. UVRS is implemented with
\(\kappa=15g\sqrt{N/2}\), while RVRS uses \(\kappa=15g\), with
\(g=10~\mathrm{kHz}\). Local decoherence is included through
Eq.~\eqref{eq:LI_local}, with
\(\gamma_{e\downarrow}=7.5~\mathrm{kHz}\) in both regimes.}
    \label{fig:spin_squeezingdecoh_8panel}
\end{figure*}

In the HP/OAT regime, local dephasing limits the usable evolution time by reducing the coherence
and increasing the noise. A compact analytic approximation that incorporates this effect is \cite{Chu2021,Barberena2023,JeremyOAT}
\begin{equation}
\xi^{2}(t)\approx e^{\gamma_d t}\left[
\frac{e^{\gamma_d t}+N\tilde{\Gamma}t}{N^{2}\tilde{\chi}^{2}t^{2}}
+\frac{N^{2}\tilde{\chi}^{4}t^{4}}{6}
\right],
\label{eq:Spin_squeez_deph}
\end{equation}
which makes explicit how a finite \(\gamma_d\) shifts the optimum to earlier times and caps the achievable gain.
With the local dephasing scale fixed by Eq.~(\ref{eq:gamma_d_scaling}), we now
classify how the optimum squeezing is modified by the
interplay of coherent twisting, collective dephasing, and
single-particle dephasing. Because this optimization
includes local dephasing, the resulting optimal operating point in the
\((\Delta_c,\alpha_{\rm in})\) plane can differ from the ideal OAT optimum
obtained in the absence of single-particle decoherence. Eq.~(\ref{eq:Spin_squeez_deph}) provides the basis for
the regime classification summarized in Table~\ref{tab:regime_summary}.
The labels I--III denote asymptotic
limits of the time-optimized solution, not distinct dynamical phases. The
conditions in the second column are the self-consistency conditions obtained
after substituting the corresponding optimal time back into
Eq.~\eqref{eq:Spin_squeez_deph}.

At a high level, the three regimes distinguish which processes have accumulated
by the time the squeezing optimum is reached. Regime~I is the nearly ideal-OAT
limit, where the optimum is reached before either local dephasing or collective
photon leakage has appreciably acted. Regime~II is local-dephasing limited: the
usable evolution time is cut off by \(1/\gamma_d\), while collective dephasing
remains perturbative. Regime~III is a mixed dephasing regime, where collective
photon leakage also accumulates on the local-dephasing timescale and contributes
at leading order to the optimum squeezing.

The distinction between Regimes~II and~III is the role played by
\(\tilde{\Gamma}\) on the local-dephasing timescale. In Regime~II, the usable
evolution time is set by \(1/\gamma_d\), but collective dephasing remains
subleading, so the optimum is essentially contrast-limited. In Regime~III, the
usable time is still set by \(1/\gamma_d\), but sector-resolving photon leakage
has accumulated enough collective dephasing to contribute at leading order. The
optimum is then set by a mixed floor from local contrast loss and collective
dephasing, rather than by local dephasing alone. A derivation of this
intermediate limit is given in Appendix~\ref{sec:intermediate_dephasing_regime}.

\subsection{Optimization and scaling with local decoherence}
\label{sec:opt_with_local_decoh}

We first show regime-classification maps for experimentally motivated UVRS and
RVRS parameter choices, which need not have identical single-particle
cooperativity. These maps illustrate how the local-decoherence regimes appear
in each implementation. We then perform a separate matched-cooperativity
comparison.
Fig.~\ref{fig:spin_squeezingdecoh_8panel} summarizes the optimal squeezing and optimal time across parameter space, together with the corresponding regime classification from Table~\ref{tab:regime_summary}.

Purple, gray, and pink denote regimes I, II, and III, respectively. The color transparency highlights the gradual crossover between neighboring regimes, reflecting that the boundaries are not sharply defined and that each region is classified according to the dominant mechanism limiting the optimized squeezing.
Our goal is to determine how local decoherence reshapes the optimal squeezing and preparation time across parameter space and to classify which mechanism ultimately terminates the squeezing dynamics in the different operating regimes.

For each parameter pair
\((\Delta_c/(N\Gamma_0),\alpha_{\text{in}}/\alpha_{\rm ref})\),
we evaluate the HP coefficients \((\tilde{\chi},\tilde{\Gamma})\) on the stable spin-polarized branch and the single-particle dephasing $\gamma_d$ then minimize \(\xi^2(t)\) over time to extract both the optimal squeezing \(\xi^2_{\min}\) and the corresponding preparation time \(t_{\min}\), reported as \(\log_{10}(N\Gamma_0 t_{\min})\). Parameter regions beyond the stable spin-polarized branch, i.e.\ beyond the branch-switching discussed above, are excluded from the optimization. The different regimes for a given choice of parameters are identified by determining which regime's squeezing estimate is closest (in dB).

In the absence of significant decoherence, one-axis twisting reaches the familiar curvature-limited minimum \cite{Kitagawa1993}
\begin{equation}
\xi^2_{\mathrm{OAT}}\simeq \frac{3^{2/3}}{2N^{2/3}},
\label{eq:xi_oat_floor_here}
\end{equation}
which is recovered from Eq.~\eqref{eq:Spin_squeez_deph} when both local and collective dephasing are neglected. Once decoherence is relevant, however, the optimum is no longer set solely by the OAT curvature, but by the competition between coherent shear and the accumulation of collective and local noise \cite{Tucker2020}. The optimal time is therefore selected as the earliest time at which the twisting generated by \(|\tilde{\chi}|\) yields substantial squeezing before dephasing degrades the correlations.

A useful way to characterize this balance is through the accumulated noise scales $w_{\mathrm{opt}} \equiv \gamma_dt_{\mathrm{min}}$, $
u_{\mathrm{opt}} \equiv N\tilde{\Gamma}t_{\mathrm{min}}$
and the two expressions
$A\equiv \frac{3^{2/3}}{2N^{2/3}},
D\equiv \frac{\gamma_d\tilde{\Gamma}}{N|\tilde{\chi}|^{2}}$. Here \(A\) is the ideal OAT benchmark, while \(D\) measures the dephasing penalty relative to the coherent shear. Increasing \(|\tilde{\chi}|\) improves the squeezing 
by shifting the optimum to earlier times, thereby reducing both \(w_{\rm opt}\) and \(u_{\rm opt}\). By contrast, increasing either \(\tilde{\Gamma}\) or \(\gamma_d\) raises the noise accumulated before the OAT-like minimum can be reached.

The top row of Fig.~\ref{fig:spin_squeezingdecoh_8panel} identifies which process limits the optimum in each region, while the bottom row shows the corresponding squeezing times. 
In the UVRS columns, the cavity response is effectively broadband, so \(\tilde{\chi}\) and \(\tilde{\Gamma}\) vary smoothly with detuning and drive. As a result, the optimal-time landscape is comparatively smooth: stronger shear shifts the optimum to earlier times, whereas in weaker-interaction regions the optimum drifts toward the single-spin  coherence time and the achievable squeezing deteriorates accordingly. In this regime, the optimization is controlled primarily by the balance between coherent twisting and the local dephasing.

In the RVRS columns, by contrast, the resolved polariton structure makes the dynamics much more sensitive to detuning. Regions of strong cavity response can enhance \(\tilde{\chi}\), but they can also generate larger transient photon populations and therefore stronger collective dephasing through photon loss. As a result, the favorable
operating region is not simply reduced in width; rather,
the branch-switching boundary compresses and shifts the
optimal region toward lower drive strengths, sharpening
the structure of the maps relative to the UVRS case. In
the optimal region, the coherent shearing rate remains large while collective
dephasing remains sufficiently weak. This competition helps explain the tilted
structure of the RVRS maps. Along representative cuts, the effective ratio
\(|\tilde\chi|/\tilde\Gamma\) changes as the low-photon branch approaches the
mean-field switching boundary. The comparatively weak variation of
\(t_{\min}\) with detuning in some regions indicates that the optimized time is
not controlled by this ratio alone, but also by the absolute scale of
\(|\tilde\chi|\) and by the parametrization of the accessible low-photon
operating window.

The regime boundaries can be understood in terms of what terminates the squeezing dynamics, following Table~\ref{tab:regime_summary}. When the squeezing minimum is reached at times short compared with both the local coherence time \(\gamma_d^{-1}\) and the collective dephasing time \((N\tilde{\Gamma})^{-1}\), decoherence is effectively negligible during state preparation and the optimum remains close to the unitary OAT minimum. As parameters are varied and the optimum shifts to later times, the evolution becomes limited by the local coherence window: in principle the twisting would continue to improve the squeezing, but the protocol must be stopped at \(t\sim \gamma_d^{-1}\) before single-particle dephasing destroys the correlations. Finally, when collective dephasing is already appreciable on that same timescale, the optimal time is still set by local decoherence, but the squeezing is further degraded by the additional photon-loss-induced collective noise.

\begin{figure}[t]
  \centering

  \begin{tikzpicture}
    \node[anchor=south west, inner sep=0] (img) at (0,0)
    {\includegraphics[width=0.98\linewidth]{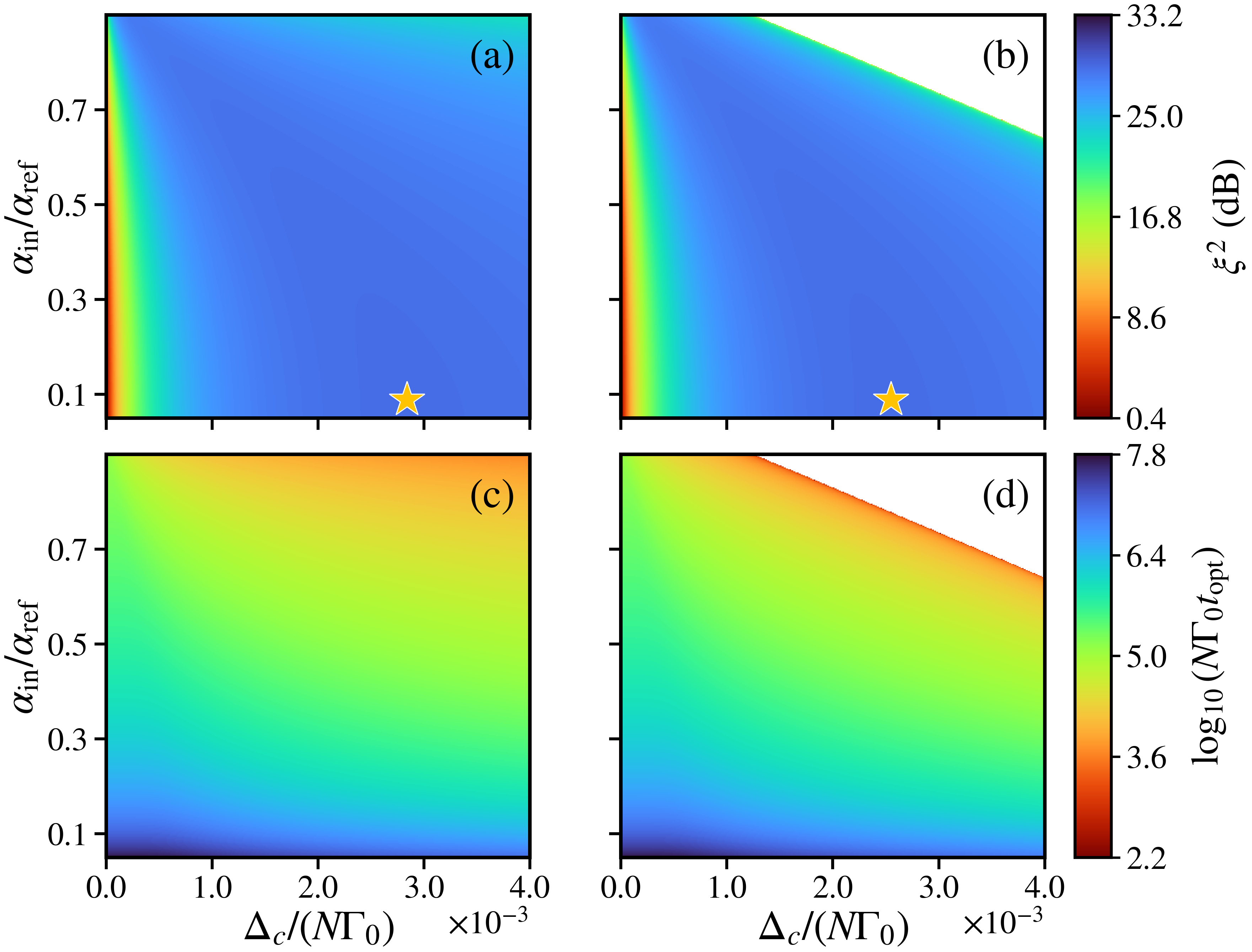}};

    \begin{scope}[x={(img.south east)}, y={(img.north west)}]
      \node[font=\bfseries\small, anchor=center] at (0.245,1.035) {UVRS};
      \node[font=\bfseries\small, anchor=center] at (0.665,1.035) {RVRS};
    \end{scope}
  \end{tikzpicture}

  \caption{{\bf Comparing the performance of the protocol in the UVRS and RVRS regimes at fixed  single-particle cooperativity for $N=10^5$}. Left column: UVRS. Right column: RVRS. Panels (a,b) show the optimized spin-squeezing parameter $\xi^2$ (dB), and panels (c,d) show the corresponding optimal normalized time $\log_{10}(N\Gamma_0 t_{\min})$, as functions of $\Delta_c/(N\Gamma_0)$ and $\alpha_{\rm in}/\alpha_{\rm ref}$.
 For the two regimes, $C_0 = 4g^2/(\kappa\gamma_{e\downarrow}) \simeq 0.356$. The stars mark the optimal spin-squeezing conditions within the displayed
window. When an optimum lies on the lower boundary
\(\alpha_{\rm in}/\alpha_{\rm ref}=0\), the corresponding star is plotted
slightly above the boundary to enhance visibility; its displayed vertical position
does not indicate a finite optimal drive. In the normalized variables shown here, the RVRS optimum is concentrated in narrower detuning bands and the branch-switching boundary reduces the useful region. In laboratory detuning units, however, this does not necessarily imply a smaller tolerance around the optimum, because the horizontal axis is scaled by $N\Gamma_0$ and $\Gamma_0$ is much larger in RVRS.}
  \label{fig:xi2_maps_xicuad_2x2}
\end{figure}

Figure~\ref{fig:xi2_maps_xicuad_2x2} summarizes the matched-cooperativity
comparison for \(N=10^{5}\). The top row shows the optimized squeezing gain
\(\xi^2_{\min}\), while the bottom row shows the corresponding optimal time
\(\log_{10}(N\Gamma_0 t_{\min})\). The stars mark the global optimum within the displayed window after this
local-dephasing-limited time optimization; as in
Figs.~\ref{fig:spin_squeezingcoll} and
\ref{fig:spin_squeezingdecoh_8panel}, stars associated with optima on the lower
boundary \(\alpha_{\rm in}/\alpha_{\rm ref}=0\) are plotted slightly above the
boundary to enhance visibility. The masked RVRS wedge denotes the region beyond the
mean-field branch-switching boundary, which is excluded because the maximum
squeezing does not occur there and the HP approximation is not controlled near
the boundary.

At fixed cooperativity, the UVRS and RVRS maps show comparable optimized
squeezing and dimensionless optimal times over much of the stable
spin-polarized branch. The RVRS optimum appears more compressed in
\(\Delta_c/(N\Gamma_0)\), but this normalized width should not be directly
interpreted as a narrower laboratory detuning tolerance because
\(\Gamma_0=4g^2/\kappa\) differs substantially between the two regimes.

\begin{figure}[t]
  \centering

  \begin{minipage}[b]{0.49\linewidth}
    \centering
    \includegraphics[width=\linewidth]{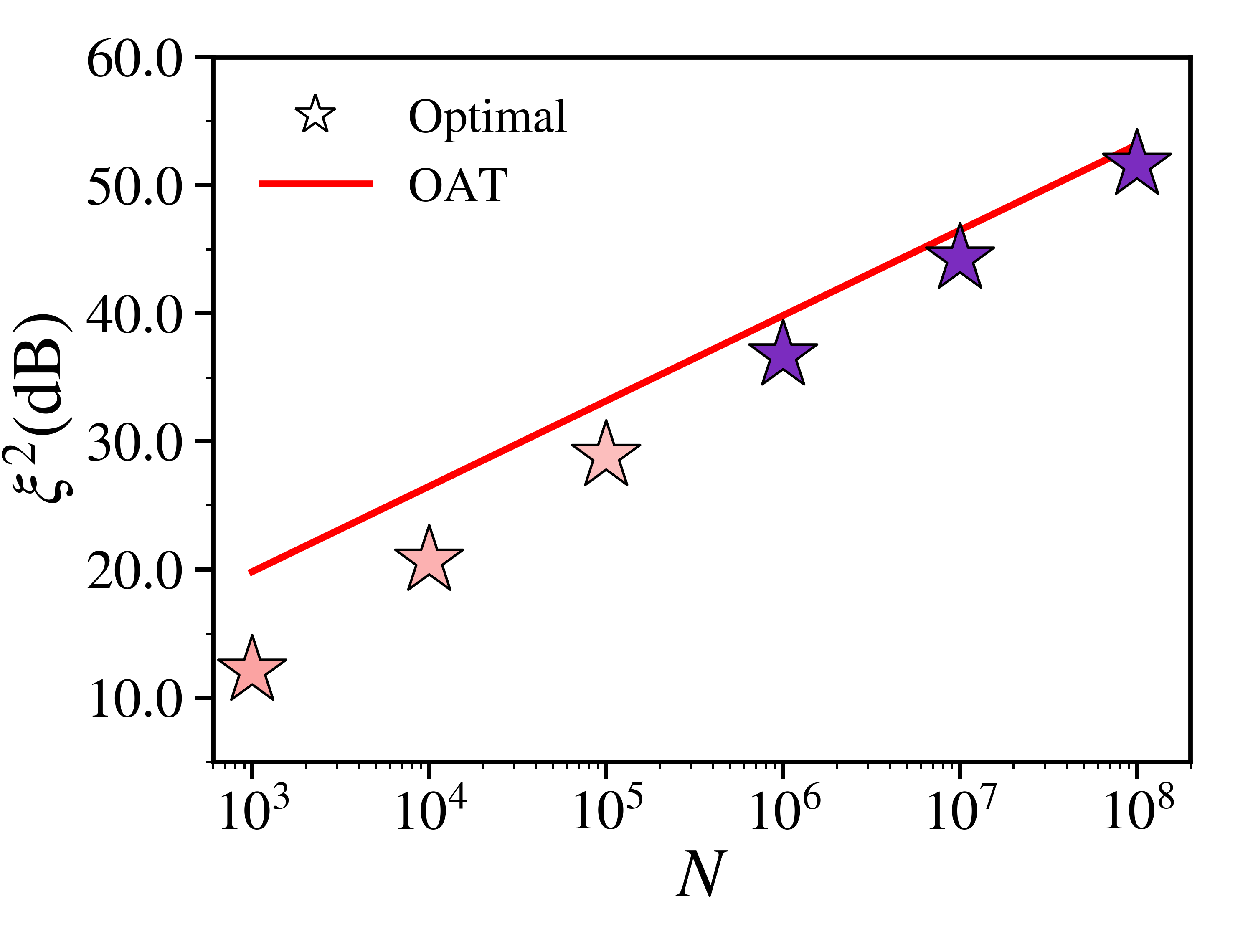}

    \vspace{0.3em}
    {\small (a)}
  \end{minipage}\hfill
  \begin{minipage}[b]{0.49\linewidth}
    \centering
    \includegraphics[width=\linewidth]{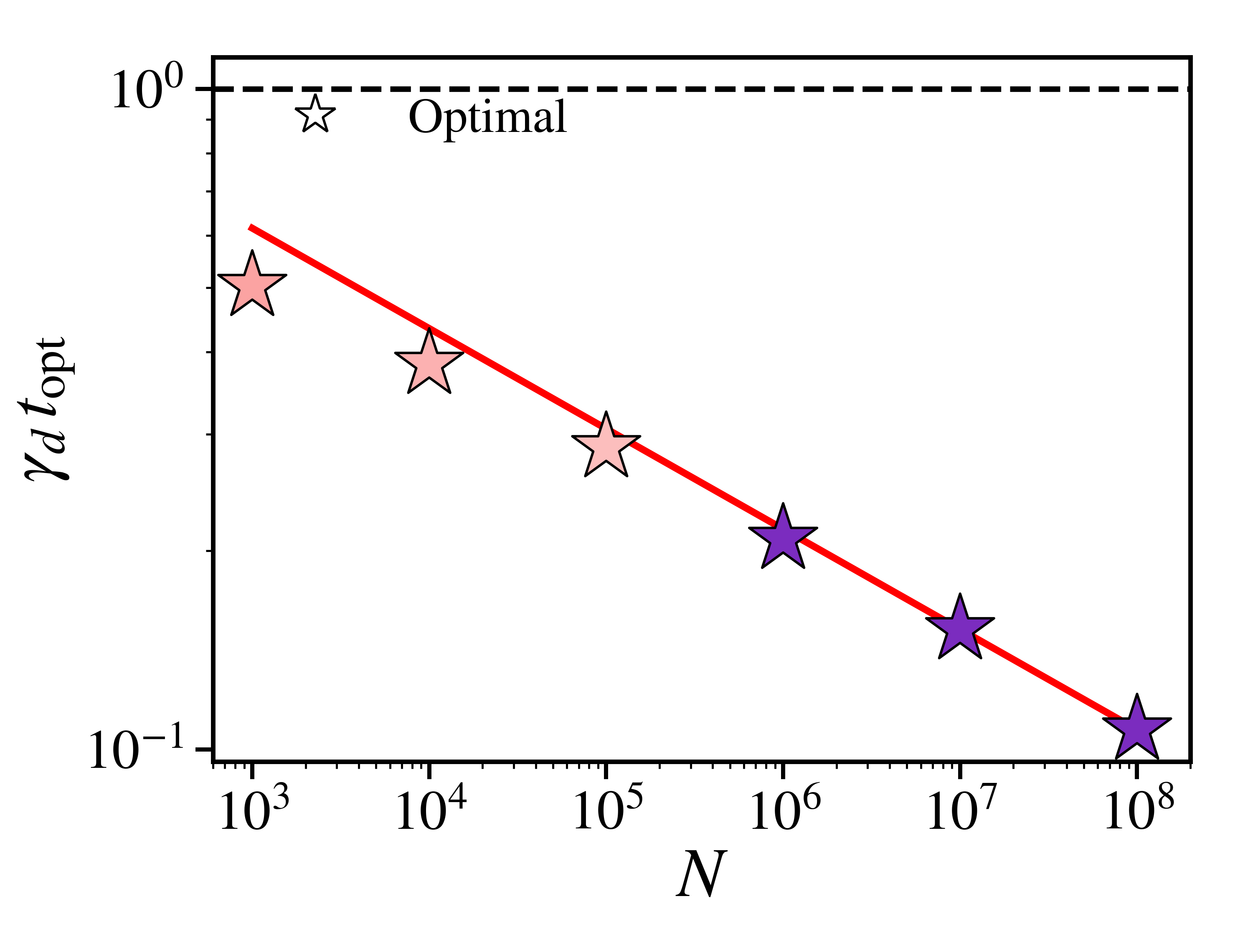}

    \vspace{0.3em}
    {\small (b)}
  \end{minipage}
    
    \caption{{\bf
    Scaling of globally optimized squeezing and preparation time with system size in the presence of single-particle dephasing}. (a) Optimal spin squeezing $\xi^2_{\min}$ (dB) as a function of $N$. Colored stars denote the global optimum obtained at each $N$ after scanning over $\alpha_{\rm in}/\alpha_{\rm ref}$, $\Delta_c/(N\Gamma_0)$, and evolution time. The color matches the classification scheme used in Fig. 7. The red curve shows the OAT benchmark used for comparison. (b) Optimal scaled time $\gamma_d t_{\rm min}$ as a function of $N$ for the same globally optimized points. The red line is a fit to the large-$N$ trend. The decrease of \(\gamma_d t_{\rm min}\) with increasing \(N\) indicates that
larger systems reach their optimal squeezing on a timescale increasingly short
compared with \(1/\gamma_d\).}
  \label{fig:xi2_maps_xicuad_1x2}
\end{figure}

We next ask how the matched-cooperativity comparison changes with system size.
For each atom number \(N\), we minimize \(\xi^2\) over the evolution time and
over the stable spin-polarized operating window
\((\Delta_c/(N\Gamma_0),\alpha_{\text{in}}/\alpha_{\rm ref})\). We denote the
resulting optimum by \(\xi_{\mathrm{opt}}^2(N)\), and the time at which it is
attained by \(t_{\mathrm{min}}(N)\). This procedure isolates how the best
achievable metrological gain changes with system size once parameter tuning
and single-particle dephasing are both taken into account.

The resulting scaling is shown in Fig.~\ref{fig:xi2_maps_xicuad_1x2}. Panel (a) shows that the globally optimized squeezing improves systematically with increasing \(N\) and remains close to the OAT benchmark over the explored range. This indicates that, after optimization over \((\Delta_c,\alpha_{\text{in}}/\alpha_{\text{ref}},t)\), the growth of collective shear continues to outpace the associated single-particle-dephasing penalty. Panel (b) shows the corresponding optimal time in units of the local-dephasing scale, \(\gamma_d t_{\mathrm{min}}\). Its decrease with increasing \(N\) shows that the optimum occurs on a timescale
increasingly short compared with \(1/\gamma_d\), so the best squeezed state is
generated before local dephasing can significantly erode the correlations. The large-\(N\) improvement therefore reflects the benefit of  stronger collective interactions. 
Figure~\ref{fig:xi2_maps_xicuad_1x2}(b) highlights the main practical distinction
between the two regimes. Although the optimized squeezing and the scaled time
remain close to the OAT trend over the explored range of \(N\), the two regimes
are not equivalent in laboratory units: the RVRS implementation reaches
comparable squeezing on a substantially shorter physical timescale given its shorter timescale $\gamma_d^{-1}$.

\section{Conclusions and Outlook}
\label{outlook}

We have generalized the strong-symmetry protocol for dissipative spin squeezing
from the UVRS limit of Ref.~\cite{JeremyOAT} into the RVRS regime relevant to
alkaline-earth
cavity-QED platforms \cite{Winchester2017,Cline2025,Eric_CRF},
where cavity photons remain active dynamical degrees of freedom and cavity
backaction can no longer be neglected. In this regime, the exact strong
symmetry continues to conserve the sector label \cite{Buca2012,SanchezMunoz2019},
but the inter-sector coherence required for the geometric/OAT picture is no
longer automatically protected because the dynamical cavity field can partially
resolve the sector dependence \cite{Barberena2024}. Our main result is that there nevertheless exists a stable low-photon operating
window, characterized by moderate drive, small detunings
\(\Delta_c,\delta_\sigma\), and smooth turn-on and turn-off ramps, in which
nonadiabatic cavity dynamics and sector-resolving leakage remain sufficiently
controlled for the protocol to generate and store significant squeezing.

Within this operating window, squeezing can still be generated and transferred directly into the long-lived clock manifold without additional control pulses. Supported by Holstein--Primakoff theory and large-scale DissTWA simulations, our results show that experimentally relevant \({}^{87}\mathrm{Sr}\) cavity-QED parameters can yield strong metrologically useful squeezing, reaching $26$--$28$ dB at $N=10^5$ and approaching the characteristic one-axis twisting scaling for larger system sizes. At fixed cooperativity, UVRS and RVRS support broadly comparable optimized squeezing and comparable dimensionless optimal times. The main practical difference appears in laboratory units: RVRS reaches the optimum within a substantially shorter preparation time. This shorter physical timescale can make the RVRS implementation less sensitive to additional slow decoherence mechanisms beyond the fundamental sources considered here.

Looking forward, the resolved-regime setting raises several questions that are
specific to having a dynamical cavity field. One important direction is to
assess the robustness of the strong-symmetry mechanism to inhomogeneous
atom--cavity couplings, which can allow the output field to acquire information
beyond the conserved sector label \(N_J\). A second question is why abrupt
quenches, which can preserve protected information in other strong-symmetry
settings, degrade the dynamics here. Our results suggest that in RVRS a quench
can transiently populate sector-dependent cavity fields, allowing photon
leakage to record which-sector information before the system returns to the
low-photon branch. Understanding how this quench-induced information leakage depends on system
size and protocol design is an important direction for future work.
It would also be useful to analyze technical noise in this geometric protocol,
including which forms of laser phase or amplitude noise are filtered by the
trajectory and which are converted into direct dephasing. This may connect to
ideas from holonomic control, where geometric structure can provide robustness
against certain control imperfections \cite{Berry1984,Simon1983,WilczekZee1984,Zanardi1999H,Zhang2023G}.

More broadly, these results provide a framework for understanding
cavity-mediated interactions in regimes where atomic and photonic degrees of
freedom are intrinsically intertwined, rather than cleanly separated by
adiabatic elimination. Such regimes are often avoided in practice because the
cavity field both mediates useful interactions and acts as a channel through
which unwanted information can be lost. Here we show that this competition can
be diagnosed and controlled: the relevant requirement is not the absence of
photons, but the suppression of distinguishable information in the output
field. This perspective opens a route to quantum simulation and quantum
metrology in unconventional spin--photon regimes, where the dynamical cavity is
not simply eliminated, but becomes an active and controllable part of the
many-body evolution. It also points to resolved cavity QED as a natural setting
for studying strong-symmetry dynamics and phase transitions beyond effective
spin-only descriptions.

\begin{acknowledgments}
We acknowledge helpful feedback on the manuscript from Yang Yang, L.~Lenstra, T.~van Meer, and  prior discussions with Asier Pi$\tilde{\rm n}$eiro-Orioli. We acknowledge  funding support from the Vannevar-Bush Faculty Fellowship,
 the National Science Foundation under Grant Numbers 1734006 (Physics Frontier Center) and  OMA-2016244 (QLCI Q-SEnSE). This work is also supported by the U.S. Department of Energy, Office of Science, National Quantum Information Science Research Centers, Quantum Systems Accelerator and NIST. D.B. acknowledges support from a Simons Investigator Award (Grant No. 511029)
and the Engineering and Physical Sciences Research Council
[grant numbers EP/V062654/1 and EP/Y01510X/1]. J.T.Y. was supported by the NWO Talent
Programme (project number VI.Veni.222.312), which is
(partly) financed by the Dutch Research Council (NWO).
\end{acknowledgments}

\appendix

\section{Spin master equation via adiabatic elimination of the cavity photons}
\label{app:uvrs_reduction}

This appendix derives the spin-only UVRS master equation quoted in Eq.~\eqref{eq:Hamil_spin}. We choose the phase convention in which \(\alpha_{\rm in}\) is real. We start from the driven spin--cavity Hamiltonian in the frame rotating at the drive frequency,
\begin{equation}
\hat H=
 g(\hat J^+\hat a+\hat J^-\hat a^\dagger)
-i\sqrt{\kappa}\alpha_{\rm in}(\hat a-\hat a^\dagger)
-\Delta_c\hat a^\dagger\hat a
-\delta_\sigma \hat J^z,
\label{eq:app_uvrs_H_full}
\end{equation}
with cavity loss \(\kappa\mathcal D[\hat a]\). The cavity equation of motion is
\begin{equation}
\partial_t \hat a=
-ig\hat J^-+\left(i\Delta_c-\frac{\kappa}{2}\right)\hat a
+\sqrt{\kappa}\alpha_{\rm in}.
\label{eq:app_uvrs_a_eom}
\end{equation}
In the UVRS limit the cavity relaxation time is short compared with the collective spin dynamics. Setting \(\partial_t\hat a\simeq0\) gives the expression used in the main text,
\begin{equation}
\hat a\simeq
\frac{\sqrt{\kappa}\alpha_{\rm in}-ig\hat J^-}{\kappa/2-i\Delta_c}
\equiv \alpha+\beta\hat J^-,
\label{eq:app_uvrs_a_AE}
\end{equation}
where
\begin{equation}
\alpha=\frac{\sqrt{\kappa}\alpha_{\rm in}}{\kappa/2-i\Delta_c},
\qquad
\beta=-\frac{ig}{\kappa/2-i\Delta_c}.
\label{eq:app_uvrs_alpha_beta}
\end{equation}
Eq.~\eqref{eq:app_uvrs_a_AE} separates the coherent drive-induced field from the field radiated by the collective dipole. A consistent substitution in both the Hamiltonian and the displaced cavity-loss
channel gives
\begin{equation}
\hat H_{\rm eff}
=\chi\hat J^+\hat J^--\delta_\sigma\hat J^z
+\left(\frac{\Omega_{\rm eff}}{2}\hat J^+ + \mathrm{h.c.}\right),
\label{eq:app_uvrs_Heff}
\end{equation}
with
\begin{equation}
\chi=\frac{g^2\Delta_c}{\Delta_c^2+\kappa^2/4},
\qquad
\Omega_{\rm eff}=2g\alpha
=2g\frac{\sqrt{\kappa}\alpha_{\rm in}}{\kappa/2-i\Delta_c}.
\label{eq:app_uvrs_chi_Omega}
\end{equation}
The dissipative part follows from the same substitution in the cavity-loss channel,
\begin{equation}
\kappa\mathcal D[\hat a]\hat\rho
\simeq
\kappa |\beta|^2\mathcal D[\hat J^-]\hat\rho
+\text{coherent drive terms},
\label{eq:app_uvrs_D_substitution}
\end{equation}
where the coherent terms generated by the displacement have been absorbed into
\(\hat H_{\rm eff}\).
Defining
\begin{equation}
\kappa|\beta|^2=\frac{g^2\kappa}{\Delta_c^2+\kappa^2/4}
\equiv \Gamma,
\label{eq:app_uvrs_Gamma}
\end{equation}
we obtain the spin-only UVRS master equation
\begin{subequations}
\label{eq:app_uvrs_spin_master}
\begin{align}
\partial_t\hat\rho
&=-i[\hat H_{\rm eff},\hat\rho]+\Gamma\mathcal D[\hat J^-]\hat\rho,
\label{eq:app_uvrs_spin_master_a}\\
\hat H_{\rm eff}
&=\chi\hat J^+\hat J^--\delta_\sigma\hat J^z
+\left(\frac{\Omega_{\rm eff}}{2}\hat J^+ + \mathrm{h.c.}\right).
\label{eq:app_uvrs_spin_master_b}
\end{align}
\end{subequations}
Equations~\eqref{eq:app_uvrs_chi_Omega} and \eqref{eq:app_uvrs_Gamma} are the coefficients used in the main-text UVRS comparison.

For completeness, one can also view the cavity as a short-memory linear filter. Formally integrating Eq.~\eqref{eq:app_uvrs_a_eom} gives
\begin{equation}
\hat a(t)=\int_0^\infty d\tau\,e^{(i\Delta_c-\kappa/2)\tau}
\left[-ig\hat J^-(t-\tau)+\sqrt{\kappa}\alpha_{\rm in}\right].
\label{eq:app_uvrs_memory_kernel}
\end{equation}
When the collective dipole varies slowly over the memory time \(\kappa^{-1}\), this expression reduces to Eq.~\eqref{eq:app_uvrs_a_AE} at leading order. In RVRS this local reduction is not a controlled dynamical elimination of the cavity; in the main text it is used only as the UVRS benchmark and as the stationary algebraic response entering the mean-field branch analysis.

\section{Equivalent rotating-frame descriptions of the protocol}
\label{app:equiv_rotating_frames}

In this appendix we clarify the relation between the two rotating-frame descriptions used to interpret the multistage protocol. The first frame rotates at the drive frequency. In this frame, the drive phase
can be chosen real without loss of generality, and the control appears as an
explicit atomic detuning. This is the most convenient frame for implementing the time-dependent protocol and for diagnosing nonadiabatic dynamics via Holstein--Primakoff. We refer to this frame as the dynamical phase frame.  The second frame follows the atomic transition frequency, removing the explicit detuning and representing the same control as a time-dependent phase of the drive.

In the reduced collective-spin description, the experimental controls \(\alpha_{\rm in}(t)\) and the detunings determine the effective Rabi frequency \(\Omega_{\rm eff}(t)\).  Going to the rotating frame of the drive (the dynamical phase frame) and fixing the phase convention such that the Rabi frequency is real, the Hamiltonian may be written as
\begin{equation}
\hat H_{\delta}(t)
=
\chi(t)\hat J^{+}\hat J^{-}
+
\frac{\Omega(t)}{2}
\left(\hat J^{+}+\hat J^{-}\right)
-
\delta_\sigma(t)\hat J^{z},
\label{eq:app_H_delta_frame}
\end{equation}
where \(\Omega(t)=|\Omega_{\rm eff}(t)|\).  In this frame the phase of the drive is fixed, while \(\delta_\sigma(t)\) appears as a detuning on the driven \(\{|\downarrow\rangle,|e\rangle\}\) manifold.

We can switch to the Berry phase frame with the collective-spin rotation
\begin{subequations}
\begin{align}
\hat U_d(t)
&=
\exp\!\left[i\varphi_d(t)\hat J^z\right],
\\
\dot{\varphi}_d(t)
&=
-\delta_\sigma(t).
\end{align}
\label{eq:app_phase_following_unitary}
\end{subequations}
The transformed Hamiltonian is
\begin{align}
\hat H_B(t)
&=
\hat U_d(t)\hat H_\delta(t)\hat U_d^\dagger(t)
+
i\dot{\hat U}_d(t)\hat U_d^\dagger(t)
\nonumber\\
&=
\chi(t)\hat J^{+}\hat J^{-}
+
\frac{\Omega(t)}{2}
\left[
e^{i\varphi_d(t)}\hat J^{+}
+
\mathrm{h.c.}
\right].
\label{eq:app_H_B_raw}
\end{align}
Equivalently, writing the Hermitian conjugate explicitly,

\begin{equation}
\hat H_B(t)
=
\chi(t)\hat J^{+}\hat J^{-}
+
\frac{\Omega(t)}{2}
\left[
e^{i\varphi_d(t)}\hat J^{+}
+
e^{-i\varphi_d(t)}\hat J^{-}
\right].
\label{eq:app_H_B_frame}
\end{equation}
Equations~\eqref{eq:app_H_delta_frame} and \eqref{eq:app_H_B_frame} are therefore two representations of the same physical control.  The dynamical phase frame is useful for following the explicit dynamics, while the Berry phase frame shows that the protocol generates an azimuthal motion of the collective \(J\)-spin through the rotating phase of the drive.

For adiabatic ramps in the Berry phase frame, each fixed-\(N_J\) sector follows its instantaneous spin-polarized branch and accumulates a geometric phase.  For a closed trajectory \(\mathcal C\), this phase can be written as
\begin{equation}
\Phi_B(N_J)
=
\frac{N_J}{2}
\oint_{\mathcal C}
\left[1-\cos\theta_J(N_J,t)\right]\dot{\phi}_J(N_J,t)\,dt .
\label{eq:app_Berry_phase}
\end{equation}
More generally, during the finite-time protocol we denote by \(\phi_B(N_J,t)\) the accumulated sector phase up to time \(t\), noting that in the gauge we have effectively chosen, the Berry phase is accumulated entirely during the azimuthal motion on the Bloch sphere.  The relevant physics is not the absolute phase of a single sector, but its dependence on \(N_J\).  Expanding around the central sector \(N_J=N/2\),
\begin{align}
\phi_B(N_J,t)
&\simeq
\phi_B^{(0)}(t)
+
\phi_B^{(1)}(t)
\left(N_J-\frac{N}{2}\right)
\nonumber\\
&\quad
+
\frac{1}{2}\phi_B^{(2)}(t)
\left(N_J-\frac{N}{2}\right)^2
+\cdots ,
\label{eq:app_Berry_expansion}
\end{align}
where \(\phi_B^{(n)}(t)= \partial_{N_J}^n\phi_B(N_J,t)|_{N_J=N/2}\).  In the encoded-spin description used in the main text,
\begin{equation}
\hat S^z \simeq \frac{N}{2}-\hat N_J .
\label{eq:app_Sz_NJ_relation}
\end{equation}
Thus the unitary generated by the sector-dependent phase,
\begin{equation}
\hat U_B(t)=\exp\!\left[i\phi_B(\hat N_J,t)\right],
\end{equation}
contains, up to an overall phase,
\begin{equation}
\hat U_B(t)
\simeq
\exp\!\left[
-i\phi_B^{(1)}(t)\hat S^z
+
\frac{i}{2}\phi_B^{(2)}(t)(\hat S^z)^2
+\cdots
\right].
\label{eq:app_Berry_unitary_OAT}
\end{equation}
The linear term produces a collective rotation of the encoded spin, while the quadratic term produces the nonlinear shear associated with one-axis twisting.  Equivalently, writing
\begin{equation}
\hat H_{\rm eff}^{(B)}(t)
\equiv
-\partial_t\phi_B(\hat N_J,t),
\label{eq:app_Berry_generator}
\end{equation}
one obtains
\begin{equation}
\hat H_{\rm eff}^{(B)}(t)
=
\omega_z^B(t)\hat S^z
+
\bar{\chi}(t)(\hat S^z)^2
+\cdots ,
\label{eq:app_Berry_OAT_Hamiltonian}
\end{equation}
with
\begin{equation}
\omega_z^B(t)=\dot{\phi}_B^{(1)}(t),
\qquad
\bar{\chi}(t)=-\frac{1}{2}\dot{\phi}_B^{(2)}(t).
\label{eq:app_Berry_rates}
\end{equation}
This identifies the curvature of the accumulated sector phase with the effective OAT rate.

The frame equivalence remains useful in the RVRS regime, but its interpretation must be more careful.  The transformation above only changes the representation of the drive and detuning; it does not eliminate the cavity as a dynamical degree of freedom.  In RVRS, different \(N_J\) sectors can generate different transient cavity fields during the ramps.  Photon leakage can then carry sector information and dephase coherences between the sectors.  Smooth ramps suppress these transients and allow the system to remain close to the intended low-photon spin-polarized branch.  In that regime, the geometric phase picture and the effective OAT description remain accurate for the squeezing dynamics considered in the main text.

\section{Mean-field  phase diagram}
\label{app:mean_field_roots}

This appendix derives the mean-field steady-state relation used to construct
the operating-window maps in Fig.~\ref{fig:mean_field_phase_diagram} and summarizes the
branch structure invoked in the main text. The goal is not to replace the
full dynamical simulations, but to identify the low-photon, spin-polarized
branch used by the squeezing protocol and the branch-switching region that
should be avoided. In the main text, this information is used only
operationally: it tells us where the driven
\(\{\ket{\downarrow},\ket{e}\}\) manifold can remain polarized while the
geometric phase accumulates. Here we give the corresponding mean-field
derivation and the root structure responsible for the multivalued response.

The deterministic mean-field equations corresponding to the spin--cavity
Hamiltonian in Eq.~\eqref{eq:app_uvrs_H_full} are
\begin{subequations}
\label{eq:app_mf_eom}
\begin{align}
\dot a&=-igJ^-+\left(i\Delta_c-\frac{\kappa}{2}\right)a
+\sqrt{\kappa}\alpha_{\rm in},
\label{eq:app_mf_eom_a}\\
\dot J^-&=2igaJ^z+i\delta_\sigma J^-,
\label{eq:app_mf_eom_b}\\
\dot J^z&=-ig(aJ^+-a^*J^-).
\label{eq:app_mf_eom_c}
\end{align}
\end{subequations}
These equations neglect quantum correlations and Langevin noise, but retain
the self-consistent coupling between the intracavity field and the collective
spin in a fixed \(N_J\) sector. We take the length of the mean-field Bloch
vector to be \(J=N_J/2\).

We parametrize a spin-polarized steady state as
\begin{equation}
J^-_{\rm ss}=\frac{N_J}{2}\sin\tilde\theta_J e^{-i\tilde\phi_J},
\qquad
J^z_{\rm ss}=-\frac{N_J}{2}\cos\tilde\theta_J,
\label{eq:app_mf_J_param}
\end{equation}
with \(0\le\tilde\theta_J\le\pi/2\). Here
\(\tilde\theta_J=0\) corresponds to the south pole of the driven-manifold
Bloch sphere, while larger \(\tilde\theta_J\) corresponds to a more strongly
tilted spin with a larger excited-state component. Setting
\(\dot J^-=0\) gives the cavity amplitude associated with this steady-state
spin orientation,
\begin{equation}
a_{\rm ss}
=
\frac{\delta_\sigma}{2g}\tan\tilde\theta_J\,e^{-i\tilde\phi_J}.
\label{eq:app_mf_a_ss}
\end{equation}
Thus, away from spin resonance, a tilted spin-polarized state generally
requires a finite intracavity field. This is the mean-field expression of the
spin--photon feedback discussed in the main text: the collective spin shapes
the optical field, and the optical field in turn determines the torque acting
on the spin.

Substituting Eq.~\eqref{eq:app_mf_a_ss} into the steady-state cavity equation
\(\dot a=0\) gives the complex balance condition
\begin{equation}
\begin{aligned}
\sqrt{\kappa}\alpha_{\rm in}
&=e^{-i\tilde\phi_J}
\Bigg[
\frac{\kappa\delta_\sigma}{4g}\tan\tilde\theta_J \\
&\qquad
+i\left(
\frac{gN_J}{2}\sin\tilde\theta_J
-\frac{\Delta_c\delta_\sigma}{2g}\tan\tilde\theta_J
\right)
\Bigg].
\end{aligned}
\label{eq:app_mf_complex_condition}
\end{equation}
This equation separates the steady-state balance into two quadratures. The
first term, proportional to \(\kappa\delta_\sigma\tan\tilde\theta_J/(4g)\),
is the dissipative quadrature set by cavity loss. The second term is the
reactive quadrature: it contains the collective dipole scale
\(gN_J\sin\tilde\theta_J/2\), shifted by the detuning-dependent cavity
response \(\Delta_c\delta_\sigma\tan\tilde\theta_J/(2g)\). The azimuthal
angle \(\tilde\phi_J\) rotates these quadratures into the phase of the
applied drive.

Taking the modulus of Eq.~\eqref{eq:app_mf_complex_condition} yields
\begin{equation}
\begin{aligned}
(\sqrt{\kappa}\alpha_{\rm in})^2
&=\left(\frac{\delta_\sigma\kappa}{4g}\right)^2
\tan^2\tilde\theta_J \\
&\quad+
\left[
\frac{gN_J}{2}\sin\tilde\theta_J
-\frac{\Delta_c\delta_\sigma}{2g}\tan\tilde\theta_J
\right]^2 .
\end{aligned}
\label{eq:app_mf_potential}
\end{equation}
For a given pair \((\Delta_c,\alpha_{\rm in})\), this equation determines the
allowed steady-state polar angles \(\tilde\theta_J\). The phase of the
steady-state Bloch vector is fixed by the same complex equation,
\begin{equation}
\begin{aligned}
e^{i\tilde\phi_J}
&=\frac{1}{\sqrt{\kappa}\alpha_{\rm in}}
\left[
\frac{\kappa\delta_\sigma}{4g}\tan\tilde\theta_J\right.\\
&\qquad\left.
+i\left(
\frac{gN_J}{2}\sin\tilde\theta_J
-\frac{\Delta_c\delta_\sigma}{2g}\tan\tilde\theta_J
\right)
\right].
\end{aligned}
\label{eq:app_mf_phi_relation}
\end{equation}
Thus \(\tilde\theta_J\) is obtained from
Eq.~\eqref{eq:app_mf_potential}, while \(\tilde\phi_J\) is then fixed by
Eq.~\eqref{eq:app_mf_phi_relation}.

Two simple limits clarify the physical content of
Eq.~\eqref{eq:app_mf_potential}. First, when
\(\delta_\sigma=0\), the right-hand side reduces to
\([g(N_J/2)\sin\tilde\theta_J]^2\). Since
\(\sin\tilde\theta_J\le1\), no spin-polarized solution exists once
\(\sqrt{\kappa}|\alpha_{\rm in}|>gN_J/2\). At the mean-field level, this
marks the boundary of the resonant spin-polarized solution. Second, for
finite detuning, the terms proportional to \(\tan\tilde\theta_J\) encode the
steady-state cavity field required to balance the detuned drive. These terms
allow the system to compensate a broader range of input fields, but they also
make the response nonlinear enough to support multiple steady-state branches.

To make this branch structure explicit, define
\begin{equation}
\begin{gathered}
 x=\cos\tilde\theta_J,
\qquad
A=\frac{\delta_\sigma\kappa}{4g},
\qquad
B=\frac{gN_J}{2},\\[0.25em]
C=\frac{\Delta_c\delta_\sigma}{2g},
\qquad
D=\kappa\alpha_{\rm in}^2 .
\end{gathered}
\label{eq:app_mf_ABCD}
\end{equation}
Then Eq.~\eqref{eq:app_mf_potential} can be written as
\begin{equation}
D x^2=(1-x^2)\left[A^2+(Bx-C)^2\right].
\label{eq:app_mf_root_compact}
\end{equation}
Equivalently, the physical roots are the real roots in \(0\le x\le1\) of
\begin{equation}
\begin{aligned}
q(x)&=-B^2x^4+2BCx^3
+(B^2-A^2-C^2-D)x^2\\
&\quad -2BCx+(A^2+C^2)=0.
\end{aligned}
\label{eq:app_mf_quartic}
\end{equation}
This compact form is useful because the generic case has either one or three
admissible roots in the physical interval. A single admissible root
corresponds to a unique mean-field steady state. Three admissible roots signal
a multivalued response, with two stable branches separated by an unstable
one. Dynamically, this is the branch-switching or bistable structure shown in
the RVRS phase diagram of the main text.

\begin{figure*}[tbp]
\centering
\includegraphics[width=1.0\textwidth]{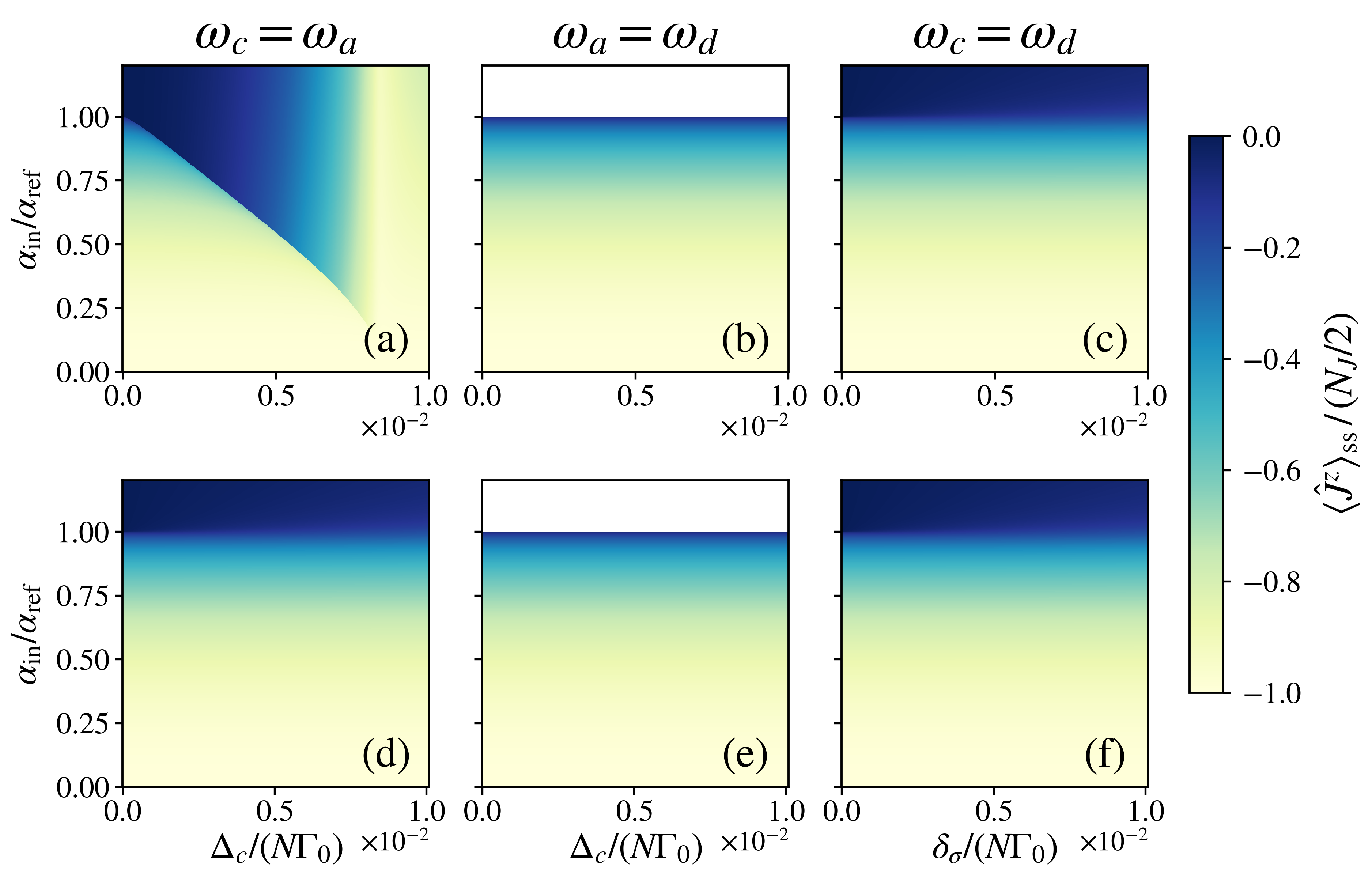}
\caption{Steady-state normalized spin projection
\(\langle \hat J^z\rangle_{\rm ss}/(N_J/2)\) for \(N=10^5\) atoms.
From left to right, the columns correspond to
\(\omega_c=\omega_a\), \(\omega_a=\omega_d\), and
\(\omega_c=\omega_d\). The top row shows the regime with
\(\kappa=15g\), while the bottom row shows the regime with
\(\kappa=15g\sqrt{N_J}\). In the \(\omega_a=\omega_d\) column, the
horizontal boundary at \(\alpha_{\rm in}/\alpha_{\rm ref}=1\) marks the
termination of the polarized branch and the onset of the normal phase. In
the \(\omega_c=\omega_d\) column, the response remains single-valued and
varies smoothly with detuning and drive. Only the atom--cavity resonant case,
\(\omega_c=\omega_a\), supports a multivalued steady-state structure; this
structure is weakly expressed in the large-\(\kappa\) regime but becomes
sharper for \(\kappa=15g\) due to dynamical cavity backaction.}
\label{fig:app_mf_resonance_comparison}
\end{figure*}

In a sweep of the drive or detuning, this multivalued structure appears as
hysteresis: the system follows one stable branch until that branch becomes
unstable, after which it jumps to the other branch. This is the standard
mechanism of optical bistability in driven cavities
\cite{Bonifacio1978,GibbsBook}, and closely related branch-switching behavior
has been studied in cavity-QED and driven Dicke-type systems
\cite{Rivero2023,Bhaseen2012,Angerer2018,Stitely2020,Gbor2023}. Additional
effects such as spontaneous emission, inhomogeneous coupling, or other
decoherence channels can broaden the sharp mean-field features
\cite{Carmichael1980,Eric_CRF}. In the present protocol, the phase diagram is
therefore used to select trajectories on the low-photon, strongly
spin-polarized branch and away from this branch-switching region.

The reference drive used to normalize the mean-field maps is
\begin{equation}
\sqrt{\kappa}\alpha_{\rm ref}
=
\sqrt{\left(\frac{\kappa\delta_\sigma}{4g}\right)^2
+\left(\frac{gN_J}{2}-\frac{\Delta_c\delta_\sigma}{2g}\right)^2}.
\label{eq:app_mf_alpha_ref}
\end{equation}
This scale is obtained from the quadrature balance in
Eq.~\eqref{eq:app_mf_complex_condition}. It combines the dissipative
quadrature set by \(\kappa\delta_\sigma/(4g)\) with the reactive quadrature
set by the collective dipole scale \(gN_J/2\), shifted by the detuning-induced
cavity response \(\Delta_c\delta_\sigma/(2g)\). When
\(\delta_\sigma=0\), it reduces to
\(\sqrt{\kappa}\alpha_{\rm ref}=gN_J/2\), the critical drive scale of the
resonant spin-polarized solution. For finite detuning,
\(\alpha_{\rm ref}\) should not be interpreted as a universal phase boundary;
rather, it provides a local drive scale that removes the leading detuning
dependence of the mean-field response and allows the UVRS and RVRS operating
regions to be compared on common axes.

For the maps shown in the main text, the plotted branch is the steady state
reached from the initial condition relevant to the protocol: an empty cavity
and a collective spin polarized at the south pole of the driven-manifold
Bloch sphere,
\begin{equation}
a(0)=0,\qquad J^-(0)=0,\qquad J^z(0)=-N_J/2.
\label{eq:app_mf_initial_condition}
\end{equation}
Operationally, we select this branch by integrating the mean-field equations,
Eq.~\eqref{eq:app_mf_eom}, to long times and extracting the resulting
steady-state spin projection. This procedure selects the branch dynamically
connected to the protocol initialization, rather than an arbitrary solution of
the quartic equation.

\subsection{Comparison of resonance conditions}

Figure~\ref{fig:app_mf_resonance_comparison} compares the steady-state
normalized spin projection
\(\langle \hat J^z\rangle_{\rm ss}/(N_J/2)\) in the
\((\Delta_c/(N\Gamma_0),\alpha_{\rm in}/\alpha_{\rm ref})\) plane for the
three resonance conditions used in the mean-field analysis. The columns
correspond to different resonance choices, while the row labels distinguish
the UVRS and RVRS regimes.

The comparison shows that the three resonance conditions lead to qualitatively distinct steady-state responses. When \(\omega_d=\omega_a\), the spin detuning vanishes, \(\delta_\sigma=0\), and the polarized branch terminates at \(\alpha_{\rm in}/\alpha_{\rm ref}=1\), as shown in Fig.~\ref{fig:app_mf_resonance_comparison}(b,e). When \(\omega_d=\omega_c\), the response remains single-valued and smooth, as shown in Fig.~\ref{fig:app_mf_resonance_comparison}(c,f). The case used in the main text is \(\omega_a=\omega_c\), shown in Fig.~\ref{fig:app_mf_resonance_comparison}(a,d). In this case, the collective dipole and the cavity-induced contribution can compete, allowing a multivalued mean-field response. This effect is weak in UVRS, where cavity dissipation smooths the response, but becomes more pronounced in RVRS, where dynamical cavity backaction is stronger.

For this reason, the optimized squeezing protocols in the main text are chosen on the low-photon, strongly spin-polarized branch and avoid the branch-switching region. The mean-field analysis therefore provides a practical guide for selecting operating points where the effective OAT description remains accurate while the cavity is still dynamically active.

\section{Holstein--Primakoff approximation}
\label{app:hp_nested_elimination}

This appendix presents the derivations of the Holstein--Primakoff equations of Sec.~\ref{sec:HP_effective}. 

\subsection{Rotating frame and fluctuation basis}

We define the sector fractions
\begin{equation}
f_J=\frac{N_J}{N},
\qquad
f_\uparrow=1-f_J.
\label{eq:app_hp_fractions}
\end{equation}
In the spin-polarized phase the optical spin in the driven \(\{|{\downarrow}\rangle,|e\rangle\}\) manifold has a common phase rotation. We remove it by working in a frame rotating at
\begin{equation}
\omega_B=\frac{\delta_\sigma}{2\cos\tilde\theta_J},
\qquad
\hat w_{\omega_B}(t)=e^{i\omega_B t}\hat w(t),
\quad \hat w\in\{\hat d,\hat e\},
\label{eq:app_hp_omegaB}
\end{equation}
so that the mean-field amplitudes are time independent. Here \(\hat d^\dagger\), \(\hat e^\dagger\), and \(\hat u^\dagger\) create collective excitations in \(|\downarrow\rangle\), \(|e\rangle\), and \(|\uparrow\rangle\), respectively, and \(\hat J^- = \hat d^\dagger\hat e\).

The HP basis is chosen so that \(\hat c\) is the macroscopically occupied mean-field mode, \(\hat s\) is the metrological mode orthogonal to \(\hat c\) in the encoded \(S\)-spin subspace, and \(\hat\jmath\) is the fast mode orthogonal to \(\hat c\) in the driven \(J\)-spin subspace:
\begin{equation}
\resizebox{\columnwidth}{!}{$
\begin{pmatrix}
\hat c^\dagger \\
\hat s^\dagger \\
\hat\jmath^\dagger
\end{pmatrix}
=
\begin{pmatrix}
\sqrt{f_J}\cos\frac{\tilde\theta_J}{2} &
\sqrt{f_J}e^{-i\tilde\phi_J}\sin\frac{\tilde\theta_J}{2} &
\sqrt{f_\uparrow} \\[4pt]
-\sqrt{f_\uparrow}\cos\frac{\tilde\theta_J}{2} &
-\sqrt{f_\uparrow}e^{-i\tilde\phi_J}\sin\frac{\tilde\theta_J}{2} &
\sqrt{f_J} \\[4pt]
\sin\frac{\tilde\theta_J}{2} &
-e^{-i\tilde\phi_J}\cos\frac{\tilde\theta_J}{2} &
0
\end{pmatrix}
\begin{pmatrix}
\hat d_{\omega_B}^\dagger \\
\hat e_{\omega_B}^\dagger \\
\hat u^\dagger
\end{pmatrix}$}
\label{eq:app_hp_basis}
\end{equation}
The HP approximation assumes
\begin{equation}
\langle \hat c^\dagger\hat c\rangle\simeq N,
\qquad
\langle \hat s^\dagger\hat s+\hat\jmath^\dagger\hat\jmath\rangle\ll N,
\label{eq:app_hp_dilute}
\end{equation}
so that
\begin{equation}
\hat c=\sqrt{N-(\hat s^\dagger\hat s+\hat\jmath^\dagger\hat\jmath)}
\simeq
\sqrt N-\frac{\hat s^\dagger\hat s+\hat\jmath^\dagger\hat\jmath}{2\sqrt N}.
\label{eq:app_hp_c_expand}
\end{equation}
We also use the quadratures
\begin{equation}
\hat x_r=\frac{\hat r+\hat r^\dagger}{\sqrt2},
\qquad
\hat p_r=\frac{\hat r-\hat r^\dagger}{i\sqrt2},
\qquad r\in\{s,\jmath\}.
\label{eq:app_hp_quadratures}
\end{equation}

\subsection{Quadratic HP Hamiltonian and cavity elimination}

Displacing the cavity field as \(\hat a=\alpha+\hat a'\), with \(\alpha\) the mean-field cavity amplitude, and keeping terms up to quadratic order in \(\hat s\), \(\hat\jmath\), and \(\hat a'\) gives
\begin{equation}
\begin{aligned}
\hat H_{\rm HP}
&\simeq g\sqrt{\!\frac{N_J}{2}\!}
\Bigg[\!
 e^{i\tilde\phi_J}
\left(\!
 i\hat p_{\jmath}
 -\cos\tilde\theta_J\hat x_{\jmath}
 -\sqrt{f_\uparrow}\sin\tilde\theta_J\hat x_s\!
\right)\hat a'
\!+\!\mathrm{h.c.}\!\Bigg]\\
&\quad-\Delta_c\hat a'^\dagger\hat a'
-\delta_\sigma\sec\tilde\theta_J\hat\jmath^\dagger\hat\jmath .
\end{aligned}
\label{eq:app_hp_H_quad}
\end{equation}
with cavity jump operator \(\hat\ell_{\rm HP}=\hat a'\). Setting \(\partial_t\hat a'\simeq0\) gives
\begin{equation}
\hat a'_{\rm AE}
=-i\frac{g\sqrt{N_J}}{2}
\frac{e^{-i\tilde\phi_J}\hat Q}{\kappa/2-i\Delta_c},
\label{eq:app_hp_a_AE}
\end{equation}
where
\begin{equation}
\hat Q=
\hat\jmath^\dagger-\hat\jmath
-\cos\tilde\theta_J(\hat\jmath+\hat\jmath^\dagger)
-\sqrt{f_\uparrow}\sin\tilde\theta_J(\hat s^\dagger+\hat s).
\label{eq:app_hp_Q}
\end{equation}
After eliminating \(\hat a'\), the intermediate two-mode Hamiltonian is
\begin{equation}
\hat H'_{{\rm HP}+{\rm AE}_a}
=\frac{g^2N_J\Delta_c}{\kappa^2+4\Delta_c^2}\hat Q^2
-\delta_\sigma\sec\tilde\theta_J\hat\jmath^\dagger\hat\jmath,
\label{eq:app_hp_H_after_a}
\end{equation}
with effective jump operator
\begin{equation}
\hat\ell_{{\rm HP}+{\rm AE}_a}
=g\sqrt{\frac{N_J}{2}}\sqrt\kappa\frac{
\cos\tilde\theta_J\hat x_{\jmath}+i\hat p_{\jmath}
+\sqrt{f_\uparrow}\sin\tilde\theta_J\hat x_s
}{\kappa/2-i\Delta_c}.
\label{eq:app_hp_l_after_a}
\end{equation}
Eqs.~\eqref{eq:app_hp_H_after_a} and \eqref{eq:app_hp_l_after_a} are the intermediate Hamiltonian and jump operator referenced in the main text.

\subsection{Elimination of the fast driven-manifold mode}

The remaining mode \(\hat\jmath\) relaxes rapidly towards a steady-state behavior relative to the squeezing dynamics in $\hat s$ and can likewise be adiabatically eliminated. Setting \(\partial_t\hat\jmath\simeq0\) gives
\begin{equation}
\hat\jmath=V_j(\hat s+\hat s^\dagger),
\qquad
\hat\jmath^\dagger=V_j^*(\hat s+\hat s^\dagger),
\label{eq:app_hp_j_slaving}
\end{equation}
with
\begin{equation}
V_j=\frac{c_ja_j^*-c_j^*b_j}{|a_j|^2-|b_j|^2},
\label{eq:app_hp_Vj}
\end{equation}
where
\begin{subequations}
\label{eq:app_hp_abcj}
\begin{align}
c_j&=\frac{g^2N_J\sin\tilde\theta_J\sqrt{f_\uparrow}}{\kappa^2+4\Delta_c^2}
\left(\kappa+2i\Delta_c\cos\tilde\theta_J\right),
\label{eq:app_hp_cj}\\
b_j&=\frac{2ig^2N_J\Delta_c\sin^2\tilde\theta_J}{\kappa^2+4\Delta_c^2},
\label{eq:app_hp_bj}\\
a_j&=i\delta_\sigma\sec\tilde\theta_J \\
&\quad
-\frac{2ig^2N_J\Delta_c(1+\cos^2\tilde\theta_J)}
{\kappa^2+4\Delta_c^2} \\
&\quad
-\frac{2\kappa g^2N_J\cos\tilde\theta_J}
{\kappa^2+4\Delta_c^2} .
\label{eq:app_hp_aj}
\end{align}
\end{subequations}
The denominator can be written as
\begin{equation}
|a_j|^2-|b_j|^2=\frac{\mathcal N_j}{(\kappa^2+4\Delta_c^2)^2},
\label{eq:app_hp_denom}
\end{equation}
with
\begin{equation}
\begin{aligned}
\mathcal N_j
&=\left[(\kappa^2+4\Delta_c^2)\delta_\sigma\sec\tilde\theta_J
-2g^2N_J\Delta_c(1+\cos^2\tilde\theta_J)\right]^2\\
&\quad+4g^4N_J^2\left(\kappa^2\cos^2\tilde\theta_J
-\Delta_c^2\sin^4\tilde\theta_J\right).
\end{aligned}
\label{eq:app_hp_Nj}
\end{equation}
The combination that controls both the coherent and dissipative pieces is
\begin{equation}
\mathcal A_j=
\left(\cos\tilde\theta_J\,\mathrm{Re}\,V_j
+\frac{\sqrt{f_\uparrow}}{2}\sin\tilde\theta_J\right)^2
+(\mathrm{Im}\,V_j)^2.
\label{eq:app_hp_Aj}
\end{equation}
Substitution of Eq.~\eqref{eq:app_hp_j_slaving} into Eq.~\eqref{eq:app_hp_H_after_a} gives the final single-mode quadratic Hamiltonian
\begin{equation}
\hat H'_{{\rm HP}+{\rm AE}_a{\rm AE}_j}
=\left[
\frac{4g^2N_J\Delta_c}{\kappa^2+4\Delta_c^2}\mathcal A_j
-\delta_\sigma\sec\tilde\theta_J|V_j|^2
\right](\hat s+\hat s^\dagger)^2,
\label{eq:app_hp_H_final_s}
\end{equation}
and the final effective jump operator
\begin{equation}
\begin{aligned}
\hat\ell_{{\rm HP}+{\rm AE}_a}^{\rm AE_j}
&=-\frac{2g\sqrt{N_J\kappa}}
{\sqrt{\kappa^2+4\Delta_c^2}}
\Bigg[
\cos\tilde\theta_J\,\mathrm{Re}\,V_j \\
&\qquad
+\frac{\sqrt{f_\uparrow}}{2}\sin\tilde\theta_J
+i\,\mathrm{Im}\,V_j
\Bigg](\hat s+\hat s^\dagger).
\end{aligned}
\label{eq:app_hp_l_final_s}
\end{equation}

\subsection{Collective-spin model}

Since the metrological quadrature is related to the encoded collective spin by
\begin{equation}
\hat x_s=\frac{\hat s+\hat s^\dagger}{\sqrt2}
\approx \sqrt{\frac{2}{N}}\,\hat S_z,
\label{eq:app_hp_xs_Sz}
\end{equation}
Eqs.~\eqref{eq:app_hp_H_final_s} and \eqref{eq:app_hp_l_final_s} reduce to the effective OAT master equation
\begin{equation}
\partial_t\hat\rho
=-i[\tilde\chi\hat S_z^2,\hat\rho]
+\tilde\Gamma\mathcal D[\hat S_z]\hat\rho.
\label{eq:app_hp_OAT_master}
\end{equation}
The exact coefficients obtained from the nested elimination are
\begin{subequations}
\label{eq:app_hp_exact_coefficients}
\begin{align}
\tilde\chi
&=\frac{4}{N}
\left[
\frac{4g^2N_J\Delta_c}{\kappa^2+4\Delta_c^2}\mathcal A_j
-\delta_\sigma\sec\tilde\theta_J|V_j|^2
\right],
\label{eq:app_hp_chi_exact}\\
\tilde\Gamma
&=\frac{16f_Jg^2\kappa}{\kappa^2+4\Delta_c^2}\mathcal A_j.
\label{eq:app_hp_Gamma_exact}
\end{align}
\end{subequations}
These are the full expressions referred to in the main text before taking the small-detuning limit.

In the regime \(|\delta_\sigma|,|\Delta_c|\ll\{g\sqrt{N_J},\kappa\}\), and keeping terms through total degree three in the detunings, Eq.~\eqref{eq:app_hp_exact_coefficients} becomes
\begin{subequations}
\label{eq:app_hp_small_detuning}
\begin{align}
\tilde\chi
&\approx
-\frac{f_\uparrow f_J}{N_J}\tan^2\tilde\theta_J\sec\tilde\theta_J\,\delta_\sigma
-\frac{f_\uparrow f_J}{g^2N_J^2}\tan^2\tilde\theta_J\sec^4\tilde\theta_J\,\delta_\sigma^2\Delta_c
\notag\\
&\quad+
\frac{f_\uparrow f_J\kappa^2}{4g^4N_J^3}\tan^2\tilde\theta_J\sec^5\tilde\theta_J\,\delta_\sigma^3
+O\!\left((\delta_\sigma,\Delta_c)^4\right),
\label{eq:app_hp_chi_small}\\
\tilde\Gamma
&\approx
\frac{f_Jf_\uparrow\kappa}{g^2N_J^2}
\tan^2\tilde\theta_J\sec^4\tilde\theta_J\,\delta_\sigma^2
+O\!\left((\delta_\sigma,\Delta_c)^4\right).
\label{eq:app_hp_Gamma_small}
\end{align}
\end{subequations}
The effect of $\Delta_c$ only appears at higher orders compared to $\delta_\sigma$. The leading coherent rate is linear in \(\delta_\sigma\), while the collective dephasing rate is quadratic in \(\delta_\sigma\). This is the origin of the large coherent-to-dissipative ratio \(|\tilde\chi|/\tilde\Gamma\) at small detuning quoted in the main text. 

This effective description is accurate only while the system remains on the low-photon, strongly spin-polarized branch and the eliminated modes \(\hat a'\) and \(\hat\jmath\) remain fast compared with the metrological dynamics of \(\hat s\).

\section{Dissipative TWA and Covariance matrix}
\label{app:disstwa_covariance}

This appendix defines the DissTWA covariance matrix used in Sec.~\ref{sec:Spin_squeezing_dyn} and explains how the same trajectory ensemble yields both the multilevel squeezing parameter and the residual spin--cavity correlation diagnostic.

The simulations evolve the full three-level atom--cavity phase space rather than an already-eliminated spin-only model. The starting master equation is
\begin{subequations}
\label{eq:app_disstwa_master}
\begin{gather}
\partial_t\hat\rho=-i[\hat H,\hat\rho]+\frac{\kappa}{2}\mathcal L_{\hat a}[\hat\rho],\\
\hat H=g(\hat J^+\hat a+\hat J^-\hat a^\dagger)
-i\sqrt\kappa\alpha_{\rm in}(\hat a-\hat a^\dagger)
-\Delta_c\hat a^\dagger\hat a
-\delta_\sigma\hat J^z,
\end{gather}
\end{subequations}
where \(\mathcal L_{\hat O}[\hat\rho]=2\hat O\hat\rho\hat O^\dagger-\{\hat O^\dagger\hat O,\hat\rho\}\). The atomic initial state is sampled using a generalized discrete truncated Wigner representation for the three-level local Hilbert space. A convenient local operator basis is formed by the six coherences
\begin{subequations}
\label{eq:app_disstwa_su3_coherences}
\begin{align}
\hat\lambda_{\alpha\beta,x}^{(i)}
&=\frac{|\alpha_i\rangle\langle\beta_i|+|\beta_i\rangle\langle\alpha_i|}{\sqrt2},
\label{eq:app_disstwa_lambda_x}\\
\hat\lambda_{\alpha\beta,y}^{(i)}
&=\frac{|\alpha_i\rangle\langle\beta_i|-|\beta_i\rangle\langle\alpha_i|}{\sqrt2 i},
\label{eq:app_disstwa_lambda_y}
\end{align}
\end{subequations}
for \((\alpha,\beta)\in\{(\downarrow,\uparrow),(\downarrow,e),(e,\uparrow)\}\), together with two independent diagonal generators, for example
\begin{equation}
\begin{aligned}
\hat\lambda_3^{(i)}&=|\downarrow_i\rangle\langle\downarrow_i|-|\uparrow_i\rangle\langle\uparrow_i|,\\
\hat\lambda_8^{(i)}&=\frac{|\downarrow_i\rangle\langle\downarrow_i|+|\uparrow_i\rangle\langle\uparrow_i|-2|e_i\rangle\langle e_i|}{\sqrt3}.
\end{aligned}
\label{eq:app_disstwa_diag_generators}
\end{equation}
For an initial product state, the local phase-space distribution factorizes over atoms. The sampled local variables are summed to form collective variables,
\begin{equation}
\Lambda_\mu=\sum_{i=1}^N\lambda_\mu^{(i)}.
\label{eq:app_disstwa_collective_vars}
\end{equation}
The cavity vacuum is sampled with the continuous Wigner function
\begin{equation}
\begin{gathered}
W(x_a,p_a)=\frac{1}{\pi}e^{-x_a^2-p_a^2},\\[0.25em]
\hat x_a=\frac{\hat a+\hat a^\dagger}{\sqrt2},
\qquad
\hat p_a=\frac{\hat a-\hat a^\dagger}{i\sqrt2}.
\end{gathered}
\label{eq:app_disstwa_cavity_sampling}
\end{equation}
The cavity-loss channel gives the It\^o increment for the complex cavity amplitude
\begin{equation}
\begin{aligned}
da\big|_{\rm loss}
&=-\frac{\kappa}{2}a\,dt
+\frac{\sqrt\kappa}{2}(dW_1+i\,dW_2),\\
\overline{dW_i dW_j}&=\delta_{ij}dt.
\end{aligned}
\label{eq:app_disstwa_cavity_sde}
\end{equation}
The deterministic evolution is generated by Eq.~\eqref{eq:app_disstwa_master}.

Trajectory averages give symmetrically ordered moments,
\begin{subequations}
\label{eq:app_disstwa_moment_rules}
\begin{align}
\langle\hat\Lambda_a(t)\rangle&\simeq\overline{\Lambda_a(t)},
\label{eq:app_disstwa_mean_rule}\\
\frac12\langle\{\hat\Lambda_a(t),\hat\Lambda_b(t)\}\rangle&\simeq\overline{\Lambda_a(t)\Lambda_b(t)},
\label{eq:app_disstwa_cov_rule}
\end{align}
\end{subequations}
where the overbar denotes the Monte Carlo average. By determining the averaged single-body density matrix, we may construct the instantaneous orthonormal basis \(\{|c(t)\rangle,|j(t)\rangle,|s(t)\rangle\}\), where \(|c(t)\rangle\) is the instantaneous mean single-particle state and \(|j(t)\rangle,|s(t)\rangle\) span the orthogonal fluctuation subspace, defined analogously to the related Holstein-Primakoff bosons. The collective fluctuation operators are
\begin{subequations}
\label{eq:app_disstwa_XY_ops}
\begin{align}
\hat X_\mu(t)&=\sum_{i=1}^N
\frac{|c_i(t)\rangle\langle\mu_i(t)|+|\mu_i(t)\rangle\langle c_i(t)|}{2},
\label{eq:app_disstwa_X}\\
\hat Y_\mu(t)&=\sum_{i=1}^N
\frac{|c_i(t)\rangle\langle\mu_i(t)|-|\mu_i(t)\rangle\langle c_i(t)|}{2i},
\qquad \mu\in\{j,s\}.
\label{eq:app_disstwa_Y}
\end{align}
\end{subequations}
Using the displaced cavity fluctuation \(\hat a'(t)=\hat a(t)-\langle\hat a(t)\rangle\), define
\begin{equation}
\hat x_{a'}=\frac{\hat a'+\hat a'^\dagger}{\sqrt2},
\qquad
\hat p_{a'}=\frac{\hat a'-\hat a'^\dagger}{i\sqrt2},
\label{eq:app_disstwa_cav_quad}
\end{equation}
and assemble the full fluctuation vector
\begin{equation}
\hat{\bm R}(t)=
\left(
\hat X_j,\hat Y_j,\hat X_s,\hat Y_s,
\hat x_{a'},\hat p_{a'}
\right)^{\mathsf T}.
\label{eq:app_disstwa_Rvec}
\end{equation}
The symmetrized covariance matrix is
\begin{equation}
\begin{aligned}
\Sigma_{\mu\nu}(t)
&=\frac12
\left\langle
\{\Delta\hat R_\mu(t),\Delta\hat R_\nu(t)\}
\right\rangle,\\
\Delta\hat R_\mu
&=\hat R_\mu-\langle\hat R_\mu\rangle .
\end{aligned}
\label{eq:app_disstwa_covariance}
\end{equation}
It has the block structure
\begin{equation}
\Sigma(t)=
\begin{pmatrix}
\Sigma_{\rm spin}(t)&\Sigma_{S:C}(t)\\
\Sigma_{S:C}^{\mathsf T}(t)&\Sigma_{\rm cav}(t)
\end{pmatrix}.
\label{eq:app_disstwa_cov_blocks}
\end{equation}
The spin block \(\Sigma_{\rm spin}\), associated with \((\hat X_j,\hat Y_j,\hat X_s,\hat Y_s)\), is the block used to compute the multilevel squeezing parameter
\begin{equation}
\begin{aligned}
\xi^2(t)
&=\lambda_{\min}
\left[
\frac{N\,\Sigma_{\rm spin}(t)}
{\langle\hat N_{c/2}(t)\rangle^2}
\right],\\
\hat N_{c/2}(t)
&=\sum_{i=1}^N
\frac{|c_i(t)\rangle\langle c_i(t)|}{2} .
\end{aligned}
\label{eq:app_disstwa_xi2}
\end{equation}

\section{Single-particle dephasing regimes}
\label{sec:intermediate_dephasing_regime}

The regimes summarized in Table~\ref{tab:regime_summary} follow from the
time-dependent squeezing estimate
\begin{equation}
\xi^{2}(t)\approx e^{\gamma_d t}\left[
\frac{e^{\gamma_d t}+N\tilde{\Gamma}t}{N^{2}\tilde{\chi}^{2}t^{2}}
+\frac{N^{2}\tilde{\chi}^{4}t^{4}}{6}
\right].
\label{eq:Spin_squeez_deph_recall}
\end{equation}
The first term describes the shearing contribution dressed by the loss of
single-particle contrast, the second term is the collective-dephasing penalty,
and the last term is the leading curvature correction of the OAT dynamics.
Regime~(I) is the familiar short-time OAT limit, where both
\(\gamma_d t_{\rm opt}\) and \(N\tilde{\Gamma}t_{\rm opt}\) are perturbative.
Regime~(II) is the standard contrast-limited limit, where the local-dephasing
time \(1/\gamma_d\) cuts off the evolution before the curvature-limited OAT
optimum can be reached, while collective dephasing remains subleading.

The additional feature of Table~\ref{tab:regime_summary} is Regime~(III). This regime
should not be interpreted as a new ideal-OAT scaling. Rather, it is an
intermediate mixed-dephasing sector in which the optimum is still controlled by
the local coherence time,
\begin{equation}
\gamma_d t_{\rm opt}=O(1),
\end{equation}
but the collective-dephasing term is no longer a perturbative correction. In
other words, the local dephasing rate \(\gamma_d\) fixes the available time
window, while \(\tilde{\Gamma}\) sets an additional floor for the optimized
squeezing within that window.

To make this structure explicit, define the accumulated local-dephasing variable
\begin{equation}
w\equiv \gamma_d t,
\end{equation}
and introduce
\begin{equation}
A\equiv \frac{3^{2/3}}{2N^{2/3}},
\qquad
D\equiv \frac{\gamma_d\tilde{\Gamma}}{N|\tilde{\chi}|^{2}} .
\label{eq:AD_defs_intermediate}
\end{equation}
Here \(A\) is the ideal OAT curvature-limited squeezing floor, while \(D\)
measures the collective-dephasing penalty evaluated on the local-dephasing
timescale. Near the crossover surface where the local-dephasing time is
comparable to the curvature-limited OAT time, the relevant optimized envelope is
\begin{equation}
\xi_{\star}^{2}(w)
=
A e^{5w/3}
+
D\frac{e^{w}}{w}.
\label{eq:xi_star_intermediate}
\end{equation}
The first term is the OAT curvature contribution dressed by local contrast
loss, while the second term is the locally dressed collective-dephasing
contribution.

The stationarity condition for Eq.~\eqref{eq:xi_star_intermediate} is
\begin{equation}
\frac{1-w_{\star}}{w_{\star}^{2}}
=
\frac{5A}{3D}\,e^{2w_{\star}/3}.
\label{eq:wstar_intermediate}
\end{equation}
This equation shows why Regime~(III) is intermediate. Since the right-hand side
is positive, the optimum satisfies
\begin{equation}
w_{\star}<1,
\end{equation}
so the collective-dephasing penalty shifts the optimum slightly earlier than the
naive local-dephasing time \(1/\gamma_d\). Nevertheless, when the collective
penalty is comparable to or larger than the OAT floor, \(D\gtrsim A\), the
optimum remains parametrically close to \(w=O(1)\). Thus
\begin{equation}
t_{\rm opt}^{(III)}\simeq \frac{w_{\star}}{\gamma_d}
\sim \frac{1}{\gamma_d}.
\end{equation}

Evaluating Eq.~\eqref{eq:xi_star_intermediate} near \(w_{\star}\sim 1\) gives
\begin{equation}
\xi_{\rm opt,III}^{2}
\approx
A e^{5/3}
+
eD,
\end{equation}
or equivalently
\begin{equation}
\xi_{\rm opt,III}^{2}
\approx
\frac{3^{2/3}e^{5/3}}{2N^{2/3}}
+
e\,\frac{\gamma_d\tilde{\Gamma}}{N|\tilde{\chi}|^{2}} .
\label{eq:xi_regimeIII_intermediate}
\end{equation}
The validity condition for this mixed regime is therefore
\begin{equation}
D\gtrsim A,
\qquad
\text{i.e.}
\qquad
\frac{\gamma_d\tilde{\Gamma}}{N|\tilde{\chi}|^{2}}
\gtrsim
\frac{3^{2/3}}{2N^{2/3}}.
\label{eq:regimeIII_condition_intermediate}
\end{equation}

This result clarifies the distinction between Regimes~(II) and~(III). In both
cases the usable squeezing time is set by the local-dephasing scale
\(1/\gamma_d\). However, in Regime~(II) the collective-dephasing contribution is
parametrically smaller than the contrast-limited squeezing floor, so the
optimized squeezing is given by $\xi_{\rm opt,II}^{2}
\simeq
e^{2}\left(\frac{\gamma_d}{N|\tilde{\chi}|}\right)^{2}$.

In Regime~(III), by contrast, the collective-dephasing channel has already
accumulated enough noise on the local-dephasing timescale to contribute at
leading order. The optimized squeezing is therefore not determined by local
contrast loss alone, but by the sum of a locally dressed OAT floor and a locally
dressed collective-dephasing floor, as in
Eq.~\eqref{eq:xi_regimeIII_intermediate}.

\bibliography{StrongSqueezing,Extra}

\end{document}